\documentclass[12pt]{article}
\usepackage{amsmath,amssymb,amsfonts}
\usepackage[dvips]{color}
\usepackage{xcolor}

\usepackage{graphicx}
\usepackage{caption}
\usepackage[longnamesfirst]{natbib}
\usepackage{rotating}
\usepackage{afterpage}
\usepackage{cancel}
\usepackage{booktabs}
\usepackage{setspace}
\usepackage{enumitem}
\usepackage{mathabx}
\usepackage{footnote}
\newcommand{\fra}[2]{\frac{\displaystyle #1}{\displaystyle #2}}
\newcommand{\bc}{\begin{center}}
\newcommand{\ec}{\end{center}}
\newcommand{\be}{\begin{equation}}
\newcommand{\ee}{\end{equation}}
\newcommand{\bea}{\begin{eqnarray}}
\newcommand{\eea}{\end{eqnarray}}
\newcommand{\bean}{\begin{eqnarray*}}
\newcommand{\eean}{\end{eqnarray*}}
\newcommand{\bt}{\begin{tabular}}
\newcommand{\et}{\end{tabular}}
\usepackage{url}  
\usepackage{xcolor}
\usepackage[margin=1in]{geometry}
\usepackage{graphicx}
\newtheorem{theorem}{Theorem}

\newtheorem{algorithm}[theorem]{Algorithm}

\newtheorem{algo}{Algorithm}

\numberwithin{theorem}{section}

\newcommand{\diag}{\operatorname*{diag}}

\newcounter{saveeqn}

\usepackage{hyperref}

\begin{document}

\title{\bf Bayesian Estimation of Panel Models under Potentially Sparse Heterogeneity}

\author{
        Hyungsik Roger Moon\\ {\small {\em University of Southern California,}} \\[-1ex]
        {\small{\em Yonsei University}}
        \and
        Frank Schorfheide\thanks{Correspondence:
                H.R. Moon: Department of Economics, Univ. of Southern California, KAP 300, Los Angeles, CA
                90089. E-mail: moonr@usc.edu. F. Schorfheide: Department of Economics, Perelman Center for Political Science and Economics, University of Pennsylvania, 133 S. 36th St.,  Philadelphia, PA 19104-6297. Email:
                schorf@ssc.upenn.edu. B. Zhang: Amazon.com, Seattle, WA. Email: zhang.boyuan@hotmail.com. We thank Dirk Kr\"uger, V\'ictor R\'ios Rull, and participants of the Wharton Macro Lunch for helpful comments and suggestions. Moon and Schorfheide gratefully acknowledge financial support from the National Science Foundation under Grants SES 1625586 and SES 1851634, respectively. This paper and its contents are not related to Amazon and do not reflect the position of the company and its subsidiaries.} \\
        {\small{\em Univ. of Pennsylvania}} \\[-1ex] {\small{\em CEPR, NBER, and PIER}}
        \and
        Boyuan Zhang\\
        {\small {\em Amazon.com}}
    }

\date{Octomber 24, 2023}
\maketitle


\begin{abstract}
We incorporate a version of a spike and slab prior, comprising a pointmass at zero (``spike'') and a Normal distribution around zero (``slab'') into a dynamic panel data framework to model coefficient heterogeneity. In addition to homogeneity and full heterogeneity, our specification can also capture sparse heterogeneity, that is, there is a core group of units that share common parameters and a set of deviators with idiosyncratic parameters. We fit a model with unobserved components to income data from the Panel Study of Income Dynamics. We find evidence for sparse heterogeneity for balanced panels composed of individuals with long employment histories.
\end{abstract}

\noindent JEL CLASSIFICATION: C11, C23, C53, E20

\noindent KEY\ WORDS: Bayesian Analysis, Forecasting, Income Dynamics, Panel Data Models, Sparsity, Spike-and-Slab Priors

\thispagestyle{empty}
\setcounter{page}{0}
\newpage

\section{Introduction}
\label{sec:intro}

Panel data models feature two types of coefficients: homogeneous coefficients that are common among all units $i=1,\ldots,N$, and heterogeneous coefficients that differ across units. In recent years, there has been a growing literature that focuses on the estimation of heterogeneous coefficients, which is challenging in environments in which the time series dimension is small. A natural way of adding information to the estimation of unit-specific coefficients is the use of prior distributions. The key insight in panel data applications is that one can extract information from the cross-section and equate the prior distribution with the (unobserved) cross-sectional distribution of unit-specific coefficients. The time series information for each unit can then be combined with the prior distribution, to generate more precise estimates of the heterogeneous coefficients. While in many models the cross-sectional coefficient distribution is in principle non-parametrically identified, in practice it is often desirable to impose parametric restrictions. In this paper, we consider a particular restriction: we assume that there is a core group of units that share a common parameter value, whereas the remaining units (deviators) have idiosyncratic values. If the core group contains a large number of units, then heterogeneity is sparse.


We assume that the heterogeneous coefficients are distributed according to a version of a spike-and-slab (S\&S) prior: with probability $1-q$ the coefficient for unit $i$ equals the predominant value (spike), and with probability $q$ it deviates from the predominant value. This setup nests two important special cases: coefficient homogeneity ($q=0$) and full heterogeneity ($q=1$). We will say that heterogeneity is sparse if $q$ is close to zero and dense if $q$ is close to one. The probability $q$ and the parameters characterizing the distribution of the deviations are hyperparameters that  are estimated from the panel data. Whether the concept of sparse heterogeneity is useful in practice, is an empirical question. We use the proposed prior distribution to estimate a dynamic model for labor earnings obtained from the Panel Study of Income Dynamics (PSID). We construct two types of samples from the PSID data: balanced panels that include individuals who have been working uninterruptedly for a long period of time, and an unbalanced panel that includes individuals with a variety of earnings and employment histories. The balanced panels do feature sparse heterogeneity, whereas the more diverse unbalanced panel is better described by full heterogeneity.


The paper makes the following contributions: first, we incorporate a version of an S\&S prior, comprising a pointmass at zero (``spike'') and a Normal distribution around zero (``slab'') into a dynamic panel data framework to model coefficient heterogeneity. Such a prior has been used by \cite{Geweke1996} and, more recently, \cite{GiannoneLenzaPrimiceri2018}, in the context of regressor selection problems. We extend the prior to handle heterogeneous variance parameters and show how to implement posterior inference using a Gibbs sampler. 
Second, we conduct a Monte Carlo study that compares the point estimation accuracy under the proposed (S\&S) prior to that obtained by (incorrectly) imposing coefficient homogeneity ($q=0$) or full heterogeneity ($q=1$). If $q$ is close to zero or one, then estimates under the S\&S prior are about as accurate as the estimates that impose the nearly-correct restriction. If $q$ is between 0.2 and 0.8 the estimates under the S\&S prior clearly dominate the alternatives. Third, we show in our application that the concept of sparse heterogeneity is empirically useful.

In the empirical application we estimate income processes based on earnings data from the PSID. There is a long-standing debate in the literature whether earnings profiles are homogeneous or heterogeneous across individuals. \cite{Guvenen2007} labeled these specifications RIP (restricted income profiles) and HIP (heterogeneous income profiles), respectively. We offer a third option in our setup: sparsely heterogeneous income profiles, which could be abbreviated as SHIP. The literature on the estimation of idiosyncratic income processes has considered many different specifications. There are a few features that all of them share: the income process typically takes the form of an unobserved components model with deterministic and stochastic components. We consider three components: a linear function in experience, an autoregressive component with AR coefficient $\rho_i$, and an $iid$ component.\footnote{The unobserved components model is fitted to residuals from a regression of earnings on time fixed effects and some demographic characteristics.} There is typically time variation and cross-sectional variation in the income risk, i.e., the stochastic components. In our model we allow for time specific variance coefficients that also vary across individuals. While most of the literature assumes that the autocorrelation parameter of the persistent income component is common across units, we allow for heterogeneity. 

The evidence for sparse heterogeneity depends on the construction of the panel data set that is used in the estimation. Using relatively small balanced samples, our posterior estimates imply that heterogeneity with respect to returns to experience and the autocorrelation of the persistent stochastic income component is sparse at best, with large core groups of individuals sharing identical coefficients and a small number of deviators. These estimates can be interpreted as evidence against a HIP version of the model. We also generate one-step-ahead forecasts for the entire cross-section of individuals using our baseline specification, a RIP specification, and a HIP specification. Using log predictive scores, the HIP version is clearly dominated, and the performance of our sparsely heterogeneous baseline specification is for most samples identical to that of the more restrictive RIP specification. Repeating the estimation on a large unbalanced panel, the parameter estimates lead to an almost fully heterogeneous specification that does not feature core groups of units with identical parameter values, which is evidence in favor of the HIP conjecture. In terms of one-step-ahead out-of-sample forecasting performance, however, even for the unbalanced panel the more parsimonious RIP specification leads to slightly better forecasts on average.

Third, we examine unit-level interval forecasts conditional on information from the full panel $Y_{1:N,0:T}$ accounting for shock and parameter uncertainty, the full panel ignoring parameter uncertainty, and the individual history $Y_{i,0:T}$ accounting for shock and parameter uncertainty, distinguishing between core group members and deviators. This comparison has an important message for the macroeconomic life-cycle model literature: subtle assumptions about the agents information set can have large effects on the income uncertainty that agents face and hence on their precautionary savings motive. 

Our paper is related to several strands of literature. We are using a variant of an S\&S prior, which was originally proposed for regressor selection problem. The prior proposed by \cite{MitchellBeauchamp1988} consisted of a point mass at zero (``spike'') and a uniform prior (``slab''). For computational reasons \cite{GeorgeMcCulloch1993,GeorgeMcCulloch1997} replaced the point mass by a Normal distribution with very small variance, and the uniform distribution by a Normal with large variance. \cite{Geweke1996} combined a pointmass at zero with a Gaussian distribution, which is the specification that was recently used by \cite{GiannoneLenzaPrimiceri2018} to examine whether regression models in economics tend to be sparse or dense. The regressor selection problem can be viewed as a special case of the fundamental problem of estimating the vector of means of a multivariate Normal distribution. \cite{JohnstoneSilverman2004} examined the large sample properties of Bayesian posteriors under a prior that is a mixture of a pointmass at zero and a continuous distribution that has fatter tails than a Normal distribution. The authors are particularly interested in sparse settings with a large number of zeros. The asymptotic concentration of posterior distributions in this environment is also studied by \cite{CastillovanderVaart2012}, albeit under a somewhat different prior distribution.

Our paper is concerned with the estimation of potentially heterogeneous slopes and variance parameters in dynamic panel data models, which involves more complicated versions of the vector of means problem. While much of the panel data literature has traditionally focused on the estimation of the common or homogeneous parameter values, treating, for instance, heterogeneous intercepts as incidental nuisance parameters, the estimation of the heterogeneous coefficients can be important for treatment effect analysis or forecasting. In the context of our application, the prediction of idiosyncratic income profiles is essential for consumption and investment decisions and is at the core of every life-cycle model in macroeconomics. 
We are building on recent Bayesian approaches by \cite{ChamberlainHirano1999}, \cite{GuKoenker2014,GuKoenkerJAE2016}, \cite{LiuMoonSchorfheide2017}, \cite{Liu2023}, \cite{LiuMoonSchorfheide2018}. The key point of departure from the earlier work is the use of an S\&S type prior that lets us pick up patterns of either sparse or dense heterogeneity, bridging homogeneity and full heterogeneity.\footnote{The notion of group heterogeneity has received a lot of attention in the recent panel data literature, see \cite{BonhommeManresa2015} for the seminal paper and \cite{Zhang2023} for a Bayesian implementation that allows users to formulate priors about group memberships. Our paper considers a specification with one core group of units. In principle it could be extended to multiple core groups, but this extension is left for future research.}

There exists a large literature at the intersection of labor economics and macroeconomics on the estimation of idiosyncratic income processes. A longstanding debate in this literature is whether income profiles are homogeneous or heterogeneous. 
The heterogeneity typically refers the return to experience. Among others, \cite{MaCurdy1982}, \cite{AboundCard1989}, and \cite{Guvenen2007,Guvenen2009} provide evidence in favor of HIP, whereas \cite{LillardWeiss1979}, \cite{Hause1980}, \cite{Baker1997}, \cite{Haider2001}, \cite{Hryshko2012}, and \cite{Hoffmann2019} favor the homogeneous RIP specification, which often involves a unit-root or highly persistent stochastic component. 
The importance of heterogeneity in income profiles, persistence of income processes, and the variance of income shocks is also emphasized by \cite{BrowningEjrnjesAlvarez2010} and \cite{BrowningEjrnjes2013}. Coefficient heterogeneity in this literature is typically modeled through random effects. The latter two papers allow for these random effects to be correlated with each other, assuming they are determined by some underlying factors.
While we do not allow for conditional heteroskedasticity in our model specification as, for instance, in \cite{MeghirPistaferri2004} or \cite{Hospido2012}, we do allow for cross-sectional heterogeneity as in \cite{GuKoenker2014} and find it to be dense. 

Traditionally, the earnings dynamics literature focuses on the estimation of (correlated) random effects distributions, characterizing the degree of coefficient heterogeneity in the population. Our Bayesian estimation approach generates such estimates along with estimates of the unit-level coefficients, under the assumption of potentially sparse heterogeneity. Whether or not the interest lies in the random effects distribution or the unit-level coefficients depends on the particular application. One novel contribution of our application below, is to also assess the fitted income-dynamics model by evaluating the pseudo-out-of-sample forecast accuracy for the individuals included in the panel data set, rather than a hypothetical population implied by the estimated model.


The remainder of the paper is organized as follows. Our approach of capturing potentially sparse heterogeneity through an S\&S type prior distribution is described in the context of a vector of means estimation in Section~\ref{sec:sparse heterogeneity}. This prior is then incorporated into two linear panel data models with heterogeneous intercept, slope coefficients, and shock standard deviations in Section~\ref{sec:model}. Results from two Monte Carlo experiments are summarized in Section~\ref{sec:MC} and the empirical analysis is presented in Section~\ref{sec:empirics}. Finally, Section~\ref{sec:conclusion} concludes. Details on the posterior samplers for the Monte Carlo study and the empirical analysis are provided in the Online Appendix. The Appendix also contains additional empirical results. 

\section{Sparse Heterogeneity}
\label{sec:sparse heterogeneity}

Throughout this paper, we pursue a Bayesian approach, place a prior distribution over the parameters, and generate draws from the posterior distribution. In the context of estimating a vector of means of a multivariate Normal distribution, our approach to sparse heterogeneity takes the following form. Consider the model
\be
  y_i = \delta_i + u_i, \quad u_i \sim iid {\cal N}(0,1), \quad i=1,\ldots,N
\ee
equipped with a prior distribution that is a mixture of a pointmass at zero and a Normal distribution:
\be
\delta_i|(q,v_{\delta}) \sim \left\{ \begin{array}{ll} {\cal N} (0,v_{\delta}) & \mbox{with prob. } q \\
	0                                          & \mbox{with prob. } 1-q \end{array} \right. \label{eq:prior.delta.alpha}.
\ee
This prior distribution has been used by \cite{Geweke1996} and, more recently, by \cite{GiannoneLenzaPrimiceri2018} in the context of regressor selection problems. The sparsity of the $\delta = [\delta_1,\ldots,\delta_N]'$ vector is determined by the hyperparameter $q$, which controls the probability of the coefficient for unit $i$ deviating from the common value. Thus, $1-q$ is the probability assigned to the pointmass at zero and hence the height of the ``spike.'' The second hyperparameter, $v_{\delta}$, controls the width of the ``slab.'' The larger $v_{\delta}$, the larger deviations from the common parameter on average. We will refer to units for which $\delta_i=0$ as the core group, whereas the other units are deviators.\footnote{ Alternatively, one could think of two groups: a $\delta_i=0$ group and a $\delta_i \not=0$ group.} The prior can be generalized in a straightforward manner by replacing the ${\cal N}(0,v_\delta)$ distribution with a ${\cal N}(\underline{\delta},v_\delta)$ distribution and treating the $\underline{\delta}$ as a third hyperparameter.

Let $Y_{1:N}$ denote $\{ y_1,\ldots, y_N\}$. Conditional on the hyperparameters $(q,v_\delta)$, the posterior distribution is also a mixture of a pointmass at zero and a Normal component:
\be
\delta_i|(Y_{1:N},q,v_{\delta}) \sim \left\{ \begin{array}{ll} {\cal N} (\delta_i^*,v_{\delta}^*) & \mbox{with prob. } q^* \\
	0                                          & \mbox{with prob. } 1-{q}^* \end{array} \right. ,
\ee
where
\be 
   \delta_i^* = \frac{1}{1/v_\delta+1} y_i, \quad
   v^*_{\delta} = \frac{1}{1/v_\delta+1}, \quad
   \frac{q^*}{1-q^*} = \frac{q}{1-q}  (v_\delta+1)^{-1/2} \exp\left\{ \frac{1}{2} \frac{v_\delta}{v_\delta+1} y_i^2\right\}.
\ee
The odds in favor of a Gaussian component are an increasing function of the squared deviations from zero, $y_i^2$. The posterior mean is given by $\mathbb{E}[\delta_i|Y_{1:N}] = q^* \delta_i^*$, which is never exactly equal to zero. However, it can be shown that the posterior median, denoted by $\mbox{med}(\delta_i|Y_{1:N})$, has the property that it can be exactly equal to zero for small values of $y_i^2$: $\mbox{med}(\delta_i|Y_{1:N}) = 0$ for $-c(\delta_i^*,q^*,v_\delta^*) \le y_i \le c(\delta_i^*,q^*,v_\delta^*)$; see, for instance \cite{JohnstoneSilverman2004}.

The hyperparameters are determined by the marginal likelihood function
\begin{eqnarray}
  \lefteqn{p(Y_{1:N}|q, v_\delta)} \\
   &=& (2 \pi)^{-N/2} \prod_{i=1}^N \left(q (1+v_\delta)^{-1/2} \exp \left\{ -\frac{1}{2(1+v_\delta)} y_i^2\right\} + (1-q) \exp \left\{ -\frac{1}{2} y_i^2\right\}  \right). \nonumber
\end{eqnarray}
We will combine this marginal likelihood with a prior $p(q,v_\delta)$ for the hyperparameters and conduct posterior inference. For the posterior computations it is convenient to introduce the auxiliary variables $z_i$ such that $z_i=0$ if $\delta_i=0$ and $z_i=1$ if $\delta_i \not=0$. We will say that unit $i$ is a member of the core group if $z_i=0$ and it is a deviator if $z_i=1$. Then,
\begin{eqnarray}
	\lefteqn{p(Y_{1:N},Z_{1:N}|q, v_\delta)} \\
	&=& p(Y_{1:N}|Z_{1:N},q,v_\delta) p(Z_{1:N}|q,v_\delta) \nonumber\\
	&=& (2 \pi)^{-N/2} \left(\prod_{i \, | \, z_i=1} (1+v_\delta)^{-1/2} \exp \left\{ -\frac{1}{2(1+v_\delta)} y_i^2\right\} \right) \left(\prod_{i \, | \, z_i=0}  \exp \left\{ -\frac{1}{2} y_i^2\right\} \right) \nonumber \\
	&& \times \prod_{i=1}^n (1-q)^{1-z_i} q^{z_i}. \nonumber
\end{eqnarray}
Given a prior for $(q,v_\delta)$, draws from the posterior distribution can be obtained via Gibbs  iterating over the conditional posteriors $Z_{1:N}|(Y_{1:N},q,v_\delta)$, $q|(Y_{1:N},Z_{1:N},v_\delta)$ and $v_\delta|(Y_{1:N},Z_{1:N},q)$.

Notice that the argmax of $p(Y_{1:N},Z_{1:N}|q, v_\delta)$ with respect to $(Z_{1:N},q,v_\delta)$ satisfies:
\begin{eqnarray*}
 \hat{q} &=& \frac{1}{N} \sum_{i=1}^N \mathbb{I}\{ \hat z_i=1\},
 \quad \hat{v}_\delta = \max \; \left\{ 0, \, \frac{\sum_{i=1}^N \mathbb{I}\{ \hat z_i=1\} y_i^2}{\sum_{i=1}^N \mathbb{I}\{ \hat z_i=1\}} -1 \right\} \\
 \hat z_i &=& \mathbb{I} \left\{ \frac{\hat{q}}{1-\hat{q}}(1+\hat{v}_\delta)^{-1/2} \exp \left\{ \frac{\hat{v}_\delta}{2(1+\hat{v}_\delta)} y_i^2\right\} \ge 1 \right\},
\end{eqnarray*} 
where $\mathbb{I}\{x=a\}$ is the indicator function that is equal to one if $x=a$ and equal to zero otherwise. Thus, $\hat{z}_i=1$ if the likelihood ratio of $\delta_i \sim {\cal N}(0,\hat{v}_d)$ versus $\delta_i=0$ exceeds one, $\hat{q}$ corresponds to the fraction of $\hat{z}_i$s that are equal to one, and $\hat{v}_d$ is the variance of the observations attributed to the Gaussian component of the model (minus one).

The prior distribution in (\ref{eq:prior.delta.alpha}) nests coefficient homogeneity ($q=1$) and full coefficient heterogeneity ($q=1$) as special cases. In practice, whether $q$ is close to zero or one is partly determined by the researcher when (s)he constructs the estimation sample. In our empirical application we use the raw panel data set to construct balanced panels in which units are fairly homogeneous and an unbalanced panel that tries to include as many units as possible. The former samples will feature sparse heterogeneity in some dimensions, whereas the latter sample will lead to $q$ estimates that are close to one.

\section{Panel Data Models with Sparse Heterogeneity}
\label{sec:model}

We consider two dynamic linear panel data models in this paper. The first model, $M_1$, which will be used in the Monte Carlo analysis in Section~\ref{sec:MC}, takes the form of a dynamic linear regression. The second model, $M_2$, is an unobserved components model that can be written in state-space form and is used for the empirical analysis. In the remainder of this section we present the specifications of $M_1$ and $M_2$ and sketch the posterior sampler for the state-space model $M_2$, which nests the simulation algorithm for the regression model $M_1$ as a special case.

\noindent {\bf Linear Dynamic Panel Regression Model $M_1$.} Under $M_1$ the dependent variable $y_{it}$ evolves according to 
\begin{equation}
y_{it} = \alpha + \delta^{\alpha}_i + (\rho + \delta^{\rho}_i )y_{it-1} + \sigma  \sqrt{\delta^{\sigma}_i}  u_{it}, \quad i=1,\ldots,N, \quad t=1,\ldots,T. 
\label{eq:model.benchmark}
\end{equation}
This specification allows for cross-sectional heterogeneity in the intercept $\alpha$, the autoregressive coefficient $\rho$, and the innovation standard deviation $\sigma$ through the discrepancies $\delta^{\alpha}_i$, $\delta^{\rho}_i$, and $\delta^{\sigma}_i$. Units $i$ for which the discrepancies are equal to zero, share a common intercept, autocorrelation parameter, and shock standard deviation, respectively. The remaining units deviate from the parameters of the core group(s) and have heterogeneous coefficients. To induce sparsity, we use the prior in (\ref{eq:prior.delta.alpha}) for $\delta_i^\alpha$ and $\delta_i^\rho$, equipping $q$ and $v_\delta$ with superscripts $\alpha$ and $\rho$. Because of the non-negativity of variances, we use a slightly different prior for $\delta_i^\sigma$:
\be
\delta_i^\sigma|q^\sigma, \nu_{\delta^{\sigma}}, \tau_{\delta^{\sigma}}  \sim \left\{ \begin{array}{ll} IG\left( \frac{\nu_{\delta^{\sigma}}}{2},  \frac{\tau_{\delta^{\sigma}}}{2} \right)  & \mbox{with prob. } q^\sigma \\
	1                                          & \mbox{with prob. } 1-q^\sigma \end{array} \right. ,
\label{eq:benchmark.prior.delta.sigma}
\ee
where $IG$ is the Inverse Gamma distribution.
We impose that $\mathbb{E}[\delta_i^\sigma]=1$ and reparameterize the prior distribution in terms of its variance $v_{\delta^\sigma}$ which leads to\footnote{The model could be restricted by requiring that $q^\alpha=q^\rho=q^\sigma=q$ and that the indicator functions $z_i^\alpha=z_i^\rho=z_i^\sigma = z_i$ are identical. This restriction implies that members of the core group have identical $\alpha$, $\rho$, and $\sigma^2$ parameter values, and deviators differ in all three dimensions.}
\be
\nu_{\delta^\sigma} = 2v^{-1}_{\delta^\sigma} + 4, \quad \tau_{\delta^\sigma} = 2v^{-1}_{\delta^\sigma} + 2.
\label{eq:benchmark.hetero.hp.delta.sigma}
\ee

We use the following priors for the common parameters $\alpha$, $\rho$, and $\sigma^2$ of the core group(s):
\be
\alpha \sim {\cal N}(0,\underline{v}_\alpha), \quad \rho \sim {\cal N}(0,\underline{v}_\rho), \quad
\sigma^2 \sim IG \left( \frac{\underline{\nu}_{\sigma}}{2},
\frac{\underline{\tau}_{\sigma}}{2} \right), \label{eq:benchmark.prior.alpha.rho.sigma}
\ee
For the hyperparameters of the S\&S priors we use
\be 
q^l \sim B(a,b), \; l\in\{\alpha,\rho,\sigma\}; \quad
v_{\delta^{l}} \sim  IG \left( \frac{\underline{\nu}_{\delta^{l}}}{2},  \frac{\underline{\tau}_{\delta^{l}}}{2} \right), \quad l\in\{\alpha,\rho,\sigma\},
\label{eq:benchmark.prior.q.vdelta}
\ee 
where $B(a,b)$ is the Beta distribution. We define the parameter vector $\theta$ as
\be
\theta(M_1) = [\alpha,\rho,\sigma^2,q^\alpha,q^\rho,q^\sigma,v_{\delta^\alpha},v_{\delta^\rho}, v_{\delta^\sigma} ]'
\ee
and stack the hyperparameters that index the prior distributions in (\ref{eq:benchmark.prior.alpha.rho.sigma}) and (\ref{eq:benchmark.prior.q.vdelta}) into the vector
\be
\lambda(M_1) = [\underline{v}_\alpha,\underline{v}_\rho, \underline{\nu}_{\sigma}, \underline{\tau}_\sigma,a,b,  \underline{\nu}_{\delta^{\alpha}}, \underline{\tau}_{\delta^{\alpha}}, \underline{\nu}_{\delta^{\rho}}, \underline{\tau}_{\delta^{\rho}}, \underline{\nu}_{\delta^{\sigma}}, \underline{\tau}_{\delta^{\sigma}}  ].'
\ee
We denote the prior density of $\theta$ conditional on the hyperparameters as $p(\theta|\lambda,M_1)$.

\noindent {\bf State-Space Model  $M_2$.} This model contains a latent variable $s_{it}$ and takes the form of a state-space model, which is given by
\begin{eqnarray} 
y_{it} &=& x'_{it} (\alpha + \delta^{\alpha}_{i} )
+ s_{it} + \sigma_{u,t} \sqrt{\delta_{i,u}^\sigma} u_{it}, \label{eq:model.ss.ME} \\
s_{it} &=& (\rho+\delta_i^\rho) s_{it-1} + \sigma_{\epsilon,t} \sqrt{\delta_{i,\epsilon}^\sigma} \epsilon_{it}.
\label{eq:model.ss.ST}
\end{eqnarray}
The main difference between (\ref{eq:model.benchmark}) and (\ref{eq:model.ss.ME}) is the presence of the latent state $s_{it}$. (\ref{eq:model.ss.ME}) can be interpreted as the measurement equation of the state-space model, and (\ref{eq:model.ss.ST}) is the state transition. Moreover, we let the shock standard deviations $\sigma_{u,t}$ and $\sigma_{\epsilon,t}$ vary over time. The parameter $\alpha$ is now a $k$-dimensional vector that interacts with the vector of regressors $x_{it}$. 

For the $\delta_i$ discrepancies we use the priors (\ref{eq:prior.delta.alpha}) and (\ref{eq:benchmark.prior.delta.sigma}) from above with the  modification that $v_\delta^\alpha$ is a covariance matrix for $k>1$. The priors for $\alpha$ and $\rho$ are the same as in (\ref{eq:benchmark.prior.alpha.rho.sigma}), with the understanding that $\underline{v}_\alpha$ is also a covariance matrix. For the shock variances we now use independent IG priors:
\be
	\sigma_{u, t}^2 \sim IG \left( \frac{\underline{\nu}_{\sigma_{u}}}{2}, \frac{\underline{\tau}_{\sigma_{u}}}{2} \right), \quad 
\sigma_{\epsilon, t}^2 \sim IG \left( \frac{\underline{\nu}_{\sigma_{\epsilon}}}{2}, \frac{\underline{\tau}_{\sigma_{\epsilon}}}{2} \right).
\ee
The prior for the hyperparameters of the S\&S prior in (\ref{eq:benchmark.prior.q.vdelta}) is modified as follows: because $\alpha$ is a vector for $k>1$ the IG distribution for $v_\delta^\alpha$ is replaced by an Inverse Wishard distribution $IW(\underline{\nu}_{\delta^\alpha},\underline{\Psi}_{\delta^\alpha})$. Moreover, the priors for $q^\sigma$ and $v_{\delta^\sigma}$ are replaced by separate priors for $q^{\sigma_u}$, $q^{\sigma_\epsilon}$, $v_{\delta^{\sigma}_u}$, and $v_{\delta^{\sigma}_\epsilon}$. Finally, we need a prior distribution for the initial state $s_0$. We assume that $s_0 \sim {\cal N}(\mu_{s_0}, v_{s_0})$, where 
\be
	\mu_{s_0} \sim {\cal N}(\underline{\mu}_{s_0}, \underline{v}_{s_0}), 
\quad v_{s_0} \sim IG  \left( \frac{\underline{\nu}_{s_0}}{2},  \frac{\underline{\tau}_{s_0}}{2} \right).
\ee
Notice that the state-transition equation (\ref{eq:model.ss.ST}) induces the distribution of $s_1,\ldots,s_T$ conditional on $s_0$. This leads to the following definitions of $\theta$ and $\lambda$:
\begin{eqnarray}
\theta(M_2) &=& [\alpha', \rho, \sigma_{u,1:T}^{2}, \sigma_{\epsilon,1:T}^{2}, q^\alpha,q^\rho, q^{\sigma_u},q^{\sigma_\epsilon}, v_{\delta^\alpha},v_{\delta^\rho}, v_{\delta_u^\sigma}, v_{\delta_\epsilon^\sigma}, \mu_{s_0}, v_{s_0}]'
\\
\lambda(M_2) &=& [\underline{v}_\alpha, \underline{v}_\rho, \underline{\nu}_{\sigma_u}, \underline{\tau}_{\sigma_u}, \underline{\nu}_{\sigma_\epsilon}, \underline{\tau}_{\sigma_\epsilon},  a, b,  \underline{\nu}_{\delta^{\alpha}},  \underline{\Psi}_{\delta^{\alpha}}, 
\underline{\nu}_{\delta^{\rho}}, \underline{\tau}_{\delta^{\rho}}, 
\underline{\nu}_{\delta_u^{\sigma}}, \underline{\tau}_{\delta_u^{\sigma}},
\underline{\nu}_{\delta_\epsilon^{\sigma}}, \underline{\tau}_{\delta_\epsilon^{\sigma}}, \\
&&
\underline{\mu}_{s_0}, \underline{v}_{s_0}, \underline{\nu}_{s_0}, \underline{\tau}_{s_0} ]' \; . \nonumber
\end{eqnarray}





\noindent {\bf Restricted Model Specifications.} For $M_1$, we consider a homoskedastic version in Experiment~1 which is obtained by setting $q^\sigma=0$ and $\delta_i^\sigma=0$. Moreover, we also estimate a model with homogeneous coefficients which is obtained by setting $q^\alpha=q^\rho = 0$ and $\delta_i^\alpha=\delta_i^\rho=0$. Under this restriction every unit $i$ belongs to the core group. A fully heterogeneous model specification is obtained by letting $q^\alpha=q^\rho=1$, meaning there no longer are core groups of units that share identical coefficients. In Experiment~2 we consider the sparsely heteroskedastic model specification and the restricted versions that are obtained by either setting $q^\alpha=q^\rho=q^\sigma=0$ (homogeneous) or $q^\alpha=q^\rho=q^\sigma=1$ (fully heterogeneous). The empirical analysis is based on model $M_2$. Here we consider a homoskedastic version, a heteroskedastic version with homogeneous $\alpha$ and $\rho$ coefficients, and a heteroskedastic version with fully heterogeneous $\alpha$ and $\rho$ coefficients. As for $M_1$ the restricted specifications are obtained by setting the appropriate $q^l$, $l \in \{\alpha,\rho,\sigma_u,\sigma_\epsilon\}$, either equal to zero or one.

\noindent {\bf Posterior Sampling.} We outline a few important aspects of the posterior sampler for $M_2$. The sampler for $M_1$ is a special case in which there is no latent variable $s_{it}$. Detailed descriptions of the posterior samplers are provided in the Online Appendix.\footnote{Posterior computations for the restricted $q=0$ and $q=1$ versions of model $M_2$ have been conducted by \cite{NakataTonetti2015}. Their paper shows through a simulation study that Bayes estimates are generally more accurate than the GMM/minimum distance estimates that are widely-used in the empirical literature.}

We use a Gibbs sampler to generate draws from the posterior distribution of the parameters  $(\theta,\delta^\alpha,\delta^\rho,\delta^{\sigma_u},\delta_{\sigma_\epsilon})$ and the latent states $S_{i,0:T} = \{s_{i0},\ldots, s_{iT} \}$. To implement the Gibbs sampler we use indicator variables $z_i^l$ such that $z_i^l=0$ if $\delta_i^l=0$ and $\delta_i^l=1$ otherwise, where $l \in \{\alpha,\rho,\sigma_u,\sigma_\epsilon\}$. Because we found $\delta_i^\alpha$ to have a strong {\em a posteriori} correlation with $S_{i,0:T}$ we do not use a Kalman filtering/simulation smoothing step. Instead, we write $M_2$ as
\be
Y_{i} = X_{i} \alpha + W_{i} \beta_i + D_{i,u} U_{i},
\ee
where $Y_i$, $X_i$, and $U_i$ are matrices that stack $y_{it}$, $x_{it}'$, and $u_{it}$, respectively. $D_i$ is a diagonal matrix with elements $\sigma_{u,t}\sqrt{\delta_{i,u}}$ and the $T \times (k+T)$ design matrix $W_i$ comprises the regressor matrix $X_i$ and time dummies. For instance, if $k=1$ and $x_{it}=1$ such that $\alpha$ is an intercept, 
\be
W_{i} =
\begin{bmatrix}
	1 & 1 & 0 & \cdots & 0\\
	1 & 0 & 1 & \cdots & 0\\
	\vdots & \vdots & \vdots & \ddots & \vdots\\
	1 & 0 & 0 & \cdots & 1\\
\end{bmatrix}
= \begin{bmatrix} X_i & I \end{bmatrix}
\ee
and $\beta_i' = [\delta^{\alpha}_{i}, \; s_{i1}, \; ... \;, s_{iT}]$ stacks the $\alpha$ discrepancy and the latent states. The prior for $\beta_i$ combines the prior for $\delta_i^\alpha$ with the prior for the latent states generated by the state-transition equation (\ref{eq:model.ss.ST}). In case of $K=1$ this leads to
\begin{eqnarray} 
	\beta_i |  (\rho, \delta^{\rho}_i, z_i^\alpha, q^{\alpha}, v_{\delta^{\alpha}}, D_{i,\epsilon}) &\sim&  {\cal N} \left( 
	\begin{bmatrix}
		0 \\
		0_{T \times 1} \\
	\end{bmatrix}, 
	\begin{bmatrix}
		z_i^\alpha v_{\delta^{\alpha}} & 0_{1 \times T} \\
		0_{T \times 1} & \underline{V}_{s_i} \\
	\end{bmatrix} 
	\right) \\
	z_i^\alpha|q^{\alpha} &=& \left\{ \begin{array}{ll}  0 & \mbox{with prob. } 1-q^\alpha\\ 1 & \mbox{with prob. } q^\alpha \end{array} \right. \; , \nonumber
\end{eqnarray}
where $D_{i,\epsilon}$ is a diagonal matrix with elements $\sigma_{\epsilon,t} \sqrt{\delta^{\sigma}_{i,\epsilon}}$.
The $(t,\tau)$ elements of the $T \times T$ prior covariance matrix $\underline{V}_{s_i}$ can be calculated as follows:
\begin{eqnarray*} 
	\underline{V}_{s_i}(t,t) &=& (\rho + \delta^{\rho}_{i})^2 \underline{V}_{s_i}(t-1,t-1) + \sigma^2_{\epsilon,t} \delta^{\sigma}_{i,\epsilon}, \quad t=1,\ldots,T \\
	\underline{V}_{s_i}(t,\tau) &=& (\rho + \delta^{\rho}_{i})^{|t - \tau|} \underline{V}_{s_i}\big(\min(t,\tau),\min(t,\tau)\big), \quad
	t=1,\ldots,T \; \mbox{and} \; \tau \not=t,
\end{eqnarray*}
with the understanding that $\underline{V}_{s_i}(0,0)=v_{s_0}$. The conditional {\em posterior} distribution of $\beta_i$ is also a mixture of Normals.

\section{Monte Carlo Simulations}
\label{sec:MC}

In the Monte Carlo experiment we generate panels of observations $Y_{1:N,0:T}$ from model $M_1$ and compute estimates $\hat{l}_i(Y_{1:N,0:T})$ of $l_i = l + \delta_i^l$, where $l \in \{\alpha,\rho\}$. We evaluate the estimates based on a quadratic loss function and report a Monte Carlo approximation of the compound risk
\be
    \int \int \left( \frac{1}{N}  \sum_{i=1}^N \big(\hat{l}_i(Y_{1:N,0:T})-l_i \big)^2 \right) p(Y_{1:N,0:T},l_{1:N}|\theta,\lambda) d l_{1:N} dY_{1:N,0:T}.
\ee
The Monte Carlo experiment is based on the following algorithm:

\begin{algo}[Monte Carlo Simulation]
	\label{algo:MC}
	\hspace*{1cm}\\[-0ex]
	Given $\theta,\lambda$, the following two steps are repeated $N_{sim}$ times:
	\begin{enumerate}
		\item {\bf Parameter Draws.} Draw $\{\delta_i^\alpha\}_{i=1}^N$, $\{\delta_i^\rho\}_{i=1}^N$, $\{\delta_i^\sigma\}_{i=1}^N$ from the priors in (\ref{eq:prior.delta.alpha}) and (\ref{eq:benchmark.prior.delta.sigma}).
		\item {\bf Data Simulation.} Draw a panel of iid error terms $u_{it}$ for $i=1,\ldots,N$ and $t=1,\ldots,T$ from the ${\cal N}(0,1)$ distribution and generate $y_{i1:T}$ based on (\ref{eq:model.benchmark}), starting from the initial values $y_{i0}=0$, $i=1,\ldots,N$ and using the parameter values generated in Step 1.
		\item {\bf Estimation and Loss Calculation.} Compute the estimators $\hat{l}_i(Y_{1:N,0:T})$ for $l \in \{\alpha,\rho\}$ and evaluate the loss
		\[
		\frac{1}{N}  \sum_{i=1}^N \big(\hat{l}_i(Y_{1:N,0:T})-l_i \big)^2.
		\]
	\end{enumerate}
\end{algo}

We compare the performance of four posterior mean estimators. The first one is the spike-and-slab (S\&S) estimator defined as
\[
   \hat{l}_i^* = \mathbb{E}[l + \delta_i^l|Y_{1:N,0:T},\lambda].
\]
This estimator treats the slab probability $q$ as unknown, but conditions on the hyperparameters $\lambda$. Second, the estimator $\hat{l}_i^0$ sets $q^l=0$, $l \in \{\alpha,\rho,\sigma\}$, and thereby imposes homogeneity on all of the coefficients. Third, we compute an estimator $\hat{l}_i^1$ for a fully heterogeneous estimator that is obtained by setting $q^l=1$, $l \in \{\alpha,\rho,\sigma\}$. Finally, we report the performance of the oracle estimator that is based on knowing the ``true'' value of  $\theta$:
\[
       \hat{l}_i^o = \mathbb{E}[l + \delta_i^l|Y_{1:N,0:T},\theta].
\]
We consider a homoskedastic design (Section~\ref{subsec:MC.homo}) and a heteroskedastic design (Section~\ref{subsec:MC.hetero}).

\subsection{Homoskedastic Design}
\label{subsec:MC.homo}

For the first Monte Carlo experiment we consider a homoskedastic design, i.e., in model (\ref{eq:model.benchmark}) we set $q^\sigma=0$ and $\delta_i^\sigma = 1$ for all $i$. Baseline choices for the  parameter vector $\theta$ and the  hyperparameters are summarized in Table~\ref{tab:MC.lambda.theta}. First, consider the elements of $\theta$. We fix $\alpha=1.0$,  set $\rho=0.6$, and let $\sigma^2=0.8$. In the experiments below we consider various values for $q^\alpha=q^\rho=q$ and $v_\delta^\alpha$, which control the height of the spike and the width of the slab of the $\delta_i^\alpha$ distribution. The parameter $v_{\delta^\rho}$ is set to 0.09. The parameter $v_{\delta^{\sigma}}$ is only relevant for the heteroskedastic design.

\begin{table}[t!]
	\caption{Parameter Values for Monte Carlo}
	\label{tab:MC.lambda.theta}

	\begin{center}

		Parameter Vector $\theta$\\

		\begin{tabular}{ccccccc} \hline \hline
			$\alpha$ & $\rho$ & $\sigma^2$ & $q$ & $v_{\delta^\alpha}$ & $v_{\delta^\rho}$ & $v_{\delta^{\sigma}}$ \\ \hline
			1.0 & 0.6 & 0.8 & \multicolumn{2}{c}{various} & .09 & 1 \\ \hline
		\end{tabular}

		\bigskip

		Hyperparameter Vector $\lambda$\\

		\begin{tabular}{ccccccc} \hline \hline
			$\underline{v}_\alpha$ & $\underline{v}_\rho$ & $(\underline{\nu}_{\sigma},\underline{\tau}_\sigma)$ & $(a,b)$ &
			$(\underline{\nu}_{\delta^\alpha},\underline{\tau}_{\delta^\alpha})$ &
			$(\underline{\nu}_{\delta^\rho},
			\underline{\tau}_{\delta^\rho})$ &
			$(\underline{\nu}_{\delta^{\sigma}}, \underline{\tau}_{\delta^{\sigma}}) $ \\ \hline
			1.0 & .25 & (12,10) & (1.0,1.0) &  (6.0,4.0) & (6.0,2.0) &  (12, 10) \\ \hline
		\end{tabular}
	\end{center}
\end{table}

Second, consider the elements of $\lambda$. We set the variance of $\alpha$, denoted by $\underline{v}_\alpha$, equal to one. Moreover, we let $\underline{v}_\rho =0.25$. The hyperparameters for the prior for $\sigma^2$ imply that $\mathbb{E}[\sigma^2]=1$ and $V[\sigma^2]=1/4$. Our choices of $(a,b)$ imply that $q \sim U[0,1]$.
We choose $\nu_{\delta^\alpha}=6$ and $\tau_{\delta^\alpha}=4$, which implies $\mathbb{E}[v_{\delta^\alpha}]=1$ and $\mathbb{V}[v_{\delta^\alpha}]=1$. This distribution covers the range of $v_{\delta^\alpha}$ values in Table~\ref{tab:mc1.mse.N500.v1}. Suppose we choose $\nu_{\delta^\rho}=6$ and $\tau_{\delta^\alpha}=2$. Then, $\mathbb{E}[v_{\delta^\alpha}]=1/2$ and $\mathbb{V}[v_{\delta^\alpha}]=1/4$. The true value of $v_{\delta^\rho}=0.09$ can be viewed as a draw from the left tail of this distribution.

\begin{table}[t!]
	\caption{Monte Carlo Experiment 1: Homoskedasticity}
	\label{tab:mc1.mse.N500.v1}
	\begin{center}
		\scalebox{0.90}{
			\begin{tabular}{llllllll} \hline \hline
				& $\text{Estimators}$ & \multicolumn{2}{l}{$q^\alpha=q^\rho=q$}  & & & \\
				\cline{3-8} \noalign{\smallskip}
				&            &  0 & 0.2 & 0.4 & 0.6 & 0.8 & 1.0 \\
				\midrule
				$v_{\delta^\alpha}$ & \multicolumn{7}{c}{MSE for Estimates $\hat{\alpha}_i$ of $\alpha_i = \alpha + \delta^\alpha_i$} \\ \midrule				
				$\text{0.05}$ & $\text{S\&S}$ & $0.001$ & $0.012$ & $0.021$ & $0.030$ & $0.038$ & $0.047$\\
				$\text{ }$ & $\text{q = 0}$ & $0.001$ & $0.521$ & $0.954$ & $1.204$ & $1.362$ & $1.481$\\
				$\text{ }$ & $\text{q = 1.0}$ & $0.052$ & $0.038$ & $0.037$ & $0.040$ & $0.045$ & $0.050$\\
				$\text{ }$ & $\text{Oracle}$ & $0.000$ & $0.009$ & $0.018$ & $0.026$ & $0.034$ & $0.042$\\
				\midrule
				$\text{0.5}$ & $\text{S\&S}$ & $0.001$ & $0.050$ & $0.093$ & $0.136$ & $0.174$ & $0.204$\\
				$\text{ }$ & $\text{q = 0}$ & $0.001$ & $0.691$ & $1.194$ & $1.537$ & $1.750$ & $1.945$\\
				$\text{ }$ & $\text{q = 1.0}$ & $0.052$ & $0.070$ & $0.108$ & $0.144$ & $0.175$ & $0.202$\\
				$\text{ }$ & $\text{Oracle}$ & $0.000$ & $0.048$ & $0.088$ & $0.126$ & $0.161$ & $0.188$\\
				\midrule
				$\text{1.0}$ & $\text{S\&S}$ & $0.001$ & $0.067$ & $0.130$ & $0.198$ & $0.255$ & $0.300$\\
				$\text{ }$ & $\text{q = 0}$ & $0.001$ & $0.857$ & $1.433$ & $1.870$ & $2.169$ & $2.460$\\
				$\text{ }$ & $\text{q = 1.0}$ & $0.052$ & $0.118$ & $0.183$ & $0.229$ & $0.263$ & $0.289$\\
				$\text{ }$ & $\text{Oracle}$ & $0.000$ & $0.063$ & $0.120$ & $0.176$ & $0.224$ & $0.258$\\
				\midrule
				$v_{\delta^\alpha}$ & \multicolumn{7}{c}{MSE for Estimates $\hat{\rho}_i$ of $\rho_i = \rho + \delta^\rho_i$} \\ \midrule
				$\text{0.05}$ & $\text{S\&S}$ & $0.000$ & $0.008$ & $0.014$ & $0.018$ & $0.021$ & $0.024$\\
				$\text{ }$ & $\text{q = 0}$ & $0.000$ & $0.156$ & $0.292$ & $0.380$ & $0.444$ & $0.497$\\
				$\text{ }$ & $\text{q = 1.0}$ & $0.015$ & $0.016$ & $0.018$ & $0.020$ & $0.022$ & $0.024$\\
				$\text{ }$ & $\text{Oracle}$ & $0.000$ & $0.008$ & $0.013$ & $0.017$ & $0.021$ & $0.023$\\
				\midrule
				$\text{0.5}$ & $\text{S\&S}$ & $0.000$ & $0.009$ & $0.017$ & $0.024$ & $0.031$ & $0.037$\\
				$\text{ }$ & $\text{q = 0}$ & $0.000$ & $0.175$ & $0.306$ & $0.393$ & $0.449$ & $0.499$\\
				$\text{ }$ & $\text{q = 1.0}$ & $0.015$ & $0.014$ & $0.019$ & $0.024$ & $0.030$ & $0.036$\\
				$\text{ }$ & $\text{Oracle}$ & $0.000$ & $0.009$ & $0.016$ & $0.023$ & $0.029$ & $0.035$\\
				\midrule
				$\text{1.0}$ & $\text{S\&S}$ & $0.000$ & $0.009$ & $0.018$ & $0.026$ & $0.034$ & $0.041$\\
				$\text{ }$ & $\text{q = 0}$ & $0.000$ & $0.190$ & $0.315$ & $0.400$ & $0.452$ & $0.501$\\
				$\text{ }$ & $\text{q = 1.0}$ & $0.015$ & $0.014$ & $0.021$ & $0.027$ & $0.033$ & $0.040$\\
				$\text{ }$ & $\text{Oracle}$ & $0.000$ & $0.009$ & $0.017$ & $0.024$ & $0.031$ & $0.037$\\
				\bottomrule
			\end{tabular}
		}
	\end{center}
	{\footnotesize {\em Notes:} In this simulation we use $N=500$, $T=8$, and $N_{sim}=100$. The choices for $(\theta,\lambda)$ are summarized in Table~\ref{tab:MC.lambda.theta}.}\setlength{\baselineskip}{4mm}
\end{table}

Using Algorithm~\ref{algo:MC}, we generate $N_{sim}=100$ panel data sets of size $N=500$ and $T=8$. For each panel data set we generate 5,000 parameter draws from the posterior distribution, discard the first 2,500 draws, and use the remaining draws to compute the posterior mean estimates $\hat{l}_i(Y_{1:N,0:T})$. Simulation results are summarized in Table~\ref{tab:mc1.mse.N500.v1}. In the top half of the table we report mean squared errors (MSEs) for the estimates $\hat{\alpha}_i$ of $\alpha_i = \alpha+\delta_i^\alpha$ for various combinations of $(v_{\delta^\alpha},q)$ and in the bottom half MSEs for $\hat\rho_i$. During the estimation we do not restrict $q^\alpha$ to be equal to $q^\rho$. Thus, the members of the core $\alpha$ and $\rho$ groups are not required to be the same. Formally, we do not impose the restriction that $z_i^\alpha = z_i^\rho$ for all $i$.

For $q=0$ the Bayes estimator with S\&S achieves the performance of the estimator $\hat{l}_i^0$ that imposes coefficient homogeneity, and for $q=0.8$ and $q=1$ its compound risk is approximately equal to the risk associated with the estimate $\hat{l}_i^1$ that treats $l_i$ as fully heterogeneous. For values of $q \in \{0.2, 0.4, 0.6\}$ the S\&S estimator achieves a lower MSE than the estimators that impose full homogeneity or full heterogeneity. Overall, $\hat{l}_i^*$ is equal or smaller than the risk of the best estimator among $\hat{l}_i^0$ and $\hat{l}_i^1$. The risk differentials between the fully homogeneous and fully heterogeneous estimators is increasing in $v_{\delta}^\alpha$ and the risk patterns are similar for the estimation of $l=\alpha_i$ and $l=\rho_i$. As a benchmark, we also result the risk of the oracle estimator that is obtained by conditioning on the ``true'' $\theta$. By construction the oracle estimator dominates the Bayes estimator $\hat{l}^*_i$, but overall the risk differentials are relatively small which means that $\theta$ is generally well estimated.

\begin{figure}[t!]
	\caption{Histograms of Posterior Means for $\hat{l}^*_i$ and Posterior Densities for $q$}
	\label{fig:mc1.estimates.N500.v1}
	\begin{center}
		\begin{tabular}{ccc}
			& $q=0.2$ & $q=0.8$ \\
			\rotatebox{90}{\hspace*{0.25in} Post. Mean $\hat{\alpha}_i^*$} &
			\includegraphics[width = 0.35\textwidth]{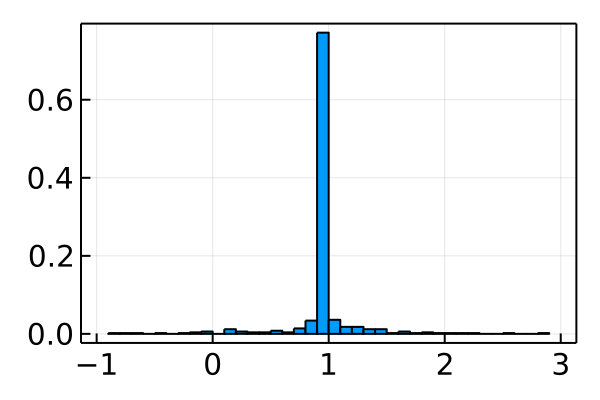} &
			\includegraphics[width=0.35\textwidth]{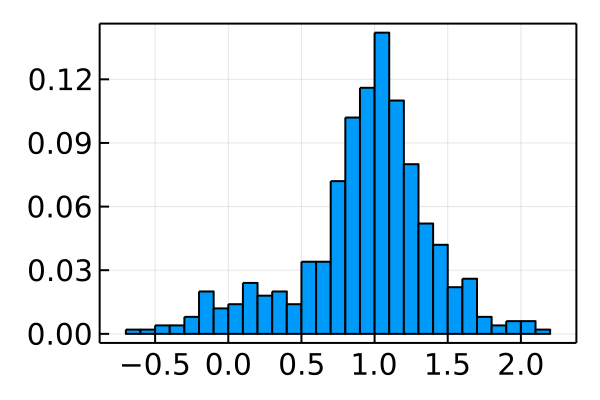} \\
			\rotatebox{90}{\hspace*{0.25in} Post. Mean $\hat{\rho}_i^*$} &
			\includegraphics[width=0.35\textwidth]{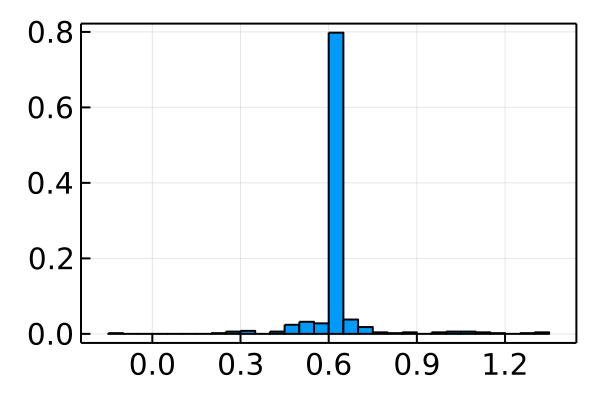} &
			\includegraphics[width=0.35\textwidth]{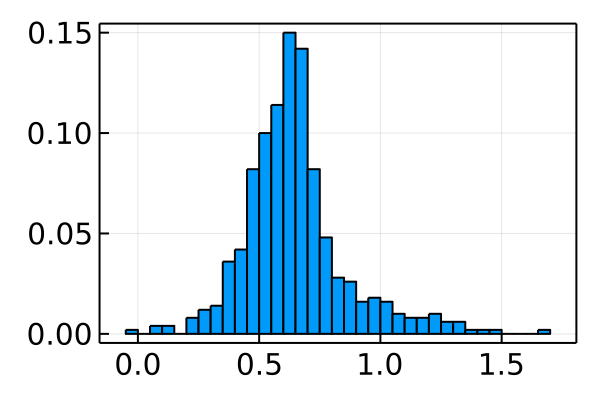} \\
			\rotatebox{90}{\hspace*{0.15in} Prior / Post. of $q$} &
			\includegraphics[width=0.35\textwidth]{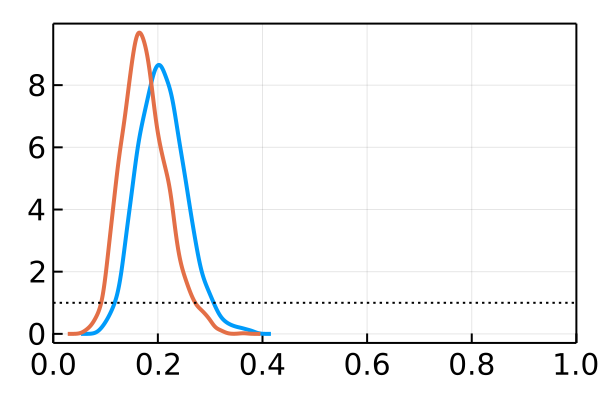} &
			\includegraphics[width=0.35\textwidth]{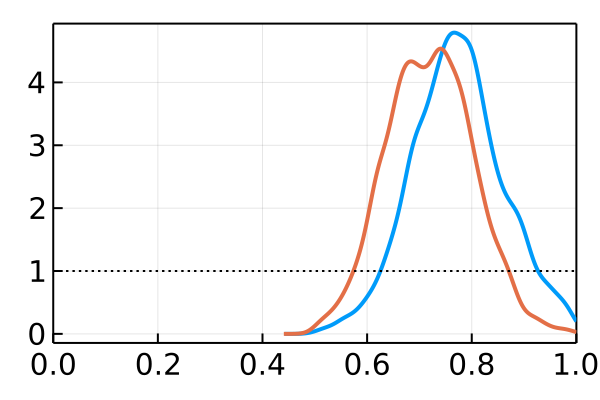}
		\end{tabular}
	\end{center}
	{\footnotesize {\em Notes}: We adjust the posterior means for $\delta^l_{i}$ in the following way: if more than 80\% of posterior draws for $\delta^l_{i}$ are 0 ($l = \alpha, \rho$), then we assume $\hat{\delta}^l_{i} = 0$ ($l = \alpha, \rho$) . The plots in the last row overlay a $B(1,1)$ prior density for $q^l_i$ (black, dotted) and posterior densities for $q^\alpha$ (blue, solid), and $q^\rho$ (red, solid). The choices for $(\theta,\lambda)$ are summarized in Table~\ref{tab:MC.lambda.theta}. We set $v_{\delta^\alpha}=0.5$.}\setlength{\baselineskip}{4mm}
\end{figure}

In the top and center panels of Figure~\ref{fig:mc1.estimates.N500.v1} we plot histograms of the cross-sectional distribution of the posterior mean estimates $\hat{\alpha}_i^*$ and $\hat{\rho}_i^*$ obtained conditional on data sets simulated with $q=0.2$ and $q=0.8$, respectively. Unlike the results reported in Table~\ref{tab:mc1.mse.N500.v1}, the histograms are generated from a single Monte Carlo iteration and not averaged across $N_{sim}$ runs. For $q=0.2$ the histogram of posterior means inherits the spike-and-slab shape of the prior distribution with a distinct spike near the ``true'' values of $\alpha$ and $\rho$, which are 1.0 and 0.6, respectively; see Table~\ref{tab:MC.lambda.theta}. For $q=0.8$ the spike in the distribution of posterior means is much less pronounced and there is a lot more mass away from $\alpha=1.0$ and $\rho=0.6$. Finally, we plot the posterior densities for the slab probabilities $q^\alpha$ and $q^\rho$. While there is considerable uncertainty about $q$, the densities peak near the ``true'' values of $0.2$ and $0.8$, respectively.

\subsection{Heteroskedastic Design}
\label{subsec:MC.hetero}

In the second Monte Carlo design we allow for heteroskedasticity by setting $q^\sigma=q$ and sampling $\delta_i^\sigma$ from (\ref{eq:benchmark.prior.delta.sigma}). Recall that we imposed $\mathbb{E}[\delta_i^\sigma] =1$. To generate the data we let $v_{\delta^\sigma}=\mathbb{V}[\delta_i^\sigma]=1$. The hyperparameters for the $v_{\delta^\sigma}$ prior imply that $\mathbb{E}[v_{\delta^\sigma}] = 1$ and $\mathbb{V}[v_{\delta^\sigma}] = 1/4$. In all other aspects, the experiment is configured in the same as the homoskedastic experiment in Section~\ref{subsec:MC.homo}. 

Results for the MSE of $\hat{\alpha}_i$ are presented in Table~\ref{tab:mc2.mse.N500.v1}. We consider two S\&S estimators: (hetsk) is based on the heteroskedastic version of the benchmark model which is the correctly specified model in light of the data generating process (DGP). (homsk) is based on the misspecified homoskedastic version of the benchmark model, i.e., $\delta_i^\sigma=0$. The results are similar to the ones obtained in the first experiment. For $q \in \{0.2,\ldots,0.8\}$ the S\&S(hetsk) estimator dominates the estimators obtained under full homogeneity, $q=0$, and full heterogeneity, $q=1$. The S\&S(homsk) generally performs poorly, except for $q=0$, when there is no heteroskedasticity. Thus, ignoring the heteroskedasticity generates imprecise estimates also of the conditional mean parameters.

\begin{table}[t!]
	\caption{Monte Carlo Experiment 2: Heteroskedasticity}
	\label{tab:mc2.mse.N500.v1}
	\begin{center}
		\scalebox{0.95}{
			\begin{tabular}{llllllll} \hline \hline
				& $\text{Estimators}$ & \multicolumn{3}{l}{$q^\alpha=q^\rho=q^\sigma=q$}  & & & \\
				\cline{3-8} \noalign{\smallskip}
				&            &  0 & 0.2 & 0.4 & 0.6 & 0.8 & 1.0 \\
				\midrule
				$v_{\delta^\alpha}$ & \multicolumn{7}{c}{MSE for Estimates $\hat{\alpha}_i$ of $\alpha_i = \alpha + \delta^\alpha_i$} \\ \midrule
				$\text{0.05}$ & $\text{S\&S(hetsk)}$ & $0.001$ & $0.012$ & $0.022$ & $0.031$ & $0.040$ & $0.049$\\
				$\text{ }$ & $\text{S\&S(homosk)}$ & $0.001$ & $0.014$ & $0.026$ & $0.041$ & $0.055$ & $0.070$\\
				$\text{ }$ & $\text{q = 0}$ & $0.001$ & $0.529$ & $0.895$ & $1.116$ & $1.313$ & $1.459$\\
				$\text{ }$ & $\text{q = 1.0}$ & $0.077$ & $0.056$ & $0.051$ & $0.051$ & $0.053$ & $0.056$\\
				$\text{ }$ & $\text{Oracle}$ & $0.000$ & $0.009$ & $0.018$ & $0.026$ & $0.034$ & $0.041$\\
				\midrule
				$\text{0.5}$ & $\text{S\&S(hetsk)}$ & $0.001$ & $0.051$ & $0.095$ & $0.135$ & $0.170$ & $0.198$\\
				$\text{ }$ & $\text{S\&S(homosk)}$ & $0.001$ & $0.053$ & $0.101$ & $0.148$ & $0.190$ & $0.224$\\
				$\text{ }$ & $\text{q = 0}$ & $0.001$ & $0.691$ & $1.138$ & $1.467$ & $1.726$ & $1.950$\\
				$\text{ }$ & $\text{q = 1.0}$ & $0.077$ & $0.078$ & $0.108$ & $0.141$ & $0.170$ & $0.195$\\
				$\text{ }$ & $\text{Oracle}$ & $0.000$ & $0.048$ & $0.088$ & $0.126$ & $0.159$ & $0.184$\\
				\midrule
				$\text{1.0}$ & $\text{S\&S(hetsk)}$ & $0.001$ & $0.067$ & $0.134$ & $0.198$ & $0.259$ & $0.295$\\
				$\text{ }$ & $\text{S\&S(homosk)}$ & $0.001$ & $0.069$ & $0.141$ & $0.215$ & $0.304$ & $0.353$\\
				$\text{ }$ & $\text{q = 0}$ & $0.001$ & $0.852$ & $1.381$ & $1.813$ & $2.156$ & $2.474$\\
				$\text{ }$ & $\text{q = 1.0}$ & $0.077$ & $0.120$ & $0.183$ & $0.229$ & $0.260$ & $0.283$\\
				$\text{ }$ & $\text{Oracle}$ & $0.000$ & $0.063$ & $0.121$ & $0.176$ & $0.222$ & $0.254$\\
				\midrule
				$v_{\delta^\alpha}$ & \multicolumn{7}{c}{MSE for Estimates $\hat{\rho}_i$ of $\rho_i = \rho + \delta^\rho_i$} \\ \midrule
				$\text{0.05}$ & $\text{S\&S(hetsk)}$ & $0.000$ & $0.008$ & $0.014$ & $0.018$ & $0.021$ & $0.023$\\
				$\text{ }$ & $\text{S\&S(homosk)}$ & $0.000$ & $0.008$ & $0.014$ & $0.019$ & $0.022$ & $0.025$\\
				$\text{ }$ & $\text{q = 0}$ & $0.000$ & $0.158$ & $0.278$ & $0.359$ & $0.433$ & $0.493$\\
				$\text{ }$ & $\text{q = 1.0}$ & $0.018$ & $0.018$ & $0.019$ & $0.021$ & $0.022$ & $0.024$\\
				$\text{ }$ & $\text{Oracle}$ & $0.000$ & $0.008$ & $0.013$ & $0.017$ & $0.020$ & $0.022$\\
				\midrule
				$\text{0.5}$ & $\text{S\&S(hetsk)}$ & $0.000$ & $0.009$ & $0.017$ & $0.025$ & $0.031$ & $0.038$\\
				$\text{ }$ & $\text{S\&S(homosk)}$ & $0.000$ & $0.010$ & $0.018$ & $0.026$ & $0.034$ & $0.041$\\
				$\text{ }$ & $\text{q = 0}$ & $0.000$ & $0.175$ & $0.293$ & $0.377$ & $0.443$ & $0.500$\\
				$\text{ }$ & $\text{q = 1.0}$ & $0.018$ & $0.015$ & $0.019$ & $0.025$ & $0.031$ & $0.037$\\
				$\text{ }$ & $\text{Oracle}$ & $0.000$ & $0.009$ & $0.016$ & $0.023$ & $0.030$ & $0.036$\\
				\midrule
				$\text{1.0}$ & $\text{S\&S(hetsk)}$ & $0.000$ & $0.010$ & $0.018$ & $0.027$ & $0.036$ & $0.043$\\
				$\text{ }$ & $\text{S\&S(homosk)}$ & $0.000$ & $0.010$ & $0.019$ & $0.029$ & $0.042$ & $0.051$\\
				$\text{ }$ & $\text{q = 0}$ & $0.000$ & $0.189$ & $0.303$ & $0.386$ & $0.448$ & $0.504$\\
				$\text{ }$ & $\text{q = 1.0}$ & $0.018$ & $0.015$ & $0.021$ & $0.028$ & $0.034$ & $0.041$\\
				$\text{ }$ & $\text{Oracle}$ & $0.000$ & $0.009$ & $0.017$ & $0.024$ & $0.032$ & $0.038$\\
				\hline
			\end{tabular}
		}
	\end{center}
	{\footnotesize {\em Notes:} In this simulation we use $N=500$, $T=8$, and $N_{sim}=100$. The hyperparameter settings are summarized in Table~\ref{tab:MC.lambda.theta}.}\setlength{\baselineskip}{4mm}
\end{table}

\begin{figure}[t!]
	\caption{Histograms of Posterior Means for $\hat{\delta}_i$ and Posterior Densities for $q$}
	\label{fig:mc2.estimates.N500.v1}
	\begin{center}
		\begin{tabular}{ccc}
			& $q=0.2$ & $q=0.8$ \\
		\rotatebox{90}{\hspace*{0.25in} Post. Mean $\hat{\alpha}_i^*$} &
		\includegraphics[width = 0.35\textwidth]{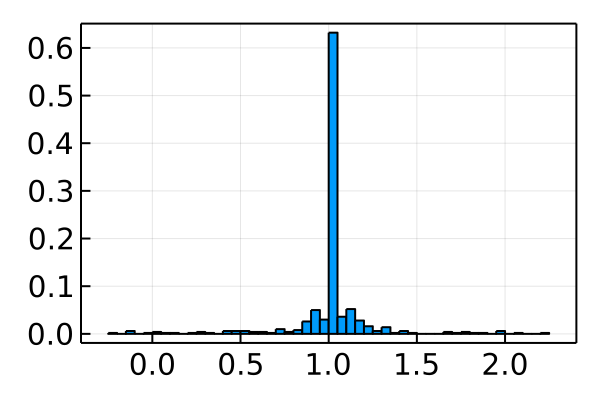} &
		\includegraphics[width=0.35\textwidth]{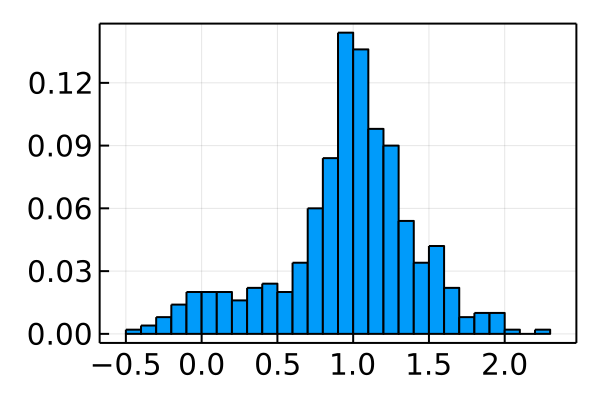} \\
        \rotatebox{90}{\hspace*{0.25in} Post. Mean $\hat{\rho}_i^*$} &
		\includegraphics[width=0.35\textwidth]{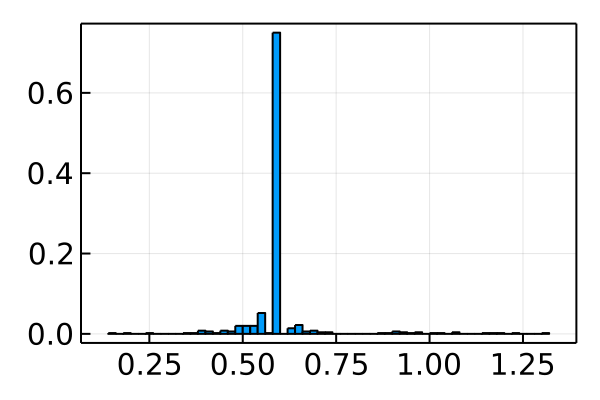} &
		\includegraphics[width=0.35\textwidth]{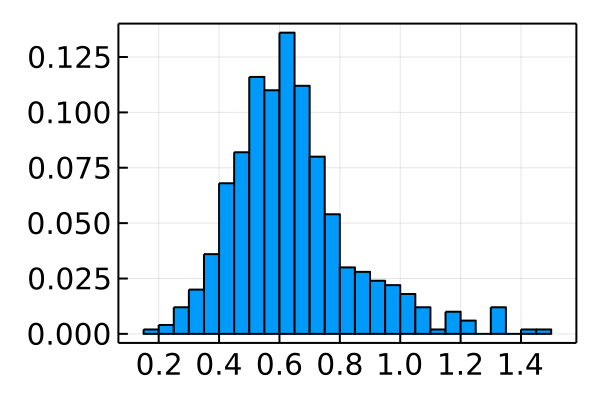} \\
		\rotatebox{90}{\hspace*{0.25in} Post. Means $\hat{\sigma}_i^{2*}$} &
		\includegraphics[width=0.35\textwidth]{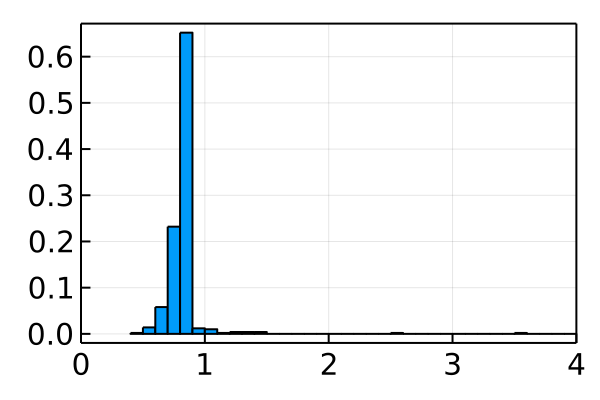} &
		\includegraphics[width=0.35\textwidth]{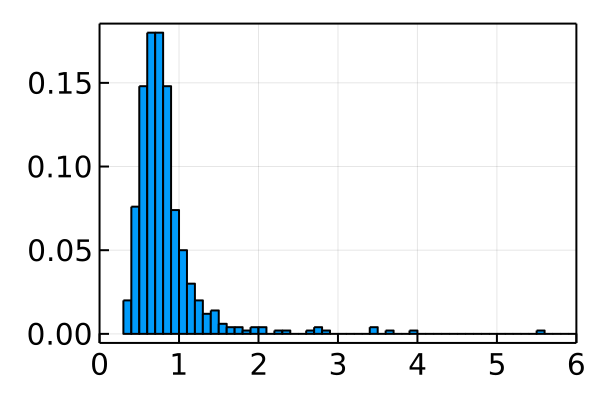} \\
	    \rotatebox{90}{\hspace*{0.15in} Prior / Post. of $q$} &
        \includegraphics[width=0.35\textwidth]{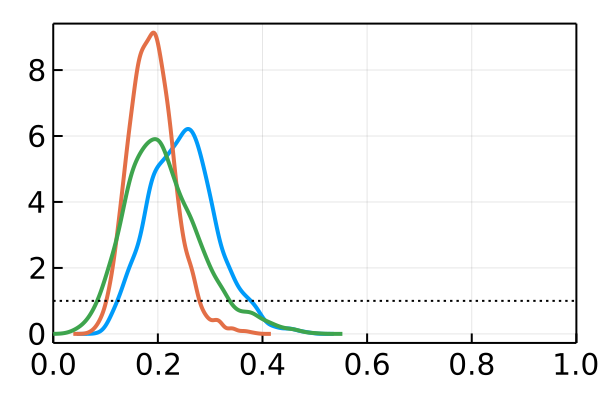} &
        \includegraphics[width=0.35\textwidth]{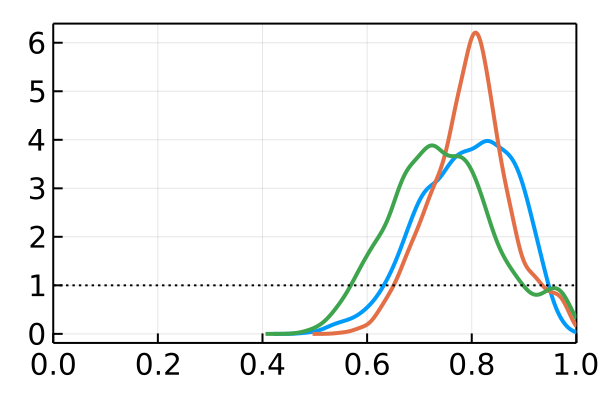}
	    \end{tabular}
    \end{center}
	{\footnotesize {\em Notes}: We adjust the posterior means for $\delta^l_{i}$ in the following way: if more than 80\% of posterior draws for $\delta^l_{i}$ are 0 ($l = \alpha, \rho$) or 1 ($l = \sigma$), then we assume $\hat{\delta}^l_{i} = 0$ ($l = \alpha, \rho$) or 1 ($l = \sigma$). The plots in the last row overlay a $B(1,1)$ prior density for $q^l_i$ (black, dotted) and posterior densities for $q^\alpha_i$ (blue, solid), $q^\rho_i$ (red, solid) and $q^\sigma_i$ (green, solid). The choices for the hyperparameter vector are summarized in Table~\ref{tab:MC.lambda.theta}.}\setlength{\baselineskip}{4mm}
\end{figure}

Figure~\ref{fig:mc2.estimates.N500.v1} extends the results shown in Figure~\ref{fig:mc1.estimates.N500.v1} to the heteroskedastic design. As before, we plot histograms of the cross-sectional distribution of the posterior mean estimates, which now include $\hat{\sigma}_i^*$. As for the homoskedastic design, for $q=0.2$ the cross-sectional distributions of the estimators inherit the spike-and-slab shape of the prior distribution, which is more pronounced for $\hat{\alpha}_i^*$ and $\hat{\rho}_i^*$ than for $\hat{\sigma}_i^{2*}$. The distributions of posterior means are generally centered near the ``true'' values of the homogeneous parameters.    

\section{Empirical Analysis}
\label{sec:empirics}

We now use model $M_2$, comprising (\ref{eq:model.ss.ME}) and (\ref{eq:model.ss.ST}), to estimate income profiles from the PSID data. Model $M_2$ captures some of the key features of the empirical specifications considered in the literature over the past four decades; see for instance the model specification discussion in the survey paper \cite{BrowningEjrnjes2013}. These features include a deterministic income component $x_{it}'\alpha$, and persistent and transitory stochastic  components, $s_{it}$ and $u_{it}$, respectively. We allow for potential heterogeneity in the $\alpha$ vector, the autocorrelation $\rho$ of the persistent component, and the shock standard deviations. 

More specifically, we let 
\be
		x_{it}=[1,h_{it}/10]',
		\label{eq:empirics.xit}
\ee
where $h_{it}$ represents experience of individual $i$ in period $t$, defined as age minus years of education minus six. We partition the $2\times 1$ vector $\alpha = [\alpha_0,\alpha_1]$. We also divide experience by 10 to scale $\alpha_1$ and improve the numerical performance of the posterior sampler.
If $\delta_i^\alpha=0$ for all $i$, then income profiles are homogeneous, which was previously labeled RIP and has been the preferred specification of \cite{LillardWeiss1979}, \cite{Hause1980}, \cite{Baker1997}, \cite{Haider2001}, \cite{Hryshko2012}, and \cite{Hoffmann2019}. The case of $\delta_i^\alpha\not=0$ for all individuals, corresponds to the HIP specification considered in the literature and is favored by, for instance, \cite{MaCurdy1982}, \cite{AboundCard1989}, and \cite{Guvenen2007,Guvenen2009}. Unlike the previous literature, our model considers a third option, under which heterogeneity is potentially sparse: $\delta_i^\alpha\not=0$ for some, but not all, individuals.  

Income has a persistent and a transitory stochastic component. The persistent component is given by an AR(1) process $s_{it}$ with autocorrelation coefficient $\rho$ and the transitory component is represented by the $iid$ process $u_{it}$.  Through $\delta_{i,u}^\sigma, \, \delta_{i,\epsilon}^\sigma \, \not=0$ we allow income risk to be heterogeneous across agents. In addition, the standard deviations $\sigma_{u,t}$ and $\sigma_{\epsilon,t}$ are time dependent, which can capture business cycle fluctuations and trends in idiosyncratic risk. Finally, we allow for idiosyncratic autocorrelations $\rho_i$ which is typically not done in the literature, an exception being  \cite{BrowningEjrnjesAlvarez2010} and \cite{BrowningEjrnjes2013}.

There are a few model features that have been explored in the literature that we do not include in $M_2$ to keep the model parsimonious: a third stochastic component that features a unit root, idiosyncratic autoregressive conditional heteroskedasticity (see, for instance, \cite{MeghirPistaferri2004} for both features), time variation in the returns to experience (e.g., \cite{Haider2001}), or age effects in the variances of the idiosyncratic shocks (e.g., \cite{Hoffmann2019}). In addition to $M_2$ itself we consider three restricted versions: one without heteroskedasticity, one with homogeneous slope parameters (RIP), and one with fully heterogeneous slope parameters (HIP). A summary is provided in Table~\ref{tab:empirics.models}.

\begin{table}[t!]
	\caption{Restricted Model Specifications}
	\label{tab:empirics.models}
	\begin{center}
		\begin{tabular}{ll} \hline \hline
			Label & Restrictions \\ \hline
			$M_2$(homosk) & $q^{\sigma_u} = q^{\sigma_\epsilon} = 1$;
			$\delta_i^{\sigma_u}=\delta_i^{\sigma_\epsilon}=0$ for $i=1,\ldots,N$ \\
			$M_2$(RIP) & $q^{\alpha} = q^{\rho} = 0$;
			$\delta_i^{\alpha}=\delta_i^{\rho}=0$ for $i=1,\ldots,N$\\
			$M_2$(HIP) & $q^{\alpha} = q^{\rho} = 1$ \\ \hline
		\end{tabular}
	\end{center}		
\end{table}



Some authors have emphasized, for instance, cohort effects on earnings, education effects on earnings profiles, or age-dependent persistence in earnings. While we are not explicitly controlling for this type of heterogeneity through sample splitting or interaction terms, it can be  captured {\em a posteriori} by estimates of $\delta_i$ that are different from zero.

Much of the previous literature has used (simulated) minimum-distance estimation of a subset of the parameters that we collect in the vector $\theta$, which involves minimizing the discrepancy between sample moments and model-implied moment. Typically, these moments are time-dependent autocovariances that are computed by cross-sectional averaging across all individuals in the panel. In a few instances, researcher have also used moments computed from unit-specific time series; see \cite{BrowningEjrnjes2013}. Our likelihood-based estimation approach aims at tracking individual earnings histories, which implicitly encompasses both time series and cross-sectional moments. This approach has several advantages. In particular, we are easily able to handle non-Normalities introduced by the S\&S prior and the cross-sectional heteroskedasticity. Morever, in addition to parameters that characterize the distribution of coefficients across agents, we can also make inference about the unit-specific coefficients themselves, and subsequently generate predictions about future income for individual units.


\subsection{Data Set}
\label{subsec:empirics.data}

The empirical analysis is based on the PSID. We use raw data from 1968 to 1997, starting from 32,465 observations for 2,052 individuals. Up to 1997 the PSID data are annual, whereas subsequently the survey was conducted only biennially. The data set used for estimation is constructed following the conventions of the literature, e.g., \cite{MeghirPistaferri2004}. The dependent variable is log real income of all sources, including: wages and salaries; bonuses, overtime, and/or commissions; income from professional practice and trade; the labor part of farm income and unincorporated business income; the labor portion of income from farming, market gardening, roomers and boarders. Nominal wages are converted into real wages, using the PCE deflator (base = 1993). We take account of the fact that the measure of income refers to the previous calender year. We eliminate observations with outlying earnings records, defined as a change in log real earnings greater than two or less than minus one.

As is common in the literature, we restrict the sample to male heads of household and keep individuals between ages 25 and 55.\footnote{If the household age is inconsistent across years, we compute the mode of the differences between reported age and year across periods. Age is then defined as the year plus the mode.} To avoid oversampling low-income households, members from the Survey of Economic Opportunity (SEO) are removed. We also drop the members of the Latino sample added in 1990 and self-employed individuals. We keep individuals who are in the labor force, have positive labor income, and have been in the sample for at least nine years. We remove individuals with missing race, marital status, region, state, and education information. In a first-stage regression we project log income on year dummies, and indicators for race, educational group,\footnote{High School dropouts: (those with less than 12 grades of schooling); High School graduates (those with at least a High School diploma, but no College degree); College graduates (those with a College degree or more)} region of residence, and marital status. The variable $y_{it}$ is the residual from this regression. We refer to the resulting panel data set as the full sample.

Based on the full sample, we construct estimation samples. The first set of estimation samples are balanced panels, denoted by ${\cal B}_{T,\tau}$, and used for the analysis in Sections~\ref{subsec:empirics.parameters} and~\ref{subsec:empirics.prediction}. The balanced panels have a time dimension of $T+1$ and include only individuals for which $T+1$ observations are available. Typically, we use the first $T$ time periods in these samples for estimation and reserve the last time period, $t=T+1$, for one-step-ahead forecast evaluation. The index $\tau$ refers to the year that corresponds to period $T+1$. For instance, the sample ${\cal B}_{20,1991}$ starts in 1971 ($t=1$) and ends in 1991 ($t=20+1$).


\begin{figure}[t!]
	\caption{Unbalanced Panel}
	\label{fig:empirics.unbal.properties}
	\begin{center}
		\begin{tabular}{cc}
			Units Per Period & Periods in Panel \\
		\includegraphics[width = 0.4\textwidth]{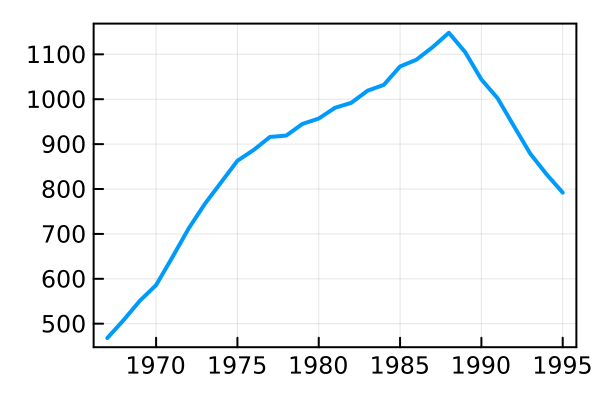} & 
		\includegraphics[width = 0.4\textwidth]{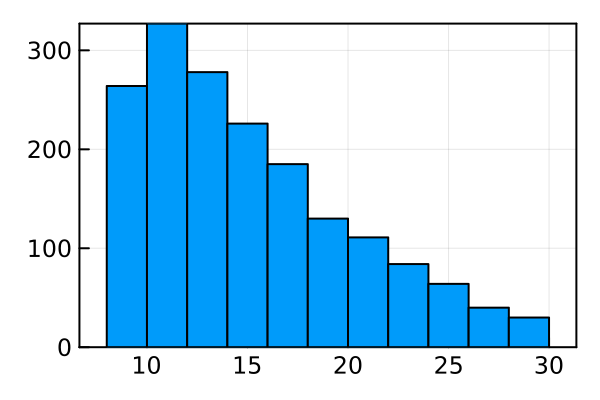}
		\end{tabular}
	\end{center}
\end{figure}

In addition to the balanced sample, we also estimate model $M_2$ and its variants based on an unbalanced sample ${\cal U}$ in Section~\ref{subsec:empirics.unbalanced}. The unbalanced sample is similar to the full sample described above, except that we require that individuals have at least nine {\em consecutive} observations.  Figure~\ref{fig:empirics.unbal.properties} provides some information about the composition of ${\cal U}$. The left panel of the figure shows the cross-sectional dimension as a function of time. In 1968 it covers approximately 500 units. The number of units rises above 1,100 in 1988 and then declines to 800 in 1995. The right panel provides information about how long units stay in the panel. The modal duration is 10-12 years, but approximately 330 units stay for more than 20 years. {\em A priori}, we expect coefficient heterogeneity to be more prevalent in the unbalanced sample ${\cal U}$ than in the balanced samples ${\cal B}_{T,\tau}$.

\subsection{Prior Distribution}
\label{subsec:empirics.prior}

\begin{table}[t!]
	\caption{Prior Distribution for $\theta$}
	\label{tab:empirics.prior}
	\begin{center}
		\begin{tabular}{llcccc} \hline \hline
			Para & Distr. & $a$ & $b$ & 5\% & 95 \% \\ \hline 
			$\alpha$ & ${\cal N}(a,b)$ & 0 & $\begin{bmatrix} 1 & 0 \\ 0 & 1  \end{bmatrix}$ & $\begin{array}{c} \text{-}1.64 \\ \text{-}1.64 \end{array}$ & $\begin{array}{c} 1.64 \\ 1.64 \end{array}$\\
			$\rho$   & ${\cal N}(a,b)$ & 0.8 & 1 & -0.84 & 2.44 \\
			$\sigma^2_{u,t}$ & $IG \big( a/2,b/2 \big)$  & 6 & 0.2 & 0.02 & 0.12 \\
			$\sigma^2_{\epsilon,t}$ & $IG \big( a/2,b/2 \big)$ & 6 & 0.2 & 0.02 & 0.12 \\
			$q^l, \; l \in \{\alpha,\rho,\sigma_u,\sigma_\epsilon\}$ & $B(a,b)$ & 1 & 1 & 0.05 & 0.95 \\
			$v_{\delta^\alpha}$ & $ IW(a,b)$ & 5.05 & $\begin{bmatrix} 0.5 &0 \\ 0 & 0.1 \end{bmatrix}$ & $\begin{array}{c} 0.05 \\ 0.01 \end{array}$ & $\begin{array}{c} 0.68 \\ 0.14 \end{array}$\\
			$v_{\delta^\rho}$ & $IG \big( a/2,b/2 \big)$ & 16.5 & 3.625 & 0.13 & 0.44 \\
			$v_{\delta_{l}^{\sigma}}, \; l \in \{u,\epsilon\}$ & $IG \big( a/2,b/2 \big)$ & 12 & 10 & 0.48 & 1.91\\ 
     		$\mu_{s_0}$ & ${\cal N}(a,b)$ & 0 & 0.05 & -0.37 & 0.37\\
			$v_{s_0}$ & $IG \big( a/2,b/2 \big)$ & 6 & 0.2 & 0.02 & 0.12 \\\hline	 
		\end{tabular}
	\end{center}
	{\footnotesize {\em Notes}: ${\cal N}(a,b)$ is the Normal distribution, $IG(a,b)$ is the Inverse Gamma distribution, $B(a,b)$ is the Beta distribution, and $IW(a,b)$ is the Inverse Wishart distribution. The last to columns contain the 5th and 95th percentile of the marginal prior distributions.  For IG priors: $(a,b) = (6,0.2)$ implies a mean of 0.05 and a variance of 0.05; $(a,b) = (16.5,3.625)$ implies a mean of 0.25 and a variance of 0.1; $(a,b) = (12,10)$ implies a mean of 1 and a variance of 0.5.}\setlength{\baselineskip}{4mm}	
\end{table}

Table~\ref{tab:empirics.prior} summarizes the marginal prior distributions for the elements of $\theta$. We are assuming that the elements are {\em a priori} independent, such that the joint distribution is the product of the marginals. In addition to the $(a,b)$ parameterization of the prior, we also report the 5th and 95th percentiles, which delimit the 90\% equal-tail probability credible intervals. 


The prior for $\alpha$ is centered at zero with a variance of one. Recall that we divided experience by ten, which means that a value of $\alpha_1=0.2$ implies a return to a year of experience of $0.2/10 = 0.02$, or two percent. Thus, the 90\% {\em a priori} credible interval covers all reasonable and some unreasonable values for the return to experience. The marginal distribution of $\rho$ is centered at 0.8, representing an {\em a priori} belief that the idiosyncratic income process $s_{it}$ is fairly persistent. The variance of one implies that the prior does not have a lot of curvature in the range of 0.6 to 1.0, which covers the estimates obtained in previous studies. We do allow for explosive dynamics {\em a priori}, but turn out to be extremely unlikely under the posterior.

The 90\% credible interval for the innovation variances reaches from 0.02 to 0.12. Note that our prior implies that they are independent across time periods, because in our panel data set we want to identify the variances from the cross-sectional information. Our prior for $\mu_{s_0}$ and $v_{s_0}$ implies that $s_{i0}$ has mean zero, which is the long-run mean of $s_{it}$ provided that $\rho$ is less than one. Moreover, $s_{i0}$ falls in the interval -1.0 to 1.0 with high probability.

The prior for the probabilities of not belonging to the core group, $q^l$, are uniform on the unit interval. The prior for the deviation variance $v_{\delta_\alpha}$ implies that the covariance matrix of the $\alpha$ deviations is diagonal with values 0.24 and 0.048. At 0.048, the 95\% credible interval for $\delta_{\alpha_1}/10$ ranges from roughly -0.044 to 0.044.\footnote{The credible interval for the off-diagonal element of $v_{\delta^\alpha}$ ranges from -0.1 to 0.1.} The prior for $v_{\delta^\sigma_l}$ implies that an individual at the 5th percentile at the prior has an idiosyncratic shock variance that is one fourth of that of someone at the 95th percentile.

\subsection{Parameter Estimates (Balanced Panels)}
\label{subsec:empirics.parameters}

\begin{figure}[t!]
	\caption{Posterior for Idiosyncratic Parameters, $M_2$, $T=20$}
	\label{fig:empirics.post.hip.hetsk.T20}
	\begin{center}
		\begin{tabular}{ccc}
			& $\tau = 1988$ & $\tau = 1991$ \\
			\rotatebox{90}{\hspace*{0.525in}  $\alpha_{0i}$} &
			\includegraphics[width = 0.3\textwidth]{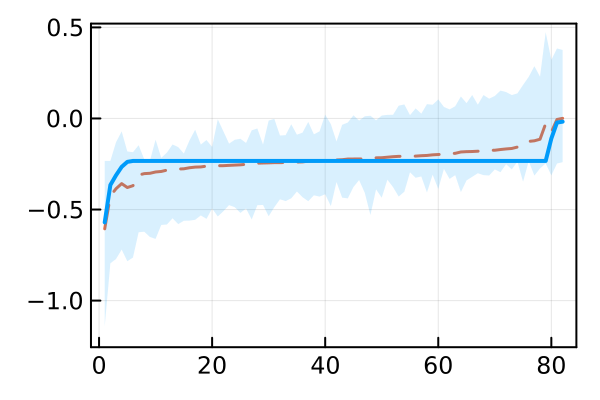} &
			\includegraphics[width=0.3\textwidth]{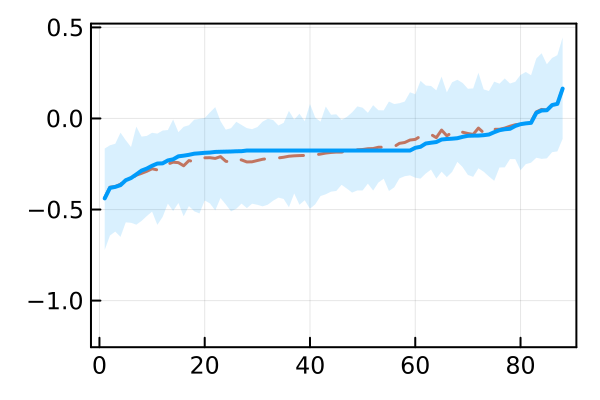} \\		
			\rotatebox{90}{\hspace*{0.525in}  $\alpha_{1i}/10$} &
			\includegraphics[width = 0.3\textwidth]{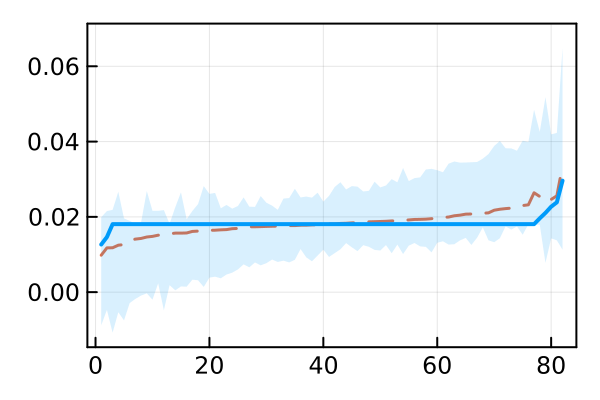} &
			\includegraphics[width=0.3\textwidth]{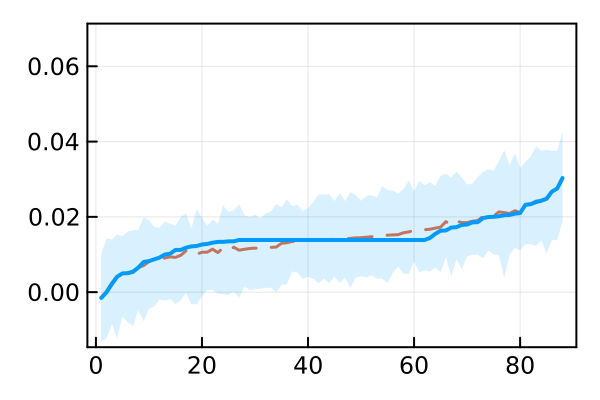} \\
			\rotatebox{90}{\hspace*{0.525in}  $\rho_i$} &
			\includegraphics[width=0.3\textwidth]{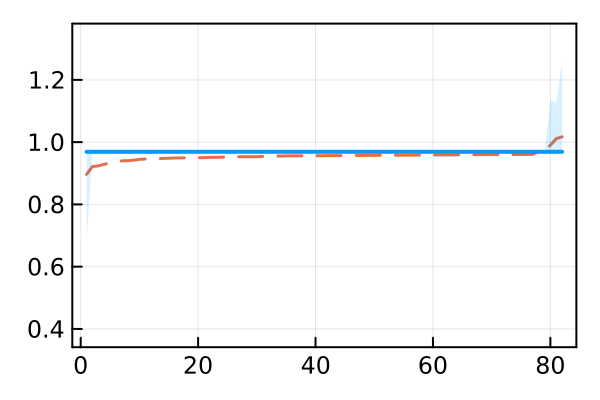} & 
			\includegraphics[width=0.3\textwidth]{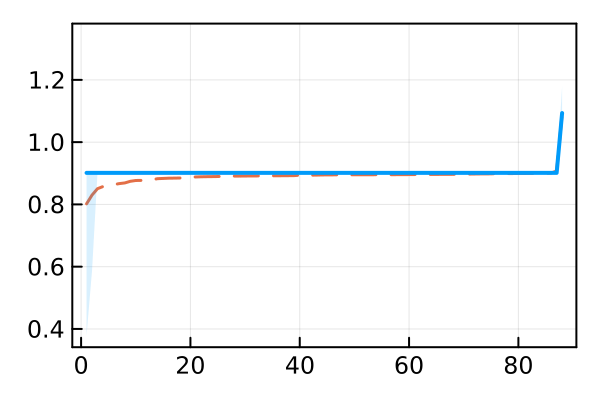} \\
			\rotatebox{90}{\hspace*{0.525in}  $\delta^{\sigma}_{i,u}$} &
			\includegraphics[width = 0.3\textwidth]{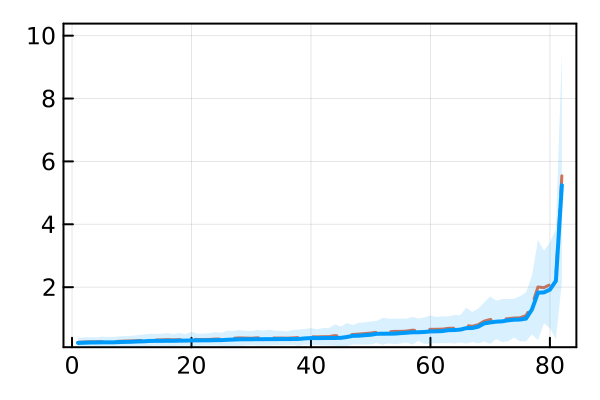} &
			\includegraphics[width=0.3\textwidth]{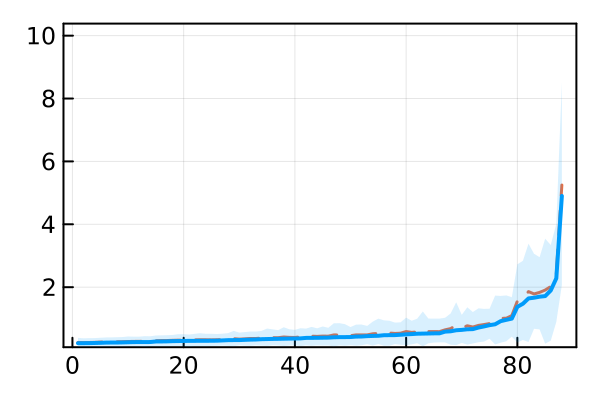} \\
			\rotatebox{90}{\hspace*{0.525in}  $\delta^{\sigma}_{i,\epsilon}$} &
			\includegraphics[width = 0.3\textwidth]{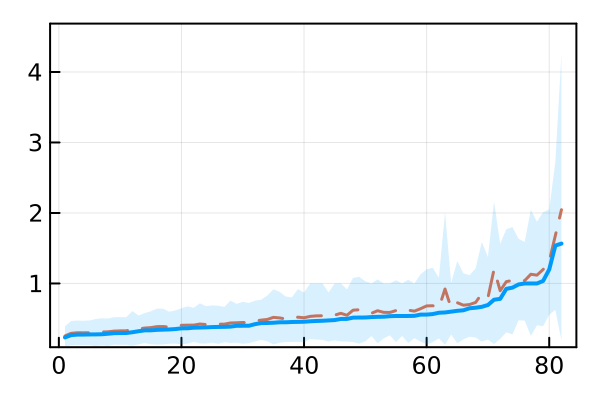} &
			\includegraphics[width=0.3\textwidth]{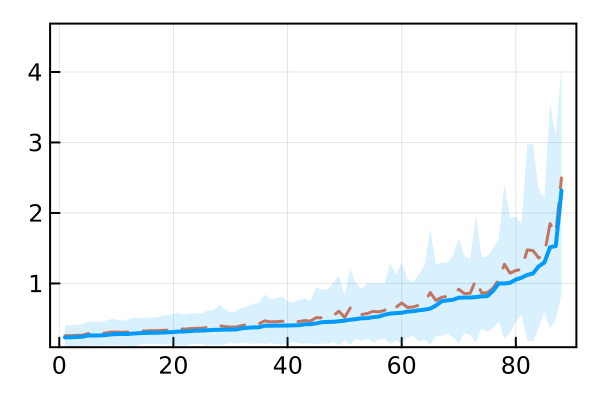} 
		\end{tabular}
	\end{center}
	{\footnotesize {\em Notes}: Solid blue lines in first three rows are $\mbox{med}(l|Y_{1:N,1:T})+\mbox{med}(\delta_i^l|Y_{1:N,1:T})$. In the bottom two rows they are $\mbox{med}(\delta_i^\sigma|Y_{1:N,1:T})$. The dashed red lines are posterior means and the light blue bands are 90\% credible bands.}\setlength{\baselineskip}{4mm}
\end{figure}

\noindent {\bf Idiosyncratic Parameters.} Of particular interest are the estimates of the idiosyncratic parameters and the degree of heterogeneity they exhibit. The results are summarized in Figure~\ref{fig:empirics.post.hip.hetsk.T20}. The figure shows posterior estimates for $\alpha_{0i}$, $\alpha_{1i}$, $\rho_i$, $\delta_{i,u}^\sigma$, and $\delta_{i,\epsilon}^\sigma$ for two samples ${\cal B}_{T,\tau}$. The panels in the left column correspond to $\tau=1988$ and the panels in the right column to $\tau=1991$. In both cases $T=20$. We depict posterior medians (solid blue lines), means (dashed red lines), and 90\% credible intervals (light blue bands). Units are sorted based on posterior median estimates. As discussed in Section~\ref{sec:sparse heterogeneity}, posterior median estimates of $\delta_i^l$ can deliver exact zeros. To highlight the potential sparsity implied by the posterior medians of $\delta_i^l$, we plot $\mbox{med}(l|Y_{1:N,1:T})+\mbox{med}(\delta_i^l|Y_{1:N,1:T})$ instead of $\mbox{med}(l+\delta_i^l|Y_{1:N,1:T})$ in the first three rows for $l \in \{\alpha_0,\alpha_1\}$. 	

Sparse heterogeneity manifests itself through a long flat section of the posterior median line. For instance, the $\tau = 1988$ sample consists of 82 observations. The number of units with $\mbox{med}(\delta_i^l|Y_{1:N,0:T})=0$ is 74 for $\alpha$. Thus, the $\alpha_i$s feature sparse heterogeneity. From examining the posterior distribution of the $\alpha_{1i}$s, which are divided by 10 to undo the rescaling of experience in (\ref{eq:empirics.xit}), we find that some units have a return experience that is up to 0.5 percentage points less than that of the core group, whereas others have a return that is up to 1.5 percentage points above that of the core group. The $\tau=1991$ sample has more heterogeneity in the $\alpha$ dimension, with only 32 out of 88 units having $\mbox{med}(\delta_i^\alpha|Y_{1:N,0:T})=0$. Unlike the posterior median curves, the posterior mean curves have generally no flat segments, because the posterior means of $\delta_i^\alpha$ are never exactly equal to zero.

For the autocorrelation and the two shock variances a different picture emerges. For $\rho$ essentially all units have $\mbox{med}(\delta_i^l|Y_{1:N,0:T})=0$, meaning that the autocorrelations are homogeneous, whereas for the variances essentially all of the units have $\mbox{med}(\delta_i^l|Y_{1:N,0:T})\not=1$ (recall the different centering of the variance discrepancies). Thus, the $\sigma^2_i$s are fully heterogeneous, meaning there is strong cross-sectional heteroskedasticity, a point that is also emphasized, for instance, by \cite{GuKoenker2014}.

\begin{figure}[t!]
	\caption{Posterior Densities for the Sizes of $\alpha$ and $\rho$ Group, $M_2$, $T=20$}
	\label{fig:empirics.post.core.size}
	\begin{center}
		\begin{tabular}{cc}
			$\tau = 1988$ & $\tau = 1991$ \\
			\includegraphics[width = 0.4\textwidth]{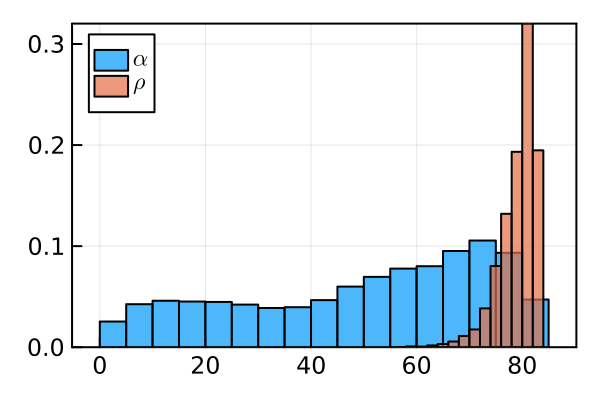} &
			\includegraphics[width=0.4\textwidth]{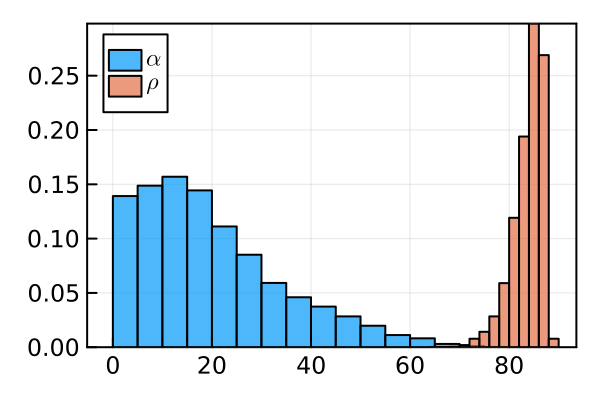} \\
		\end{tabular}
	\end{center}
	{\footnotesize {\em Notes}: The number of units is $N=82$ for the 1988 sample and $N=88$ for the 1991 sample.}\setlength{\baselineskip}{4mm}
\end{figure}

Figure~\ref{fig:empirics.post.core.size} shows posterior densities for the sizes of the $\alpha$ and $\rho$ core groups, defined as $N-\mathbb{E} \big[ \sum_{i=1}^N z_i^l \, | \, Y_{1:N,0:T} \big]$.	The posterior modes of the $\rho$ core group sizes are 80 (out of 82) and 82 (out of 88), respectively. For the 1988 sample the posterior histogram for the size of the $\alpha$ group peaks at around 70, but the probability mass is spread out between zero (homogeneity) and 82 (full heterogeneity). In the 1991 sample, the probability mass for the $\alpha$ core group size shifts to the left and much of it lies between 0 and 30. We do not plot the distribution of the $\sigma$ group sizes, as it is clear from Figure~\ref{fig:empirics.post.hip.hetsk.T20} that there are no core groups and the coefficients are fully heterogeneous. 

\begin{table}[t!]
	\caption{Posterior Distribution for $\theta$, $M_2$, $T=20$}
	\label{tab:empirics.posterior}
	\begin{center}
		\begin{tabular}{lcccc} \hline \hline
			 & \multicolumn{2}{c}{$\tau=1988$} &  \multicolumn{2}{c}{$\tau=1991$} \\
			Para & Median & HPD & Median & HPD  \\ \hline 
			$\alpha_0$ & -0.23 & [-0.30, -0.18] & -0.18 & [-0.25, 0.11] \\
			$\alpha_1/10$ & .018 & [.015, .021] & .014 & [.011, .017] \\			
			$\rho$  & 0.97 & [0.91, 1.00] & 0.90 & [0.84, 0.95] \\
			$\mu_{s_0}$ & 0.00 & [-0.06, 0.06] & -0.01 & [-0.06, 0.04] \\
			$v_{s_0}$ & 0.03 & [0.01, 0.06] & 0.02 & [0.01, 0.04] \\
			$q^\alpha$ & 0.34 & [0.00, 0.84] & 0.81 & [0.53, 1.00] \\
			$q^\rho$ & 0.04 & [0.00, 0.12] & 0.05 & [0.00, 0.12] \\
			$q^{\sigma_u}$& 0.97 & [0.94, 1.00] & 0.98 & [0.95, 1.00] \\
			$q^{\sigma_\epsilon}$ & 0.97 & [0.93, 1.00] & 0.97 & [0.92, 1.00] \\
			$v_{\delta^{\alpha_0}}$ & 0.08 & [0.03, 0.21] & 0.05 & [0.02, 0.07] \\
			$v_{\delta^{\alpha_1}}/100$ & .0002 & [.0001, .0004] & .0001 & [.0001, .0002] \\
			$v_{\delta^\rho}$ & 0.21 & [0.11, 0.35] & 0.20 & [0.10, 0.33] \\
			$v_{\delta_{u}^{\sigma}}$ & 6.97 & [2.76, 12.59] & 8.36 & [3.40, 16.27] \\
			$v_{\delta_{\epsilon}^{\sigma}}$ & 4.58 & [1.63, 8.87] & 5.95 & [2.26, 11.23] \\ \hline
		\end{tabular}
	\end{center}
	{\footnotesize {\em Notes}: We report 90\% posterior HPD intervals. The $\alpha_1$ and $v_{\delta^{\alpha_1}}$ are re-scaled to undo the division of experience by 10 in the definition of $x_{it}$, see (\ref{eq:empirics.xit}).}\setlength{\baselineskip}{4mm}	
\end{table}

\noindent {\bf Posterior Distribution of $\theta$.} Table~\ref{tab:empirics.posterior} summarizes the posterior median and 90\% highest posterior density (HPD) estimates for the elements of the $\theta$ vector. The point estimates and credible intervals for $q^l$ line up with the information in Figures~\ref{fig:empirics.post.hip.hetsk.T20} and~\ref{fig:empirics.post.core.size}. The posterior medians of $q^{\sigma_u}$ and $q^{\sigma_\epsilon}$ are close to one and those for $q^{\rho}$ are close to zero. The posteriors for $q^\alpha$ are more spread out. For the 1988 sample the 90\% credible set ranges from 0 to 0.84 and for 1991 from 0.53 to 1.

As in Figure~\ref{fig:empirics.post.hip.hetsk.T20}, we rescale the  $\alpha_1$ (and $v_{\delta^{\alpha_1}}$) entries in the table to undo the division of experience by 10 in the definition of $x_{it}$ in (\ref{eq:empirics.xit}) and to facilitate its interpretation. For the 1988 sample the posterior median estimate of the return to an additional year of experience is 1.8\%. It is 1.4\% for the 1991 sample. Both the time-series and the cross-sectional variation contributes to the identification of the return to experience. Examining the cross-sectional distribution of experience at the beginning of the 1988 sample, we find that the 25th percentile is seven years of experience, the median is ten years, and the 75th percentile is thirteen years. For the 1991 sample these numbers are very similar.	

The size of the discrepancy $\delta_i^\alpha$ from the $\alpha$ value of the core group is controlled by the matrix $v_{\delta^\alpha}$. We report posterior summary statistics for the diagonal elements. For the 1988 sample, the posterior credible interval for $v_{\delta^\alpha_0}$ ranges from 0.03 to 0.21. The interval for $v_{\delta^\alpha_1}$ (divided by 100) extends from .0001 to .0004. To put these estimates into perspective we provide a comparison to \cite{Guvenen2009}, who reports a $v_{\delta^{\alpha_0}}$ estimate of .022 with a standard error of .074, and a $v_{\delta^{\alpha_1}}$ estimate of .00038 with a standard error of 0.0008. \cite{Hryshko2012} reports $\hat{v}_{\delta^{\alpha_1}}=.0004$. These estimates are based on a different subset of the PSID data and obtained from a different estimation procedure. Our $v_{\delta^{\alpha_0}}$ estimate is slightly higher, and the $v_{\delta^{\alpha_1}}$ estimate is slightly lower than those obtained in the two earlier papers, but intervals overlap. 

The HPD interval for the autocorrelation parameter $\rho$ ranges from 0.91 to 1.00 for the 1988 sample and from 0.84 to 0.95 for the 1991 sample. \cite{Guvenen2009} obtains $\hat{\rho}=0.988$ for RIP specification with homogeneous coefficients, and 0.82 for the HIP model with fully heterogeneous coefficients. Our posterior medians of 0.97 and 0.9 are in between those estimates. The $\tau=1988$ features less heterogeneity in the experience profiles and is therefore closer to Guvenen's RIP  model, whereas the 1991 one sample exhibits more heterogeneity in the $\alpha_i$s. Thus, we are able to reproduce the finding that homogeneous coefficient specifications are associated with more persistence in the stochastic component that is persistent. Because the estimated $q^\rho$ is close to zero, the sample contains hardly any information about $v_{\delta^\rho}$ and the posterior is close to the prior.

The estimated mean of the initial state $s_{i0}$ is zero. Using the posterior median for $v_{s_0}$ from the 1988 sample and ignoring the estimation uncertainty, the standard deviation of $s_{i0}$ is 0.17. A second back-on-the-envelope calculation using posterior median estimates of $\rho$, $v_{\delta_\epsilon^\sigma}$ and the average level of $\sigma^2_{\epsilon,t}$ in Figure~\ref{fig:empirics.sigma.t} below, leads to an unconditional standard deviation of $s_{it}$ of 1.25. Thus, focusing on a particular cohort, over time the income inequality due to the persistent stochastic component increases. This also shows up in the cross-sectional dispersion of the posterior mean estimates $\hat{s}_{it}$, when comparing period $t=1$ to $t=T=20$: the estimates range from -0.2 to 0.2 at the beginning of the sample, and from -1 to 1 at the end of the sample.

%

\begin{figure}[t!]
	\caption{Effect of $\alpha$ Heterogeneity on Income Inequality, ${\cal B}_{20,1988}$ Sample}
	\label{fig:empirics.ratio.var}
	\begin{center}
		\includegraphics[width = 0.5\textwidth]{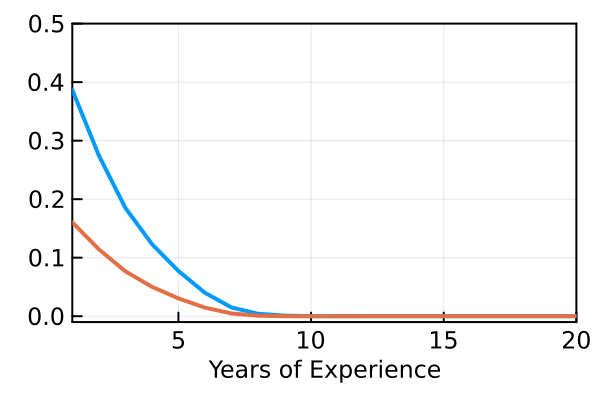} \\
	\end{center}
	{\footnotesize {\em Notes}: We plot $1 - V_t(\delta^\alpha=0)/V_t$ (solid blue line) and $1 - V_t(u=0)/V_t$ (solid red line) defined in the main text.}\setlength{\baselineskip}{4mm}
\end{figure}

Using our estimates, we can decompose income inequality in a cohort into the contribution of return to experience heterogeneity and heterogeneity in the {\em ex post} realization of the persistent stochastic process $s_{it}$. To construct the decomposition, we simulate a cohort of $N=10,000$ individuals for $T=20$ periods from the model $M_2$ using the posterior mean estimates presented in Figure~\ref{fig:empirics.post.hip.hetsk.T20} and Table~\ref{tab:empirics.posterior}. The cohort starts with an experience of $h_{i1}=1$. We subsequently conduct two more simulations in which we set either the return to experience heterogeneity $\delta_i^\alpha$  or the transitory shock variances $\sigma^2_{u,t}$ to zero, but use the same innovations as in the baseline simulation. We measure inequality as the cross-sectional variance of log income. For the baseline simulation, this variance is denoted by $V_t$, and for the alternative simulations they are denoted by $V_t(\delta^\alpha=0)$ and $V_t(u=0)$. 

The ratio $[V_t-V_t(\delta^\alpha=0)]/V_t$ for the ${\cal B}_{20,1988}$ sample is plotted in Figure~\ref{fig:empirics.ratio.var} as solid blue line. Recall that $\hat{q}^\alpha = 0.34$. Initially, return to experience heterogeneity, due to the individuals deviating from the core group, explains about 40\% of income inequality in the cohort. However, the share drops quickly and after seven years $\alpha$ heterogeneity is irrelevant and income inequality is solely due to the stochastic income components. The red line corresponds to the relative effect of the transitory shock on inequality. Initially, its is about 15\% but it also quickly converges to zero. Thus, after seven years, inequality is essentially solely due to the persistent idiosyncratic income shocks.




\begin{figure}[t!]
	\caption{Evolution of Average Idiosyncratic Volatility}
	\label{fig:empirics.sigma.t}
	\begin{center}
		\begin{tabular}{cc}
			$\sigma^2_{u,t}$ & $\sigma^2_{\epsilon,t}$ \\
			\includegraphics[width = 0.45\textwidth]{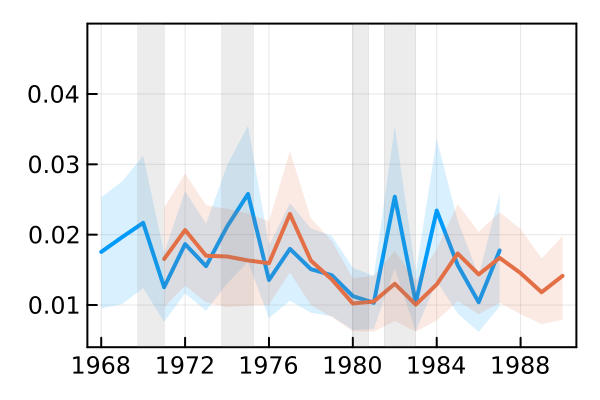} &
			\includegraphics[width = 0.45\textwidth]{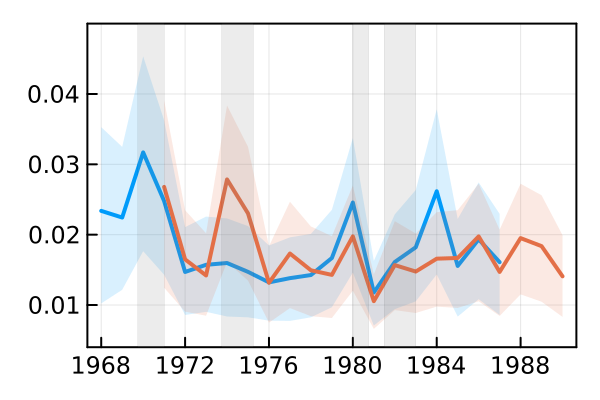}
			\\
		\end{tabular}
	\end{center}
	{\footnotesize {\em Notes}: Posterior means and 90\% credible intervals. Blue: $\tau = 1988$; Red: $\tau = 1991$. Grey shading denotes NBER-dated recessions.}\setlength{\baselineskip}{4mm}
\end{figure}

\noindent {\bf Time-path of Average Idiosyncratic Income Risk.}
Figure~\ref{fig:empirics.sigma.t} depicts posterior means and credible intervals for $\sigma^2_{u,t}$ and $\sigma^2_{\epsilon,t}$ as a function of time. We overlay the posteriors for the two samples: $\tau=1988$ and $\tau=1991$. Years with quarters that were classified by the NBER as recessions are indicated by the grey shaded areas. 
It is important to note that that the cross sections for the two panels in any particular year are not identical. This explains the variation in the estimates of the year-specific standard deviations. 

While the volatility spikes do not line up perfectly with the NBER recessions, there is a positive correlation. For the 1988 sample $\sigma^2_{u,t}$ spikes in the 1970, 1974, and 1982 recessions. On the other hand, it does not spike in the 1980 recession, but does have a pronounced peak during the recovery in 1984. The innovation variance $\sigma^2_{\epsilon,t}$ peaks in 1970 and 1980, but not in the other recessions. Note that individuals in our sample are always employed. Thus, we are not capturing unemployment risk, which contributes to the cyclical variation of income risk.


\begin{figure}[t!]
	\caption{Characteristics of Core Group vs. Deviators, ${\cal B}_{20,1991}$ Sample}
	\label{fig:empirics.group.characteristics}
    \begin{center}
    	\begin{tabular}{cc}
    	Age & Years of Education \\
    	\includegraphics[width = 0.45\textwidth]{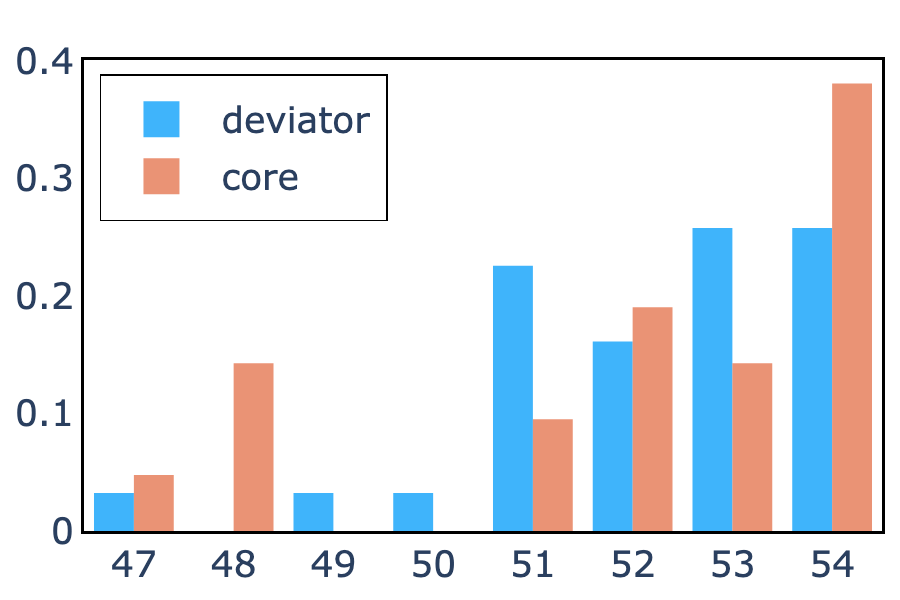} &
		\includegraphics[width = 0.45\textwidth]{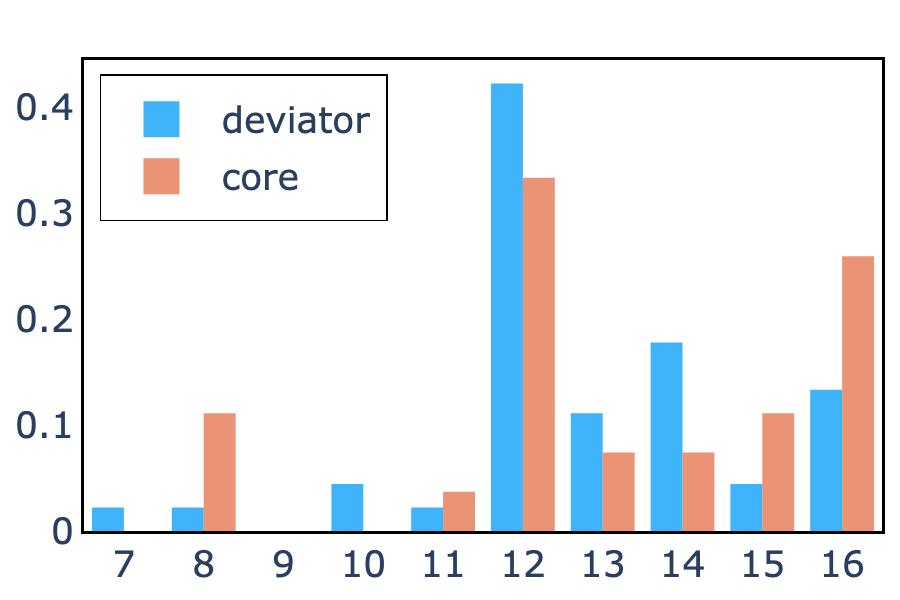} \\
		Mean of Hourly Earnings & Std. of Residual Income \\
		\includegraphics[width = 0.45\textwidth]{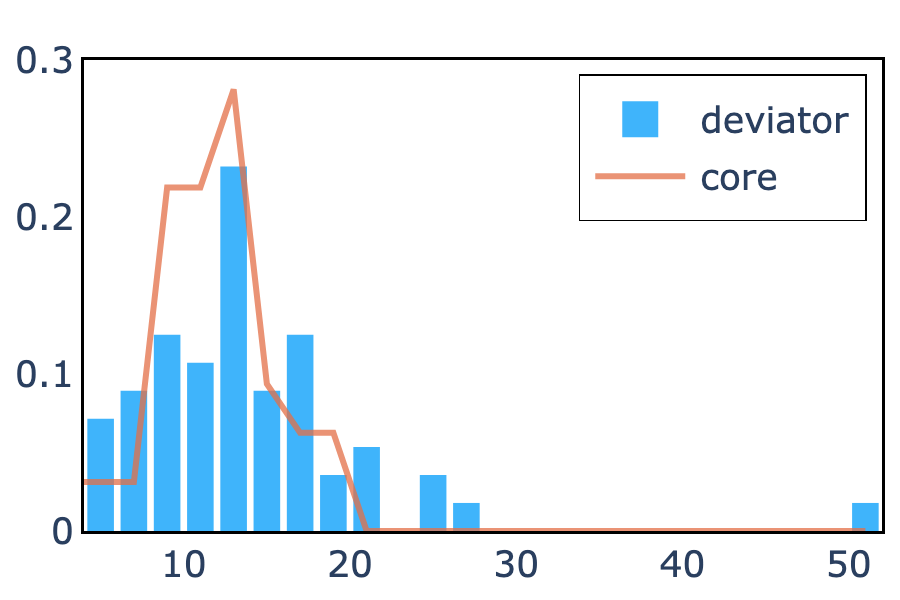} &
		\includegraphics[width = 0.45\textwidth]{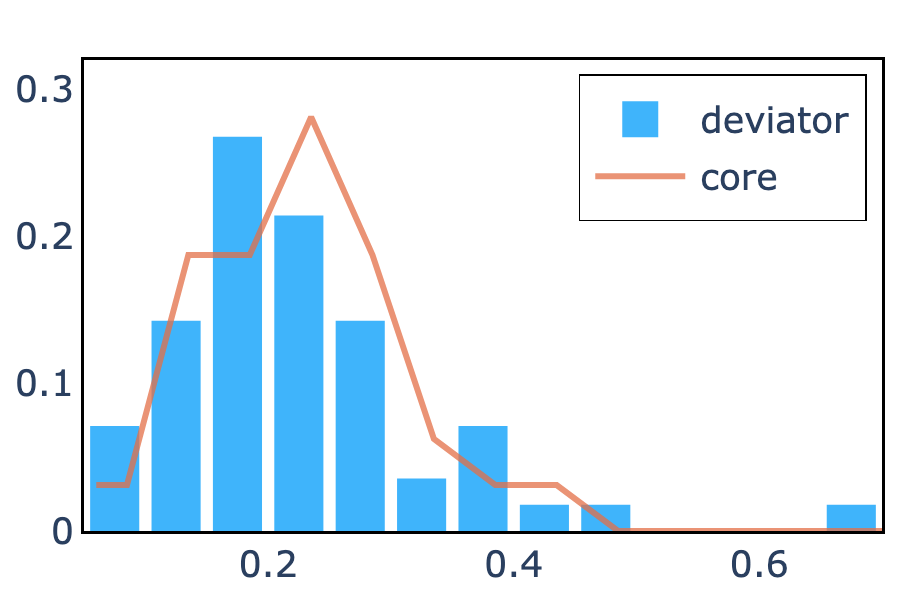} 				
    	\end{tabular}	
    \end{center}
    {\footnotesize {\em Notes}: Top panels: bar charts represent histograms for core group and deviators. Bottow panels: histograms for deviators and density estimates for core group.}\setlength{\baselineskip}{4mm}
\end{figure}

\noindent {\bf How Are Deviators Different?} In Figure~\ref{fig:empirics.group.characteristics} we compare characteristics of the core group members and the deviators. The top panels show two histograms and the bottom panels show a histogram and a density estimate. In terms of the age distribution and years of education the core group and deviators are quite similar. The core group contains slightly fewer individuals below the age of 50 and not as many individuals of age 54, but if the information is aggregated to a bin width of two or three years, many of these differences would vanish. The same is true for years of education. The distribution of time series means (computed prior to projecting onto demographics) for the deviators and the core group also look quite similar, except for the presence of a few high(er) income individuals among the deviators. A similar pattern emerges for the time series standard deviation of the residual income after projecting it unto the demographics. Overall, in terms of the distribution of age, years of education, mean earnings, and standard deviation of earnings residuals the core group members and deviators look quite similar.

\subsection{Prediction (Balanced Panels)}
\label{subsec:empirics.prediction}

We now examine the predictive properties of our estimated models. We consider two types of analyses. In the first part, we use the predictive performance as a way of assessing several restricted versions of model $M_2$. For a given sample ${\cal B}_{T,\tau}$, we estimate the model specifications based on data up to period $T$ and then we forecast income in period $T+1$ for the individuals $i=1,\ldots,N$ and evaluate the accuracy of the predictions, averaging across all individuals. In the second part of this section we examine the prediction problem from the perspective of an individual who would like to forecast her/his income to make economic decisions, e.g., a consumption/savings decision. 

\noindent {\bf Assessing Model Specifications Using Predictive Performance.} In Table~\ref{tab:empirics.models} we listed three variants of the baseline model $M_2$: $M_2$(homosk), $M_2$(RIP), and $M_2$(HIP). Our pseudo-out-of-sample approach to compare competing model specifications is different from what is typically done in the income-dynamics literature. Since a large number of authors have estimated the income models using minimum-distance techniques, e.g., by minimizing the discrepancy between sample autocovariances $\widehat{\mbox{Cov}}[y_{it},y_{i,t-h}]$ and the model-implied population analogues, it is natural to consider tests of the hypothesis $v_{\delta^\alpha}=0$. In our Bayesian setting we could in principle use posterior odds tests of $q^l=0$ versus $q^l > 0$, but because the income processes are often embedded in life-cycle models in which agents have to predict their future income to make optimal decisions, we directly focus on the pseudo-out-of-sample performance.

We estimate the model specifications for a sequence of samples ${\cal B}_{T,\tau}$, starting with $\tau=1988$ and ending with $\tau=1996$. Throughout, we consider $T=20$. The forecast evaluation approach is essentially the same as in \cite{LiuMoonSchorfheide2017}, \cite{LiuMoonSchorfheide2018}, and \cite{Liu2023}. As evaluation criteria we consider mean-squared-error (MSE) statistics and log predictive scores (LPS). The MSEs are computed as in Step~3 of \ref{algo:MC} and the LPS are defined as $\frac{1}{N} \sum_{i=1}^N \ln\hat{p}(y_{iT+1}|Y_{1:N,1:T})$, where $\hat{p}(\cdot)$ is the predictive density with a particular model specification.

\begin{figure}[t!]
\caption{One-Step-Ahead Predictive Performance Relative to $M_2$, $T=20$, various samples $\tau$}
\label{fig:empirics.panel.prediction2}
\begin{center}
	\begin{tabular}{cc}
		Relative Change in MSE [\%] & LPS Differential \\
		\includegraphics[height=2.3in]{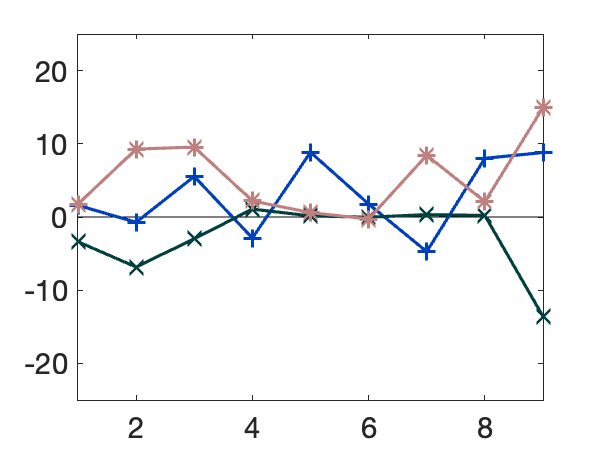} &
		\includegraphics[height=2.3in]{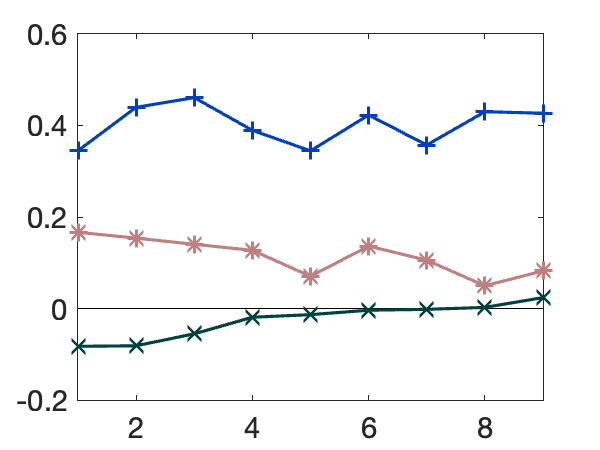} 
	\end{tabular}
\end{center}
{\footnotesize {\em Notes}: Samples $\tau$ are sorted by the value of LPS for $M_2$(RIP) in both panels. Negative values indicate an improvement. $M_2$(homosk), blue plus; $M_2$(RIP), dark green cross; $M_2$(HIP), brown asterisk. The 1988 and 1991 samples are ranked 4th and 6th, respectively. }\setlength{\baselineskip}{4mm}
\end{figure}

The results are summarized in Figure~\ref{fig:empirics.panel.prediction2}. In the left panel of the figure we show relative changes in MSEs [\%] defined as $100 \cdot \big(\mbox{MSE}(M_2(xxx))-\mbox{MSE}(M_2)\big)/\mbox{MSE}(M_2)$, where $M_2(xxx)$ is one of the alternative specifications in Table~\ref{tab:empirics.models}. Thus, a negative value indicates that model $M_2(xxx)$ attains a lower MSE than the benchmark model $M_2$. The LPS differentials are defined such that a negative value is an improvement, meaning that we report $\mbox{LPS}(M_2)-\mbox{LPS}(M_2(xxx))$. To convert them into predictive log odds, they would have to be multiplied by the cross-sectional dimension $N$, which various across samples but is between 82 and 127. 

We draw the following conclusions from Figure~\ref{fig:empirics.panel.prediction2}. First, the fully heterogeneous $M_2$(HIP) specification has a larger MSE than the baseline model $M_2$ for eight out of nine samples. Moreover, in terms of the LPS, $M_2$ clearly dominates $M_2$(HIP). Second, there are only three samples for which $M_2$(homosk) attains a lower MSE than $M_2$. In the other six samples $M_2$ delivers more accurate point forecasts. More importantly, the density forecasting performance of $M_2$(homosk) is dismal compared to $M_2$. This confirms that the cross-sectional heteroskedasticity is an important feature of the data. Finally, $M_2$(RIP) delivers a similar point forecast performance as $M_2$ in five and a similar LPS in six out of nine samples. In the remaining samples $M_2$(RIP) slightly dominates the baseline model. This is not surprising, because as we have seen in Section~\ref{subsec:empirics.parameters} regression coefficient heterogeneity is generally quite sparse and imposing a restriction that is not quite correct often favorably trades off a slight increase in bias against a variance reduction.

\begin{table}[t!]
	\caption{Prior Distribution for $\theta$}
	\label{tab:empirics.prior.ind}
	\begin{center}
		\begin{tabular}{llcc} \hline \hline
			Para & Distr. & $a$ & $b$ \\ \hline 
			$\delta^{\alpha}_i$ & ${\cal N}(a,b)$ & 0 & $\begin{bmatrix} 0.24 & 0 \\ 0 & 0.05  \end{bmatrix}$ \\
			$\delta^{\rho}_i$   & ${\cal N}(a,b)$ & 0.8 & 0.25  \\
			$\delta^{\sigma}_{i,u}$ & $IG \big( a/2,b/2 \big)$  & 4.02 & 0.101 \\
			$\delta^{\sigma}_{i,\epsilon}$  & $IG \big( a/2,b/2 \big)$ & 4.02 & 0.101 \\
			$s_{i0}$ & ${\cal N}(a,b)$ & 0 & 0.05 \\ \hline	 
		\end{tabular}
	\end{center}
	{\footnotesize {\em Notes}: ${\cal N}(a,b)$ is the Normal distribution, $IG(a,b)$ is the Inverse Gamma distribution, $B(a,b)$ is the Beta distribution, and $IW(a,b)$ is the Inverse Wishart distribution. The last to columns contain the 5th and 95th percentile of the marginal prior distributions.  For IG priors: $(a,b) = (4.02,0.101)$ implies a mean of 0.05 and a variance of 0.5. }\setlength{\baselineskip}{4mm}	
\end{table}

\noindent {\bf Forecasting From an Individual's Perspective.}
We previously generated simultaneous forecasts for all units to compare the predictive ability of competing model specifications. Now we consider the forecasting problem from the perspective on a particular individual $i$. We consider three different interval forecasts. The baseline forecast of individual $i$ conditions on the information $Y_{1:N,0:T}$ and accounts for parameter uncertainty. \cite{ChamberlainHirano1999} interpret this scenario as\footnote{This setup is also considered by \cite{GuKoenker2014}.} ``an individual seeking advice from a financial planner. The individual provides data on his earnings history and on various personal characteristics such as age and education. The planner has access to longitudinal data sets that provide data on earnings histories and personal characteristics for samples of individuals.'' (Abstract, Page 211) The second set of interval forecasts is also based on the $Y_{1:N,0:T}$ information, but we shut down the parameter uncertainty by fixing $(\delta, \theta)$ at the posterior mean estimates. 

Third, we generate forecasts based on the time series information of individual $i$: $Y_{i,0:T}$. To obtain forecasts based on $Y_{i,0:T}$, Equations (\ref{eq:model.ss.ME}) and (\ref{eq:model.ss.ST}) of model $M_2$ are re-estimated based on data from individual $i$ only. The prior distributions are summarized in Table~\ref{tab:empirics.prior.ind}. The hyperparameters for the priors are selected based on the prior specification for $M_2$, as outlined in Table \ref{tab:empirics.prior}. Specifically, the prior variance of $\delta^{\alpha}_i$ and $\delta^{\rho}_i$ are set to be equal to the prior mean of $v^{\delta^\alpha}$ and $v^{\delta^\rho}$, respectively. The prior mean of $\delta^{\sigma}_{i,u}$ and $\delta^{\sigma}_{i,\epsilon}$ are aligned with the prior means of $\sigma^2_{u,t}$ and $\sigma^2_{\epsilon,t}$. Both the prior mean and variance of $s_{i0}$ are matched with the prior mean of $\mu_{s_0}$ and $v_{s_0}$.

\begin{figure}[t!]
	\caption{Individual Earnings Forecasts, $\mathcal{B}_{20, 1991}$ Sample}
	\label{fig:empirics.multi.step}
	\begin{center}
		\begin{tabular}{cccc}
			&& ID 74   ($\delta^{\alpha} =  0$) & ID 59 ($\delta^{\alpha}\ne 0$) \\
			\rotatebox{90}{\hspace*{0.3in} $Y_{1:N,0:T}$ Info} &
			\rotatebox{90}{\hspace*{0.15in} With Para. Unc.} &
			\includegraphics[width = 0.4\textwidth]{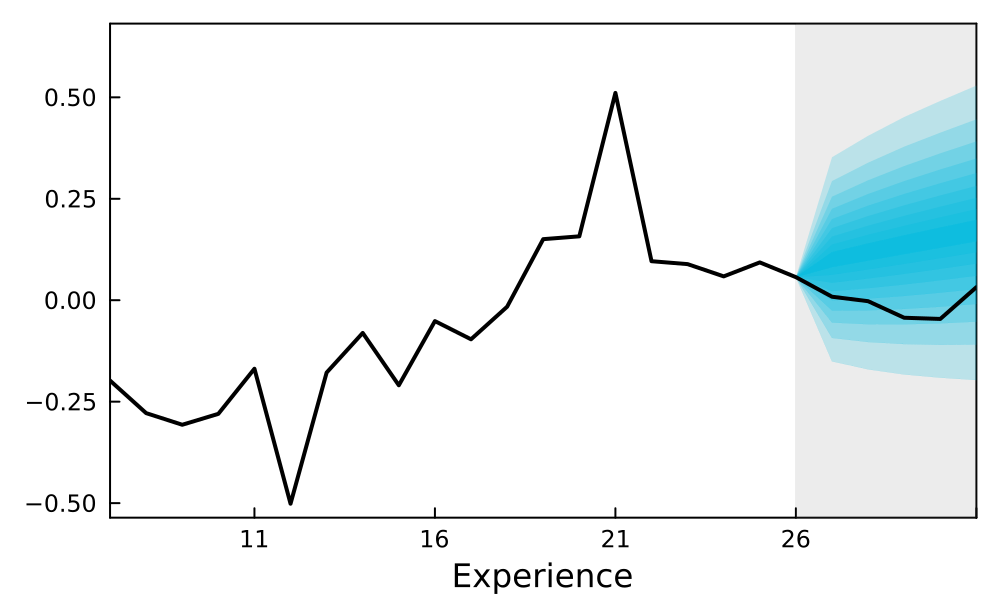} & 
			\includegraphics[width = 0.4\textwidth]{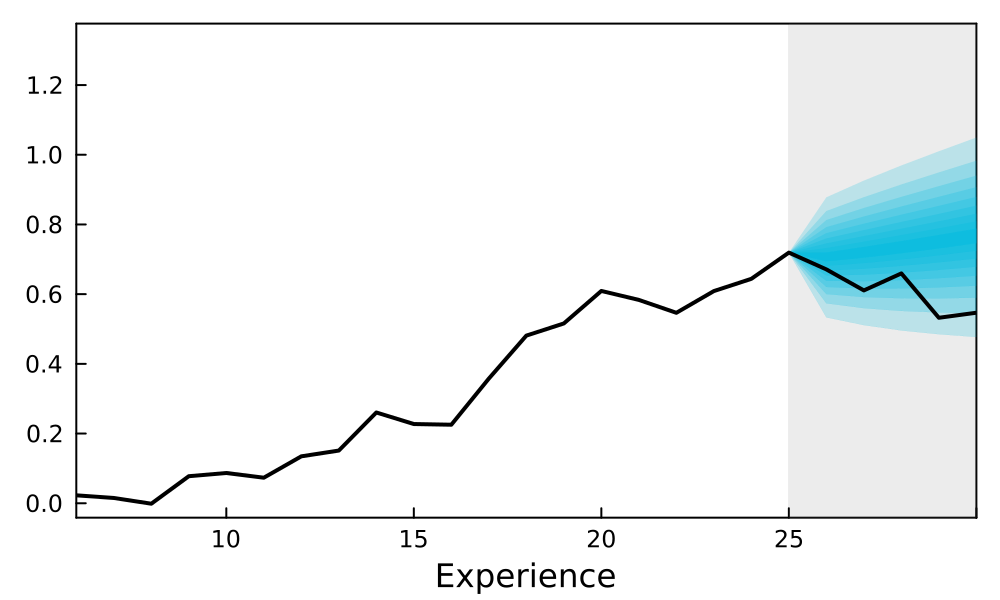} \\			
			\rotatebox{90}{\hspace*{0.3in} $Y_{1:N,0:T}$ Info} &
			\rotatebox{90}{\hspace*{0.15in} No Para. Unc.} &
  			\includegraphics[width=0.4\textwidth]{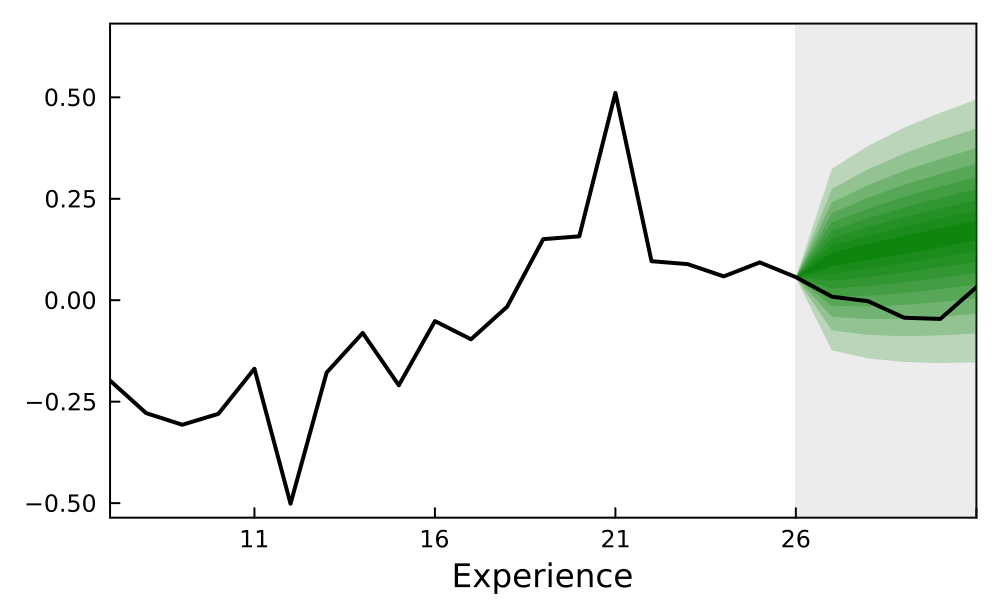} &
			\includegraphics[width=0.4\textwidth]{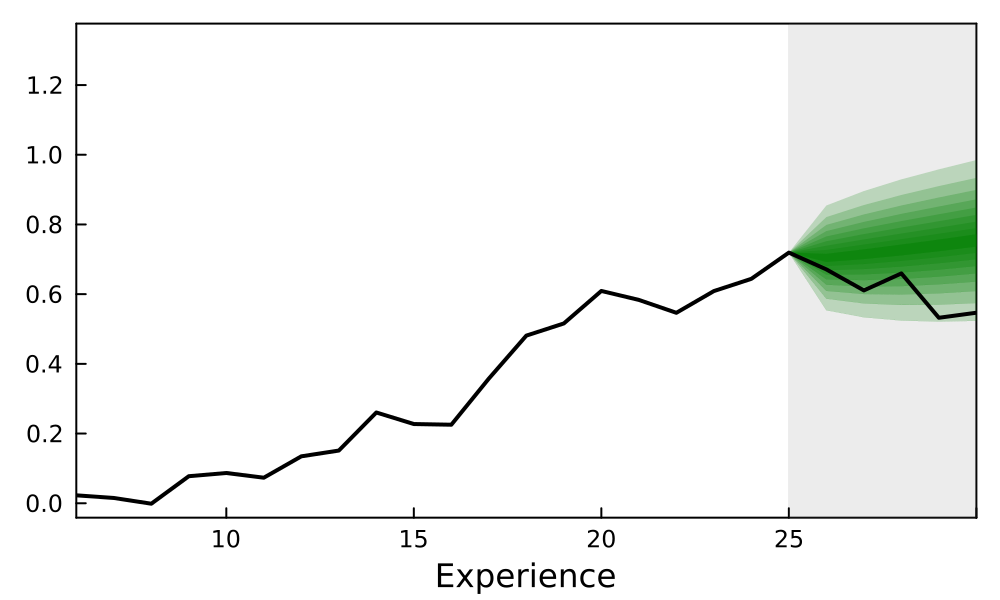} \\						
			\rotatebox{90}{\hspace*{0.3in} $Y_{i,0:T}$ Info} &
			\rotatebox{90}{\hspace*{0.15in} With Para. Unc.} &
			\includegraphics[width=0.4\textwidth]{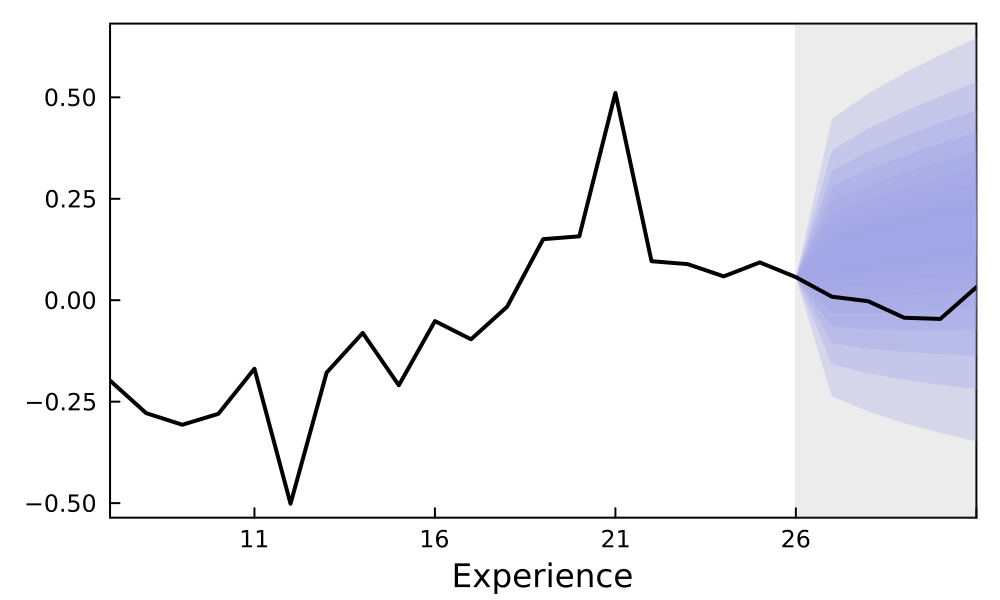} & 
			\includegraphics[width=0.4\textwidth]{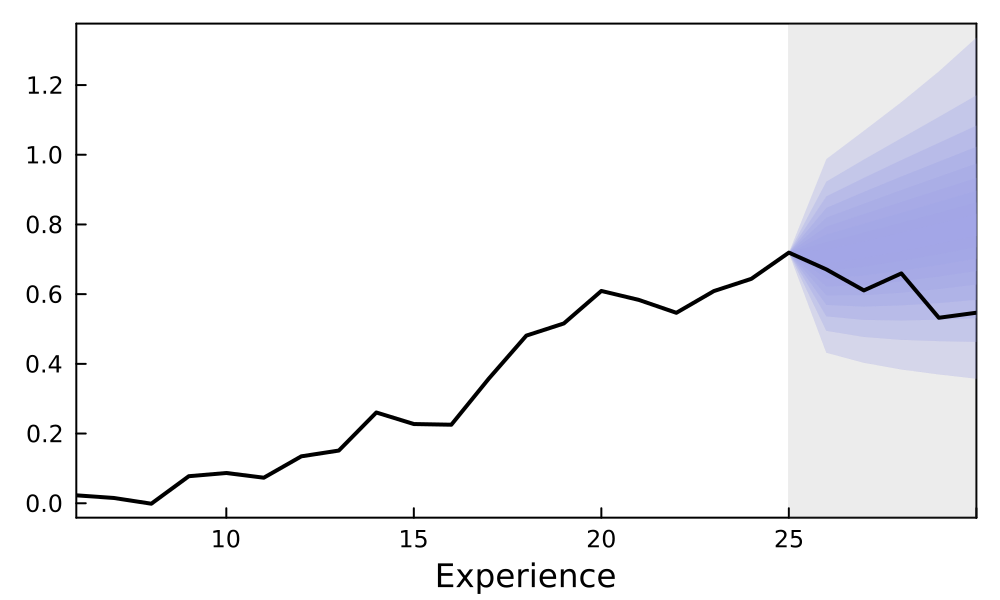}  						
		\end{tabular}
	\end{center}
	{\footnotesize {\em Notes}:  Forecast origin is period $t=T=20$. Pointwise prediction bands are shown with gradual shading indicating bands from the 0.05 to 0.95 quantiles.}\setlength{\baselineskip}{4mm}
\end{figure}

Figure~\ref{fig:empirics.multi.step} depicts fan charts to summarize interval forecasts under the three scenarios for a member of the core group, the unit with ID 74, and one of the deviators, the unit with ID 59.  We show their actual income for periods $t=1,\ldots,T=20$ and then interval forecasts for the next five years. A comparison of the panels in the first and second row of the figure suggests that shutting down parameter uncertainty shortens the predictive intervals somewhat. But, overall, the uncertainty appears to be dominated by future shocks. The forecasts in the third row of the figure are obtained by conditioning on the smaller information set $Y_{i,0:T}$, which widens the intervals relative to the baseline forecasts in the first row. The reason is that the cross-section provides information about the distribution of the idiosyncratic parameters that can be used as a prior to generate sharper time series estimates for unit $i$. 

\begin{figure}[t!]
	\caption{Ratio of Predictive Interval Widths, ${\cal B}_{20,1991}$ Sample}
	\label{fig:empirics.interval.width.ratios}
	\begin{center}
		\begin{tabular}{ccc}
			& \multicolumn{2}{c}{$Y_{1:N,0:T}$ Information with Parameter Uncertainty Versus } \\[1ex]
			& \hspace{1.2cm}$Y_{1:N,0:T}$ No Para. Unc. & \hspace{0.8cm} $Y_{i,0:T}$ with Para. Unc.\\
			\rotatebox{90}{\hspace*{0.5in} $H = 1$} &
            \includegraphics[width = 0.4\textwidth]{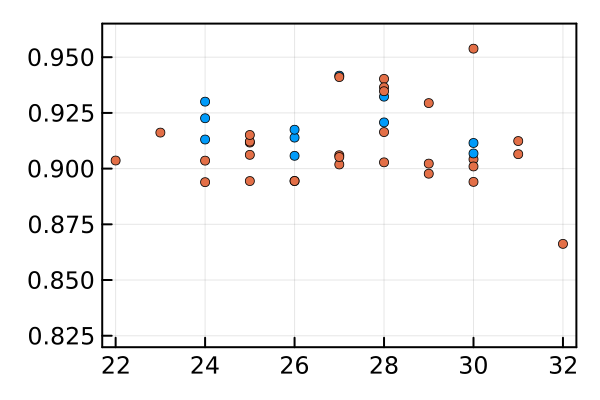} &
			\includegraphics[width= 0.4\textwidth]{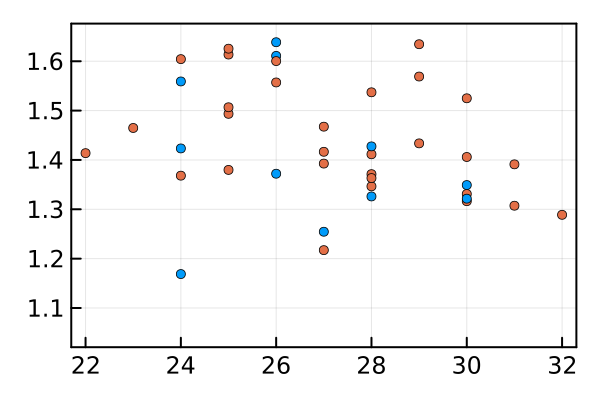} 
            \\
			\rotatebox{90}{\hspace*{0.5in} $H = 5$} &
			\includegraphics[width= 0.4\textwidth]{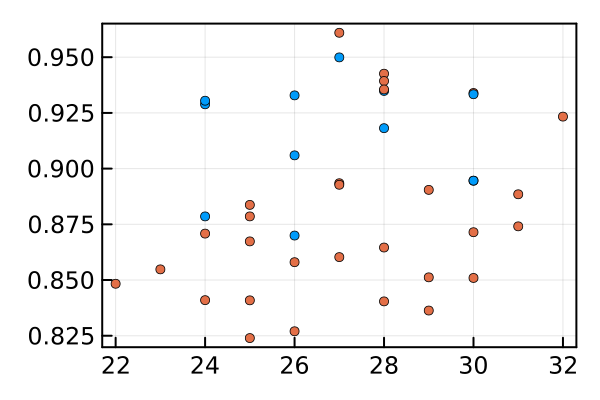}  &
			\includegraphics[width= 0.4\textwidth]{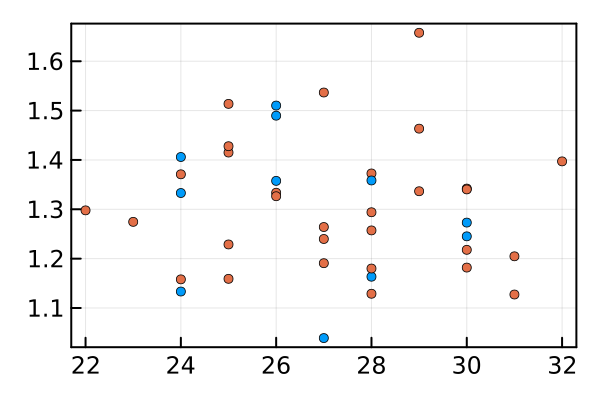} 
		\end{tabular}
	\end{center}
	{\footnotesize {\em Notes}: Ratio of width of 90\% predictive interval based on $Y_{1:N,0:T}$ information with parameter uncertainty versus two alternatives. $x$-Axis: years of experience. Core group: blue dots. Deviators: orange dots.}\setlength{\baselineskip}{4mm}
\end{figure}

To conduct a more systematic analysis of how parameter uncertainty and the information set used for generating the forecasts affect interval and density forecasts, we plot 90\% interval width ratios for all individuals $i$ in the ${\cal B}_{20,1991}$ sample in Figure~\ref{fig:empirics.interval.width.ratios} as a function of years of experience.  The denominators of the ratios are obtained from the forecasts based on $Y_{1:N,0:T}$ information with parameter uncertainty (baseline). The numerators for the left panels are the interval widths associated with the no-parameter uncertainty forecasts, whereas the numerators for the right panels are constructed from the $Y_{i,0:T}$. As suggested by Figure~\ref{fig:empirics.multi.step}, the ratios on the left are below one, whereas the ratios on the right are above one. 

Ratios for the core group are indicated through blue dots and values for the deviators by red dots. For the no-parameter-uncertainty-to-baseline ratios in the left panels, we expect the average ratio for the deviators to be smaller than for the core group. Because the parameter estimates for the deviators do not benefit from pooling observations across units, parameter uncertainty should play a more important role for the predictive distribution. For $h=1$ the average for the deviators is 0.91 and for the core group is 0.92. For $h=5$ the discrepancy is more pronounced as the average of the deviators drops to 0.88. However, none of these differences are statistically significant. 

For the $Y_{i,0:T}$-to-$Y_{1:N,0:T}$ information ratios plotted in the right panel, we expect the core group members to benefit more from including the information from other individuals. How much more depends on the size of the deviations, captured by the $v_\delta$ parameters. In principle, this translates into higher ratios for core group members than for deviators. It turns out, this effect is not reflected in the average ratios. For $h=1$ ($h=5$) the average core group ratio is 1.4 (1.3), whereas the average deviator ratio is 1.45 (1.31). As for the ratios in the left panels, the differences are not statistically significant. One reason for the insignificance could be that the classification of units into core members and deviators is not sharp. In general there is posterior uncertainty about the memberships, which might explain the insignificance.


Uncertainty about future income plays an important role in life-cycle models because it determines the magnitude of precautionary savings. Many models assume full information rational expectations, which in the context of our analysis means that there is no uncertainty about the parameters of the income process. The only source of uncertainty is the realization of future shocks. Some authors, e.g., \cite{Guvenen2007}, explicitly model the Bayesian learning about heterogeneous parameters, and account for parameter uncertainty in addition to shock uncertainty. Our estimates highlight, that the degree of uncertainty varies considerably based on the assumptions how the income forecasts are generated. The no-parameter uncertainty case could be interpreted as rational expectations. Accounting for parameter uncertainty increases the width of the 90\% predictive intervals by approximately 10\%. Using a sparser information set that only includes the income trajectory of individual $i$ herself, raises the width of the intervals by an additional 40\%.

%

\subsection{Results From an Unbalanced Panel}
\label{subsec:empirics.unbalanced}

\begin{figure}[t!]
	\caption{Posterior Idiosyncratic Parameters, Unbalanced Panel}
	\label{fig:app.post.hip.hetsk.unbalanced_v5}
	\begin{center}
		\begin{tabular}{cc}
			$\alpha_{i0}$ & $\alpha_{i1}/10$ \\
			\includegraphics[width = 0.4\textwidth]{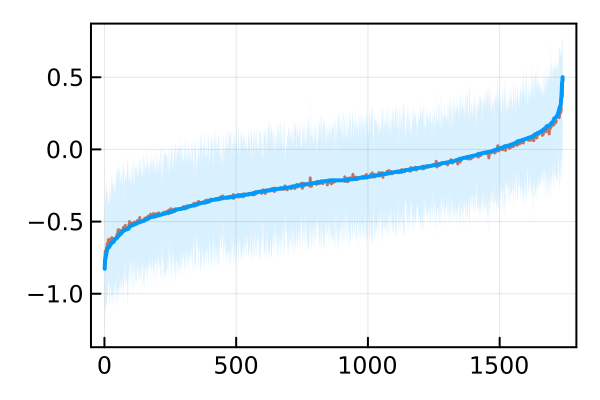} &
			\includegraphics[width=0.4\textwidth]{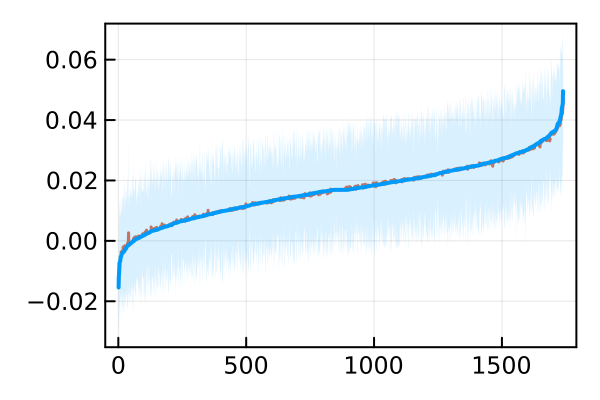} \\
			
			$\rho_i$ & $\delta^{\sigma}_{i,\epsilon}$ \\
			\includegraphics[width=0.4\textwidth]{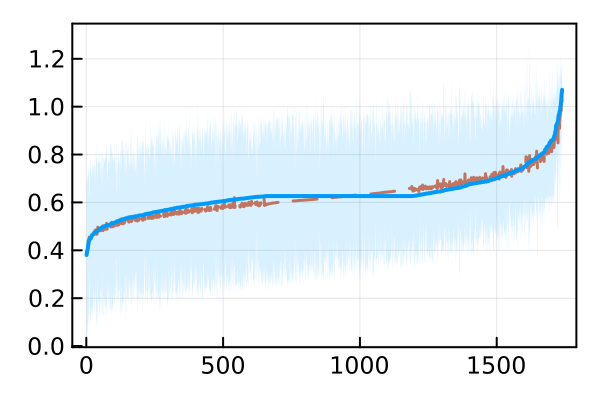} & 
			\includegraphics[width=0.4\textwidth]{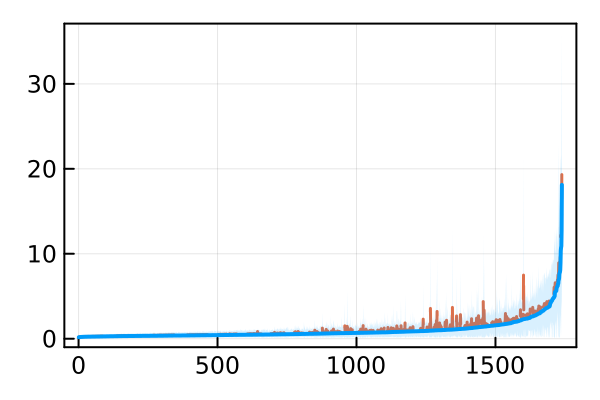} 			
		\end{tabular}
	\end{center}
	{\footnotesize {\em Notes}: Solid blue lines in $\hat{\alpha}^*_{i0}$, $\hat{\alpha}^*_{i1}/10$, and $\hat{\rho}_i^*$ panels are $\mbox{med}(l|Y_{1:N,1:T})+\mbox{med}(\delta_i^l|Y_{1:N,1:T})$. In the bottom-right panel they are $\mbox{med}(\delta_i^\sigma|Y_{1:N,1:T})$. The dashed red lines are posterior means and the light blue bands are 90\% credible bands.}\setlength{\baselineskip}{4mm}
\end{figure}

\noindent {\bf Idiosyncratic Parameters.} We now re-estimate model $M_2$ based on the unbalanced panel ${\cal U}$. Posterior estimates of the idiosyncratic parameters $\alpha_{i0}$, $\alpha_{i1}/10$, $\rho_i$, and $\delta_{i,\epsilon}^\sigma$ are provided in Figure~\ref{fig:app.post.hip.hetsk.unbalanced_v5}. In each panel, we sort the units by the posterior median estimates. As before, heteroskedasticity, i.e., full heterogeneity of $\delta_i^\sigma$ is a key feature of the panel. Unlike in the case of the balanced panels in Figure~\ref{fig:empirics.post.hip.hetsk.T20}, there is no evidence of sparsity in the $\delta_{i}^\alpha$ estimates in this larger panel data set. The sorted posterior median estimates of $\alpha_{i0}$ and $\alpha_{i1}/10$ do not exhibit any flat segments. The only parameter that does have a flat segment in the posterior median function is $\rho_i$, but the flat segment is not as pronounced as under the ${\cal B}_{20,1988}$ and ${\cal B}_{20,1991}$ samples. Thus, as expected, expanding the cross-sectional dimension of the panel by including more individuals, pushes the estimates toward full heterogeneity.

\begin{table}[t!]
	\caption{Posterior Distribution for $\theta$,  Unbalanced Panel}
	\label{tab:empirics.posterior.unbalanced}
	\begin{center}
		\scalebox{0.95}{
		\begin{tabular}{l cc cc cc} \hline \hline
			& \multicolumn{2}{c}{$M_2$} & \multicolumn{2}{c}{$M_2$(HIP)} & \multicolumn{2}{c}{$M_2$(RIP)} \\
			\cline{2-3} \cline{4-5} \cline{6-7} 
			Para & Median & HPD  & Median & HPD & Median & HPD \\ \hline 
			$\alpha_0$ & -0.21 & [-0.23, -0.20] & -0.21 & [-0.23, -0.20] & -0.37 & [-0.39, -0.35]  \\
			$\alpha_1/10$ & .017 & [.016, .018] & .017 & [.016, .018] & .020 & [.019, .021]  \\			
			$\rho$  & 0.63 & [0.60, 0.65] & 0.63 & [0.61, 0.65] & 0.97 & [0.97, 0.98]   \\
			$\mu_{s_0}$ & -0.23 & [-0.26, -0.20] & -0.23 & [-0.23, -0.20] & -0.02 & [-0.04, -0.00]   \\
			$v_{s_0}$ & 0.26 & [0.23, 0.29] & 0.27 & [0.24, 0.30] & 0.14 & [0.13, 0.15] \\
			$q^\alpha$ & 0.99 & [0.97, 1.00] & - & -  & - & - \\
			$q^\rho$ & 0.89 & [0.77, 1.00] & - & - & - & - \\
			$q^{\sigma_u}$ & 1.00 & [0.99, 1.00] & 1.00 & [0.99, 1.00] & 1.00 & [0.99, 1.00]\\
			$q^{\sigma_\epsilon}$ & 1.00 & [0.99, 1.00] & 1.00 & [0.99, 1.00] &1.00 & [0.99, 1.00]\\
			$v_{\delta^{\alpha_0}}$ & 0.09 & [0.08, 0.10] & 0.08 & [0.07, 0.10] & - & -  \\
			$v_{\delta^{\alpha_1}}/100$ & .002 & [.002, .002]  & .002 & [.002, .002] & - & - \\
			$v_{\delta^\rho}$ & 0.05 & [0.04, 0.05] & 0.04 & [0.04, 0.05] & - & -  \\
			$v_{\delta_{u}^{\sigma}}$ & 42.56 & [17.75, 78.24] & 43.03 & [18.72, 79.81] & 42.37 & [19.04, 82.14]  \\
			$v_{\delta_{\epsilon}^{\sigma}}$ & 35.85 & [15.23, 67.73] & 35.56 & [16.31, 68.49] & 30.07 & [12.90, 53.81]  \\ \hline
			LPS Diff. & \multicolumn{2}{c}{-} & \multicolumn{2}{c}{-.000} & \multicolumn{2}{c}{-.007}  \\
			RMSE Diff. (\%) & \multicolumn{2}{c}{-} & \multicolumn{2}{c}{0.18} & \multicolumn{2}{c}{-1.44}  \\ \hline
		\end{tabular}
	}
	\end{center}
	{\footnotesize {\em Notes}: For the parameters we report 90\% posterior HPD intervals. $\alpha_1$ and $v_{\delta^{\alpha_1}}$ are re-scaled to undo the division of experience by 10 in the definition of $x_{it}$, see (\ref{eq:empirics.xit}). LPS and RMSE differentials for the out-of-sample predictions are computed with respect to the baseline model $M_2$.}\setlength{\baselineskip}{4mm}	
\end{table}

\noindent {\bf Posterior Distribution of $\theta$.} Parameter estimates from the baseline model and the HIP and RIP versions are reported in Table~\ref{tab:empirics.posterior.unbalanced}. Consistent with Figure~\ref{fig:app.post.hip.hetsk.unbalanced_v5}, the estimates of $q^\alpha$ and $q^\rho$ are very close to one, implying full heterogeneity. The posterior median estimates of $\alpha_0$ and the return to experience $\alpha_1/10$ are with -0.21 and and .017, respectively, very close to the estimates for the balanced samples ${\cal B}_{20,1988}$ and ${\cal B}_{20,1991}$ reported in Table~\ref{tab:empirics.posterior}. Due to the larger sample size, the credible intervals for those parameters are much smaller for the unbalanced sample.

An important difference between the estimates from the ${\cal U}$ and ${\cal B}_{T,\tau}$ samples manifests itself in regard to the autocorrelation of the persistent income component $\rho$ and the return-to-experience heterogeneity captured by $v_{\delta^{\alpha_1}}$. For the unbalanced sample $\hat{\rho} =0.63$, which is considerably lower than the estimates obtained for the balanced samples and the numbers generally reported in the literature. In regard to $v_{\delta^{\alpha_1}}$, recall that the estimates for the two balanced panels reported in Table~\ref{tab:empirics.posterior} were .0002 and .0001. In the unbalanced panel the estimate is an order of magnitude larger: $\hat{v}_{\delta^{\alpha_1}} = .002$. Qualitatively, the low estimate of $\rho$ is consistent with the larger return-to-experience heterogeneity, as these two parameters have found to be negatively related in the existing literature.

%
%

The low estimate of $\rho$ and the high estimate of $\hat{v}_{\delta^{\alpha_1}}$ implies that return to experience heterogeneity and transitory shocks play a more important role for income inequality within cohorts. For the ${\cal B}_{20,1988}$ sample in Figure~\ref{fig:empirics.ratio.var} we found that with five years of experince, $\alpha$ heterogeneity generates about 8\% of the income inequality and the transitory income shock about 4\%. With eight years of experience the contribution is essentially zero. Using the estimates from the unbalanced panel instead, $\alpha$ heterogeneity  and transitory shock explain 40\% and 10\% of the income inequality within the cohort after five years of work experience and 25\% and 5\% after ten years.

Comparing the $v_{\delta^\sigma}$ estimates from the unbalanced panel to those from the balanced panels in Table~\ref{tab:empirics.posterior} one also notices a substantial increase in the shock variances. For instance, for ${\cal B}_{20,1988}$ the posterior median estimate of $v_{\delta^\sigma_\epsilon}$ is 4.58, whereas it equals 35.85 for the unbalanced panel. To interpret the magnitude of $\delta_{i,\epsilon}^\sigma$ (and its variance), recall that it has to be multiplied by $\sigma^2_{\epsilon,t}$ which we plotted in Figure~\ref{fig:empirics.sigma.t} for the balanced panels. The values for the unbalanced panel are slightly smaller. Overall, the cross-sectional averages of $\sigma^2_{\epsilon,t} \delta_{i,\epsilon}^\sigma$ are quite similar across samples.

The remaining columns of Table~\ref{tab:empirics.posterior.unbalanced} contain estimates for $M_2$(HIP) and $M_2$(RIP). The HIP estimates, generated conditional on $q^\alpha=q^\rho=1$, are essentially equal to the baseline estimates, which feature $\hat{q}^\alpha =0.99$ and $\hat{q}^\rho = 0.89$. The RIP estimates feature a substantially larger $\hat{\rho} = 0.97$. Moreover, the estimated return to experience is about three percentage points larger than for $M_2$ and $M_2$(HIP). In the bottom rows of Table~\ref{tab:empirics.posterior.unbalanced} we report results for the one-step-ahead out-of-sample predictive performance. As previously in Figure~\ref{fig:empirics.panel.prediction2}, we compute LPS differentials and relative RMSE differentials (in percent) with respect to the baseline model $M_2$. Negative values indicate an improvement. The differentials between $M_2$ and $M_2$(HIP) are essentially zero. The RIP specification with the more persistent income process leads to a small improvement in terms of point (RMSE) and density (LPS) forecasts. The RMSE gain is about 1.5\%. Multiplying the LPS differential by the cross-sectional sample size of approximately 800 observations at the forecast origin, we obtain log odds of about 5.6 in favor of the RIP specification. Here the parsimony of the RIP specification seems to improve the out-of-sample forecast performance if we average across all the units.


\subsection{Summary}
\label{subsec:empirics.summary}

Using balanced panels of individuals with a long employment history, we found that heterogeneity with respect to returns to experience and the autocorrelation of the persistent stochastic income component is sparse, with large core groups of individuals sharing identical coefficients and a small number of deviators. For a large unbalanced panel, we found evidence for full coefficient heterogeneity, with the caveat that the specification with homogeneous coefficients performs better in terms of one-step-ahead out-of-sample forecasting. We also examined forecasts for specific individuals conditional on information from the full panel $Y_{1:N,0:T}$ accounting for shock and parameter uncertainty, the full panel ignoring parameter uncertainty, and the individual history $Y_{i,0:T}$ accounting for shock and parameter uncertainty. This comparison has an important message for the macroeconomic life-cycle model literature: subtle assumptions about the agents information set can have large effects on the income uncertainty that agents face and hence on their precautionary savings motive.

\section{Conclusion}
\label{sec:conclusion}

We incorporated a version of a spike and slab prior, comprising a pointmass at zero (``spike'') and a Normal distribution around zero (``slab'') into a dynamic panel data framework to model coefficient heterogeneity. Our framework nests coefficient homogeneity and full heterogeneity as a special case, but can also capture sparse heterogeneity, meaning that there is a core group of units that share the same coefficients. The remaining units deviate from the core group. There is a straightforward extension of our framework to correlated random effects and one could also treat the slab part of the $\delta_i$ distribution nonparametrically as in \cite{Liu2023}. However, in many empirical applications simple parametric specifications tend to work well and we hope that the proposed framework will prove to be useful beyond the application presented in this paper.





\bibliography{ref}

%
%
%
%


\clearpage

\renewcommand{\thepage}{A.\arabic{page}}
\setcounter{page}{1}

\begin{appendix}

\markright{Online Appendix -- This Version: \today }
\renewcommand{\theequation}{A.\arabic{equation}}
\setcounter{equation}{0}

\renewcommand*\thetable{A-\arabic{table}}
\setcounter{table}{0}
\renewcommand*\thefigure{A-\arabic{figure}}
\setcounter{figure}{0}

\begin{center}

	{\large {\bf Online Appendix: Bayesian Estimation of Sparsely Heterogeneous Panel Models}}

	{\bf Hyungsik Roger Moon, Frank Schorfheide, and Boyuan Zhang}
\end{center}

\noindent The Online Appendix consists of the following parts:

\begin{itemize}
	\item[A.] Posterior Sampling for Variants of $M_1$
	\item[B.] Posterior Sampling for Variants of $M_2$
	\item[C.] Additional Empirical Results
\end{itemize}

\section{Posterior Sampling for Variants of $M_1$}
\label{appsec:posterior.sampling}

\subsection{Homoskedastic Model}
\label{appsubsec:posterior.sampling.benchmark.homo}

We first consider a homoskedastic model  with the restriction $\delta_{i}^{\sigma} = 1$.  Let $x_{it} = [1 \; y_{it-1}]'$, the model (\ref{eq:model.benchmark}) can be rewritten as
\begin{equation}
	y_{it} = x_{it}'(\beta + \delta_i)+ \sigma u_{it} \label{eq:model.benchmark.homo}
\end{equation}
or in its matrix form
\begin{equation}
	Y_{i} = X_{i}(\beta + \delta_i)+ \sigma u_{i}
\end{equation}
Assume the hyperprior for $q$ is $\text{Beta}(a,b)$, and $\sigma^2$ follows IG$\left( \frac{\underline{\nu}_\sigma}{2},\frac{\underline{\tau}_\sigma}{2} \right)$.
Rewrite the model using a set of latent variable $Z_{1:N} = [z_1,z_2,...,z_N]'$ and $z_i = [z_i^{\alpha}, z_i^{\rho}]'$. $z_i^{\alpha}$ equals 1 if $\delta_i^{\alpha} \ne 0$ and zero whenever $\delta^{\alpha}_i=0$. Similarly, $z_i^{\rho} = 1$ if $\delta_i^{\rho} \ne 0$ and equals zero when $\delta^{\rho}_i=0$. We define $Y_{i}$ to be a $T \times 1$ vector: $Y_{i}=[y_{i1},y_{i2},...,y_{iT}]'$, and $X_{i}$ to be a $T \times 2$ matrix  with rows $x_{it}'$. Denote $\beta = [\alpha \; \rho]'$, $\delta_i = [\delta_i^{\alpha} \; \delta_i^{\rho}]'$, and $v_{\delta} = [v_{\delta^\alpha} \; v_{\delta^\rho}]'$. We present the sampler for the case in which  $\delta^\alpha$ and $\delta^\rho$ are independent. This is the algorithm used to generate the Monte Carlo results in Section~\ref{sec:MC} of the paper. 
We also present samplers for the homogeneous ($q=0$) and the fully heterogeneous ($q=1$) specifications, which are special cases of the $q\not=0$ samplers.

\subsubsection{Spike-and-Slab Prior: $\delta^\alpha$ and $\delta^\rho$ are Independent}
\label{appsubsubsec:appsubsec:posterior.sampling.benchmark.homo.independent}

The posterior of the unknown objects of this model is:
\begin{eqnarray*}
	\lefteqn{p(\beta, \delta, \sigma^2, Z_{1:N}, q, v_{\delta}| Y_{1:N})} \\
	&\propto & p(Y_{1:N}|\beta,\delta, \sigma^2) p(\delta | Z_{1:N}, v_{\delta}) p(\beta)  p(Z_{1:N}|q) p(q) p(\sigma^2) p(v_{\delta}) \\
	&\propto & \prod_{i=1}^{N} \left( \fra{1}{2\pi \sigma^2} \right)^{\frac{T}{2}} \exp \left \{- \fra{1}{2\sigma^2} \left[Y_{i}-X_{i}(\beta + \delta_i) \right]' \left[Y_{i}-X_{i}(\beta + \delta_i) \right]\right\}\\
	& & \times \prod_{i=1}^N \left[ \left( \fra{1}{2\pi v_{\delta^{\alpha}}} \right)^{\frac{1}{2}} \exp \left( -\frac{\delta_i^{\alpha^2}}{2v_{\delta^{\alpha}}}\right)\right]^{z_i^{\alpha}} \mathbb{I}\{\delta^\alpha_i = 0\}^{1-z_i^{\alpha}}
	\left[ \left( \fra{1}{2\pi v_{\delta^{\rho}}} \right)^{\frac{1}{2}} \exp \left( -\frac{\delta_i^{\rho^2}}{2v_{\delta^{\rho}}}\right)\right]^{z_i^{\rho}} \mathbb{I}\{\delta^\rho_i = 0\}^{1-z_i^{\rho}} \\
	& & \times \fra{1}{2 \pi } \left| \underline{v}_{\beta} \right|^{-\frac{1}{2}} \exp \left(-\frac{1}{2} \beta' \underline{v}_{\beta}^{-1} \beta \right)\\
	& & \times \prod_{i=1}^N (q^{\alpha})^{z_i^{\alpha}}(1-q^{\alpha})^{1-z_i^{\alpha}} (q^{\rho})^{z_i^{\rho}}(1-q^{\rho})^{1-z_i^{\rho}} \\
	& & \times (q^{\alpha})^{a-1}(1-q^{\alpha})^{b-1} \cdot (q^{\rho})^{a-1} (1-q^{\rho})^{b-1} \\
	& & \times \left(\frac{1}{\sigma^2}\right)^{\frac{\underline{\nu}_{\sigma}}{2}+1} \exp \left(-\fra{\underline{\tau}_\sigma}{2\sigma^2}\right) \\
	& &
	\times  \left(\frac{1}{v_{\delta^{\alpha}}}\right)^{\frac{\underline{\nu}_{\delta^{\alpha}}}{2}+1} \exp \left(-\fra{\underline{\tau}_{\delta^{\alpha}}}{2v_{\delta^{\alpha}}}\right)
	\left(\frac{1}{v_{\delta^{\rho}}}\right)^{\frac{\underline{\nu}_{\delta^{\rho}}}{2}+1} \exp \left(-\fra{\underline{\tau}_{\delta^\rho}}{2v_{\delta^{\rho}}}\right),
\end{eqnarray*}
where $\underline{v}_{\beta} = \mbox{diag}(\underline{v}_\alpha,\underline{v}_\rho)$ and $\mathbb{I}\{x=a\}$ is the indicator function that is equal to one if $x=a$ and equal to zero otherwise.
We can sample from the joint posterior of $(\beta, \delta, q, Z_{1:N}, \sigma^2, v_{\delta})$ using a Gibbs sampling algorithm with five blocks.

\noindent {\bf Conditional Posterior of $\beta$:} Let $\widetilde{Y}_i = Y_{i} - X_{i} \delta_i$. Then
\begin{eqnarray*}
	p(\beta| Y_{1:N}, \delta, \sigma^2)
	& \propto & \exp \left[- \fra{1}{2\sigma^2} \sum_{i=1}^N \left(\widetilde{Y}_i-X_{i}\beta \right)'  \left(\widetilde{Y}_i-X_{i}\beta \right) \right] \exp \left( - \frac{1}{2} \beta' \underline{v}_{\beta}^{-1} \beta \right)\\
	& \propto & \exp \left\{ - \frac{1}{2} ( \beta -  \bar{\beta})' \bar{v}_{\beta}^{-1} ( \beta -  \bar{\beta} ) \right\},
\end{eqnarray*}
where
\[
\bar{v}_{\beta} = \left( \underline{v}_{\beta}^{-1} + \sigma^{-2} \sum_{i=1}^{N} X_{i}' X_{i} \right)^{-1}, \quad \bar{\beta} = \bar{v}_{\beta} \sigma^{-2} \sum_{i=1}^{N} X'_i \widetilde{Y}_i.
\]
This implies
\be
\beta| (Y_{1:N}, \delta,\sigma^2) \sim N \left( \bar{\beta}, \bar{v}_{\beta} \right).
\label{appeq:benchmark.homo.post.beta}
\ee

\noindent {\bf Conditional Posterior of $(q^\alpha,q^\rho)$} is given by
\[
p(q^l| Z_{1:N}) \; \propto \; (q^l)^{a+\psi(z^l)-1}(1-q^l)^{b+N-\psi(z^l)-1},
\]
where $\psi(z^l) = \sum_{i=1}^N z_i^{l}$ is the total number of non-zero elements in $\delta^{l}$. This implies
\be
  q^l | Z_{1:N} \sim \text{Beta}(a+\psi(z^l),b+N-\psi(z^l)), \quad l \in \{\alpha,\rho\}.
  \label{appeq:benchmark.homo.post.q}
\ee

\noindent {\bf Conditional Posterior of $(v_{\delta^\alpha}, v_{\delta^\rho})$} is given by
\begin{eqnarray*}
	p(v_{\delta^l} | Z_{1:N}, \delta)
	& \propto & \prod_{i=1}^N \left[ \left( \fra{1}{v_{\delta^{l}}} \right)^{\frac{1}{2}} \exp \left( -\frac{\delta_i^{l^2}}{2v_{\delta^{l}}}\right)\right]^{z_i^{l}}  \left(\frac{1}{v_{\delta^{l}}}\right)^{\frac{\underline{\nu}_{\delta^{l}}}{2}+1} \exp \left(-\fra{\underline{\tau}_{\delta^{l}}}{2v_{\delta^{l}}}\right) \\
	& \propto & \left(\frac{1}{v_{\delta^{l}}}\right)^{\frac{\underline{\nu}_{\delta^{l}} + \psi(z^l)}{2}+1} \exp \left(-\frac{\underline{\tau}_{\delta^{l}} + \sum_{i | z_i^l = 1} \delta_{i}^{l^2}}{2v_{\delta^{l}}}\right).
\end{eqnarray*}
This implies
\be
v_{\delta^l} | (Z_{1:N}, \delta) \sim \text{IG} \left(\frac{\bar{\nu}_{\delta^l}}{2},\frac{\bar{\tau}_{\delta^l}}{2} \right), \quad l \in \{\alpha,\rho\},
\label{appeq:benchmark.homo.post.vdetla}
\ee
where
\begin{eqnarray*}
	\bar{\nu}_{\delta^l} & = & \underline{\nu}_{\delta^l} + \psi(z^l) \\
	\bar{\tau}_{\delta^l} & = & \underline{\tau}_{\delta^l} + \sum_{i | z_i^l = 1} \delta_{i}^{l^2}.
\end{eqnarray*}

\noindent {\bf Conditional Posterior of $(Z,\delta)$:} Notice that we assume cross-sectional independence. To derive the posterior $(Z_i,\delta_i)$ conditional on $(Y_{i}, \beta, q, \sigma^2, v_{\delta})$, we use Gibbs sampler to draw $(z_{i}^\alpha,\delta_{i}^\alpha)$ conditional on $(Y_{i}, \beta, q^\alpha, \sigma^2, z_{i}^\rho,\delta_{i}^\rho, v_{\delta^{\alpha}})$ and then draw $(z_{i}^\rho,\delta_{i}^\rho)$ conditional on $(Y_{i}, \beta, q^\rho, \sigma^2, z_{i}^\alpha,\delta_{i}^\alpha, v_{\delta^{\rho}})$.

We start with $z_{i}^\alpha,\delta_{i}^\alpha | Y_{i}, \beta, q^\alpha, \sigma^2, z_{i}^\rho,\delta_{i}^\rho, v_{\delta^{\alpha}}$. The posterior odds of $z_i^\alpha=1$ versus $z_i^\alpha=0$ are given by
\be
K_i^\alpha = \frac{\mathbb{P}(z_i^\alpha=1|Y_{i},\beta,q^\alpha,\sigma^2,\delta_{i}^\rho, v_{\delta^{\alpha}})}{\mathbb{P}(z_i^\alpha=0|Y_{i},\beta,q^\alpha, \sigma^2, \delta_{i}^\rho, v_{\delta^{\alpha}})} = \frac{q^\alpha}{1-q^\alpha} \frac{p(Y_{i}|\beta,\sigma^2, \delta_{i}^\rho, v_{\delta^{\alpha}}, z_i^\alpha=1)}{p(Y_{i}|\beta, \sigma^2, \delta_{i}^\rho, v_{\delta^{\alpha}}, z_i^\alpha=0)}.
\ee
Under $z_i^\alpha = 1$, $\delta_i^\alpha$ is set to follow a normal distribution. The marginalized likelihood of $Y_{i}|(\beta,\sigma^2, \delta_{i}^\rho, v_{\delta^{\alpha}}, z_i^\alpha=1)$ can be obtained by rearranging Bayes Theorem:
\be
p(Y_{i}|\beta, \sigma^2, \delta_{i}^\rho, v_{\delta^{\alpha}}, z_i^\alpha=1) =
\frac{p(Y_{i}|\beta, \sigma^2, \delta_i^{\alpha}, \delta_{i}^\rho, v_{\delta^{\alpha}}, z_i^\alpha=1)p(\delta_i^{\alpha}| v_{\delta^{\alpha}}, z_i^\alpha=1)}{p(\delta_i^\alpha | Y_{i}, \beta, \sigma^2, \delta_{i}^\rho, v_{\delta^{\alpha}}, z_i^\alpha=1)}.
\ee
It is given by
\begin{eqnarray*}
	p(Y_{i}|\beta, \sigma^2, \delta_{i}^\rho, v_{\delta^{\alpha}}, z_i^\alpha=1)
	& = & \frac{ (2\pi \sigma^2)^{-\frac{T}{2}} \exp \left ( - \frac{1}{2\sigma^2} \sum_{t=1}^T \widecheck{y}_{it}^{\alpha^2} \right) (2 \pi v_{\delta^\alpha} )^{-\frac{1}{2}}  \exp \left( - \frac{0^2}{2v_{\delta^\alpha}} \right) }
	{\left(2 \pi \bar{v}_{\delta^\alpha_i} \right)^{-\frac{1}{2}} \exp \left( - \frac{\bar{\delta}_i^{\alpha^2}}{2\bar{v}_{\delta^\alpha_i}} \right) } \\
	& = & (2\pi \sigma^2)^{-\frac{T}{2}} \left( \frac{v_{\delta^\alpha}}{\bar{v}_{\delta^\alpha_i}} \right)^{-\frac{1}{2}} \exp \left( - \frac{1}{2\sigma^2} \sum_{t=1}^T \widecheck{y}_{it}^{\alpha^2} + \frac{\bar{\delta}_i^{\alpha^2}}{2\bar{v}_{\delta^\alpha_i}} \right),
\end{eqnarray*}
where $\widecheck{y}^\alpha_{it} = y_{it} - \alpha - (\rho + \delta_i^\rho) y_{it-1}$, $\bar{\delta}_i^\alpha$ and $\bar{v}_{\delta^\alpha_i}$ are posterior mean and variance of $\delta_i^\alpha$:
\[
	\bar{v}_{\delta^\alpha_i} =  \left(v_{\delta^\alpha}^{-1} + T\sigma^{-2} \right)^{-1}, \quad
\bar{\delta}_i^\alpha  =  \bar{v}_{\delta^\alpha_i} \sigma^{-2} \sum_{t=1}^T \widecheck{y}_{it}^\alpha.
\]
For $z_i^\alpha=0$ we have $\delta_{i}^\alpha = 0$ and
\begin{eqnarray*}
	p(Y_{i}|\beta,\sigma^2, \delta_{i}^\rho, v_{\delta^{\alpha}}, z_i^\alpha=0)
	& = & (2\pi \sigma^2)^{-\frac{T}{2}} \exp \left( - \frac{1}{2\sigma^2} \sum_{t=1}^T \widecheck{y}_{it}^{\alpha^2} \right).
\end{eqnarray*}

We conclude that the posterior odds of $z_i^\alpha = 1$ vs. $z_i^\alpha = 0$ are
\be
	K_i^\alpha
	 =  \frac{q^\alpha}{1-q^\alpha} \left( \frac{v_{\delta^\alpha}}{\bar{v}_{\delta^\alpha_i}} \right)^{-\frac{1}{2}} \exp \left( \frac{\bar{\delta}_i^{\alpha^2}}{2\bar{v}_{\delta^\alpha_i}}\right).
\ee
We can draw $z_{i}^\alpha$ as follows:
\be
z_{i}^\alpha|(Y_{i},\beta,q^\alpha,\sigma^2,\delta_{i}^\rho, v_{\delta^{\alpha}}) = \left\{ \begin{array}{ll} 1 & \mbox{with prob. } \frac{K_i^\alpha}{K_i^\alpha+1} \\
	0 & \mbox{with prob. } \frac{1}{K_i^\alpha+1} \end{array} \right. .
\label{appeq:benchmark.homo.post.z.alpha}
\ee

If $z_{i}^\alpha =1$, the conditional posterior of $\delta_{i}^\alpha$ is given by
\begin{eqnarray*}
	p(\delta_i^\alpha|Y_{i},z_i^\alpha,\beta,\sigma^2,\delta_i^\rho, v_{\delta^{\alpha}}) & \propto & \exp \left[ -\frac{1}{2\sigma^2} \sum_{t=1}^T \left( \widecheck{y}_{it}^\alpha-\delta_i^\alpha \right)^2 \right] \exp \left( -\frac{\delta_i^{\alpha^2}}{2 v_{\delta^\alpha}}\right)\\
	& \propto & \exp \left[-\frac{1}{2\left(v_{\delta^\alpha}^{-1} + T\sigma^{-2}\right)^{-1}}\left( \delta_i^\alpha - \frac{\sigma^{-2} \sum_{t=1}^T \widecheck{y}_{it}^\alpha}{ v_{\delta^\alpha}^{-1} + T\sigma^{-2} }\right)^2 \right] .
\end{eqnarray*}
If $z_i^\alpha = 0$ then $\delta_i^\alpha = 0$. Overall, we obtain
\be
\delta_i^\alpha|(Y_{i},z_i^\alpha,\beta,\sigma^2,\delta_i^\rho, v_{\delta^{\alpha}}) \sim
\left\{ \begin{array}{ll} N \left( \bar{\delta}_i^\alpha,  \bar{v}_{\delta^\alpha_i} \right)
	& \mbox{if } z_i^\alpha = 1 \\
	0 & \mbox{if } z_i^\alpha = 0 \end{array} \right. .
\label{appeq:benchmark.homo.post.delta.alpha}
\ee

Sampling from the conditional distribution $z_{i}^\rho,\delta_{i}^\rho | (\beta, q^\rho, \sigma^2, z_{i}^\alpha,\delta_{i}^\alpha, v_{\delta^{\rho}})$ proceeds in the same manner.
We define $\widecheck{y}^\rho_{it} =y_{it} - \alpha - \delta_i^\alpha - \rho y_{it-1}$ and let $\bar{\delta}_i^\rho$ and $\bar{v}_{\delta_i^\rho}$ be the posterior mean and variance of $\delta_i^\rho$:
\[
	\bar{v}_{\delta_i^\rho}  =  \left( v_{\delta^\rho}^{-1} + \sigma^{-2}\sum_{t=1}^T y_{it-1}^2 \right)^{-1}, \quad
	\bar{\delta}_i^\rho  =  \bar{v}_{\delta_i^\rho} \sigma^{-2} \sum_{t=1}^T y_{it-1} \widecheck{y}_{it}^\rho.
\]
The posterior odds of $z_i^\rho = 1$ vs. $z_i^\rho = 0$ are
\be
	K_i^\rho
	= \frac{q^\rho}{1-q^\rho} \left( \frac{v_{\delta^\rho}}{\bar{v}_{\delta_i^\rho}} \right)^{-\frac{1}{2}} \exp \left(  \frac{\bar{\delta}_i^{\rho^2}}{2\bar{v}_{\delta_i^\rho}} \right).
\ee
We can draw $z_{i}^\alpha$ as follows:
\be
z_{i}^\rho|(Y_{i},\beta,q^\rho,\sigma^2,\delta_{i}^\alpha, v_{\delta^{\rho}}) = \left\{ \begin{array}{ll} 1 & \mbox{with prob. } \frac{K_i^\rho}{K_i^\rho+1} \\
	0 & \mbox{with prob. } \frac{1}{K_i^\rho+1} \end{array} \right. .
\label{appeq:benchmark.homo.post.z.rho}
\ee
Moreover,
\be
    \delta_i^\rho|(Y_{i},z_i^\rho,\beta, \sigma^2, \delta_{i}^\alpha, v_{\delta^{\rho}}) \sim
\left\{ \begin{array}{ll} N \left( \bar{\delta}_i^\rho,\bar{v}_{\delta_i^\rho} \right)
	& \mbox{if } z_i = 1 \\
	0 & \mbox{if } z_i = 0 \end{array} \right. .
\label{appeq:benchmark.homo.post.delta.rho}
\ee

\noindent {\bf Conditional Posterior of $\sigma^2$} is given by
\begin{eqnarray*}
	\lefteqn{p(\sigma^2| Y_{1:N},\beta,\delta)}\\ & \propto & \left(\frac{1}{\sigma^2}\right)^{\frac{\underline{\nu}_{\sigma}}{2}+1} \exp \left(-\fra{\underline{\tau}_\sigma}{2\sigma^2}\right) \prod_{i=1}^{N} \left( \fra{1}{\sigma^2} \right)^{\frac{T}{2}} \exp \left \{- \fra{1}{2\sigma^2} \left[Y_{i}-X_{i}(\beta + \delta_i) \right]' \left[Y_{i}-X_{i}(\beta + \delta_i) \right]\right\} \nonumber \\
	& = & \left(\frac{1}{\sigma^2}\right)^{\frac{\underline{\nu}_{\sigma}+NT}{2}+1} \exp \left \{- \fra{1}{2\sigma^2} \left \{ \underline{\tau}_\sigma + \sum_{i=1}^N \left[Y_{i}-X_{i}(\beta + \delta_i) \right]' \left[Y_{i}-X_{i}(\beta + \delta_i) \right] \right\} \right\}
\end{eqnarray*}
This implies
\be
\sigma^2| (Y_{1:N},\beta,\delta) \sim \text{IG} \left(\frac{\bar{\nu}_{\sigma}}{2},\frac{\bar{\tau}_\sigma}{2} \right),
\label{appeq:benchmark.homo.post.sigma}
\ee
where
\begin{eqnarray*}
	\bar{\nu}_{\sigma} & = & \underline{\nu}_{\sigma} + NT \\
	\bar{\tau}_\sigma & = & \underline{\tau}_\sigma + \sum_{i=1}^N \left[Y_{i}-X_{i}(\beta + \delta_i) \right]' \left[Y_{i}-X_{i}(\beta + \delta_i) \right].
\end{eqnarray*}

\subsubsection{A Homogeneous Version}

The model is obtained from (\ref{eq:model.benchmark}) by setting $\delta^\alpha_i = \delta^\rho_i = 0$ while maintaining $\alpha$ and $\rho$ as unknown parameters.
\begin{equation}
y_{it} = \alpha + \rho y_{it-1} + \sigma u_{it} = x_{it}'\beta + \sigma u_{it}, \label{eq:model.benchmark.3.2}
\end{equation}
where $\beta = [\alpha \; \rho]'$, and $x_{it} = [1 \; y_{it-1}]'$. The posterior computations reduce to a Gibbs sampler with two blocks.

\noindent {\bf Conditional Posterior of $\beta = [\alpha \; \rho]'$}: Same as (\ref{appeq:benchmark.homo.post.beta}) with $\delta_i=0$.

\noindent {\bf Conditional posterior of $\sigma^2$:} Same as (\ref{appeq:benchmark.homo.post.sigma}) with $\delta_i=0$.

\subsubsection{A  Fully Heterogeneous Version}

The model is obtained from (\ref{eq:model.benchmark}) by setting $q=1$ which in turn implies that $z_i=1$ for all $i$. The posterior computations reduce to a Gibbs sampler with four blocks.

\noindent {\bf Conditional Posterior of $\beta = [\alpha \; \rho]'$}: Same as (\ref{appeq:benchmark.homo.post.beta}).

\noindent {\bf Conditional Posterior of $v_\delta = [v_{\delta^\alpha} \; v_{\delta^\rho}]'$}: Same as (\ref{appeq:benchmark.homo.post.vdetla}).

\noindent {\bf Conditional Posterior of $\delta = [\delta_\alpha \; \delta_\rho]'$:} 
\be
\delta_i | (Y_{i},\beta,\sigma^2, v_\delta) \sim
 N \left( \bar{\delta}_i,  \bar{v}_{\delta_i} \right)
 \label{appeq:benchmark.homo.post.delta}
\ee
where we define $\widecheck{Y}_{i} = Y_{i} - X_{i}\beta $, and $\bar{v}_{\delta_i}$ and $\bar{\delta}_i$ are posterior mean and variance of $\delta_i$:
\[
	\bar{v}_{\delta_i} =   \left( v_{\delta}^{-1} + \sigma^{-2} X'_i X_{i} \right)^{-1} , \quad
\bar{\delta}_i  =  \bar{v}_{\delta_i} \sigma^{-2} X'_i \widecheck{Y}_{i}.
\]

\noindent {\bf Conditional Posterior of $\sigma^2$:} Same as (\ref{appeq:benchmark.homo.post.sigma}).

\subsection{Heteroskedastic Model}
\label{appsubsec:posterior.sampling.benchmark.hetero}

We now consider (\ref{eq:model.benchmark}) without the restriction $\delta_{i}^{\sigma} = 1$. 

%
%
%

\subsubsection{Spike-and-Slab Prior: $q^\alpha$, $q^\rho$ and $q^\sigma$ are Independent}

The posterior of the unknown objects of this model is:
\begin{eqnarray*}
	\lefteqn{p(\beta, \delta, q, Z_{1:N}, \sigma^2, v_{\delta} | Y_{1:N})} \\
	&\propto & p(Y_{1:N}|\beta,\delta, \sigma^2) p(\delta^{\alpha} | z^{\alpha}, v_{\delta^{\alpha}}) p(\delta^{\rho} | z^{\rho}, v_{\delta^{\rho}}) p(\delta^{\sigma} | z^{\sigma}, v_{\delta^\sigma} ) p(\beta)  p(Z_{1:N}|q) p(q) p(\sigma^2) p(v_{\delta^{\alpha}}) p(v_{\delta^{\rho}}) p(v_{\delta^\sigma} )\\
	&\propto & \prod_{i=1}^{N} \left( \fra{1}{2\pi \sigma^2 \delta_i^{\sigma} } \right)^{\frac{T}{2}} \exp \left \{- \fra{1}{2\sigma^2 \delta_i^{\sigma}} \left[Y_{i}-X_{i}(\beta + \delta_i) \right]' \left[Y_{i}-X_{i}(\beta + \delta_i) \right]\right\}\\
	& & \times \prod_{i=1}^N \left[ \left( \fra{1}{2\pi v_{\delta^{\alpha}}} \right)^{\frac{1}{2}} \exp \left( -\frac{\delta_i^{\alpha^2}}{2v_{\delta^{\alpha}}}\right)\right]^{z_i^{\alpha}} \mathbb{I}\{\delta_i^{\alpha} = 0\}^{1-z_i^{\alpha}}
	\left[ \left( \fra{1}{2\pi v_{\delta^{\rho}}} \right)^{\frac{1}{2}} \exp \left( -\frac{\delta_i^{\rho^2}}{2v_{\delta^{\rho}}}\right)\right]^{z_i^{\rho}} \mathbb{I}\{\delta_i^{\rho} = 0\}^{1-z_i^{\rho}} \\
	& & \quad \quad \;  \left[ \frac{ (v^{-1}_{\delta^{\sigma}} + 1)^{\left(v^{-1}_{\delta^{\sigma}} + 2\right)} }{ \Gamma(v^{-1}_{\delta^{\sigma}} + 2) }  \left(\frac{1}{\delta^{\sigma}_i} \right)^{v^{-1}_{\delta^{\sigma}} +  3} \exp \left(-\fra{ v^{-1}_{\delta^{\sigma}} +  1}{\delta^{\sigma}_i }\right) \right]^{z^{\sigma}_i} \mathbb{I}\{ \delta^{\sigma}_i = 0 \}^{1-z^{\sigma}_i} \\
	& & \times \fra{1}{2 \pi } \left| \underline{v}_{\beta} \right|^{-\frac{1}{2}} \exp \left(-\frac{1}{2} \beta' \underline{v}_{\beta}^{-1} \beta \right)\\
	& & \times \prod_{i=1}^N (q^{\alpha})^{z_i^{\alpha}}(1-q^{\alpha})^{1-z_i^{\alpha}} (q^{\rho})^{z_i^{\rho}}(1-q^{\rho})^{1-z_i^{\rho}} (q^{\sigma})^{z_i^{\sigma}}(1-q^{\sigma})^{1-z_i^\sigma}\\
	& & \times (q^{\alpha})^{a-1}(1-q^{\alpha})^{b-1} \cdot (q^{\rho})^{a-1} (1-q^{\rho})^{b-1} \cdot (q^{\sigma})^{a-1} (1-q^{\sigma})^{b-1} \\
	& & \times \left(\frac{1}{\sigma^2}\right)^{\frac{\underline{\nu}_{\sigma}}{2}+1} \exp \left(-\fra{\underline{\tau}_\sigma}{2\sigma^2}\right) \\
	& &
	\times  \left(\frac{1}{v_{\delta^{\alpha}}}\right)^{\frac{\underline{\nu}_{\delta^{\alpha}}}{2}+1} \exp \left(-\fra{\underline{\tau}_{\delta^{\alpha}}}{2v_{\delta^{\alpha}}}\right)
	\left(\frac{1}{v_{\delta^{\rho}}}\right)^{\frac{\underline{\nu}_{\delta^{\rho}}}{2}+1} \exp \left(-\fra{\underline{\tau}_{\delta^\rho}}{2v_{\delta^{\rho}}}\right)
	\left( \frac{1}{v_{\delta^\sigma}} \right)^{ \frac{\underline{\nu}_{\delta^\sigma}}{2} + 1 } \exp \left(-\fra{\underline{\tau}_{\delta^\sigma}}{2v_{\delta^\sigma}}\right),
\end{eqnarray*}
where $\underline{v}_{\beta} = \mbox{diag}( \underline{v}_\alpha, \underline{v}_\rho)$ and $\delta_i = [\delta_i^{\alpha} \; \delta_i^{\rho}]'$. Note that, as we impose that $\mathbb{E}[\delta_i^\sigma] = 1$, the ``slab" part of the prior of $\delta_{i}^{\sigma}$ can be reparameterized in terms of $v_{\delta^\sigma}$ which lead to,
\[
	\nu_{\delta^\sigma} = 2v^{-1}_{\delta^\sigma} + 4, \quad \tau_{\delta^\sigma} = 2v^{-1}_{\delta^\sigma} + 2,
\]
and 
\[ 
	\delta_i^{\sigma} | q^{\sigma}, \nu_{\delta^{\sigma}}, \tau_{\delta^{\sigma}} \sim IG \left(v^{-1}_{\delta^\sigma} + 2, v^{-1}_{\delta^\sigma} + 1\right) \; \text{ with prob. } q^{\sigma}.
\]

We can sample from the joint posterior of $(\beta, q, \delta, Z_{1:N},  v_{\delta^{\alpha}}, v_{\delta^{\rho}}, v_{\delta^\sigma},  \sigma^2)$ using a Gibbs sampling algorithm with six blocks.

\noindent {\bf Conditional posterior of $\beta$.} Let $\sigma_i =  \sigma \sqrt{\delta^{\sigma}_i}$. The posterior is the same as in (\ref{appeq:benchmark.homo.post.beta}), except for a slight change in the posterior mean and variance formulas reflecting the heteroskedasticity:
\[
\bar{v}_{\beta} = \left( \underline{v}_{\beta}^{-1} + \sum_{i=1}^{N} \sigma_i^{-2} X_{i}' X_{i} \right)^{-1}, \quad  \bar{\beta} = \bar{v}_{\beta} \sum_{i=1}^{N} \sigma_i^{-2} X'_i \widetilde{Y}_i.
\]

\noindent {\bf Conditional Posterior of $(q^\alpha,q^\rho,q^\sigma)$}: Same as (\ref{appeq:benchmark.homo.post.q}), for $l \in \{\alpha,\rho,\sigma\}$.

\noindent {\bf Conditional Posterior of $(v_{\delta^\alpha}, v_{\delta^\rho})$}: Same as (\ref{appeq:benchmark.homo.post.vdetla}).

\noindent {\bf Conditional Posterior of $v_{\delta^{\sigma}}$}: To economize on notation, we represent $v_{\delta^{\sigma}}$ by $\omega$. The conditional posterior of $\omega$ is given by
\begin{eqnarray*}
	p(\omega |  z^{\sigma}, \delta^{\sigma})
	& \propto &
	\prod_{i | z_i^{\sigma} = 1}  \left[ \frac{ (\omega^{-1} + 1)^{\left(\omega^{-1} + 2\right)} }{ \Gamma(\omega^{-1} + 2) }  \left( \frac{1}{\delta^{\sigma}_i}\right)^{ \omega^{-1} +  3 } \exp \left(-\fra{ \omega^{-1} +  1}{\delta^{\sigma}_i }\right) \right]
	\left(\frac{1}{\omega}\right)^{\frac{\underline{\nu}_{\delta^\sigma}}{2} + 1 }
	\exp \left(-\fra{\underline{\tau}_{\delta^\sigma}}{2\omega}\right)
	\\
	& \propto &
	\frac{ (\omega^{-1} + 1)^{\psi(z^{\sigma}) \left(\omega^{-1} + 2\right)} }{ \Gamma(\omega^{-1} + 2)^{\psi(z^{\sigma})} }
	\left( \prod_{i | z_i^{\sigma} = 1} \delta^{\sigma}_{i} \right)^{- \omega^{-1} }
	\exp \left( - \omega^{-1}\sum_{i | z_i^{\sigma} = 1} \frac{1}{\delta^{\sigma}_{i} } -\fra{\underline{\tau}_{\delta^\sigma}}{2\omega}\right)
	\omega^{-\left( \frac{\underline{\nu}_{\delta^\sigma}}{2} + 1\right)},
\end{eqnarray*}
where $\psi(z^\sigma) = \sum_{i=1}^N z_i^{\sigma}$ is the total number of non-zero elements in $\delta^{\sigma}$. The log posterior of $\omega$ is given by
\begin{eqnarray*}
	\ln p(\omega | z^{\sigma}, \delta^{\sigma})
	& \propto & \psi(z^{\sigma}) \left(\omega^{-1} + 2\right) \ln \left(\omega^{-1} + 1\right)  - \psi(z^{\sigma}) \ln \Gamma(\omega^{-1} + 2) \\
	& &   - \left(\frac{\underline{\nu}_{\delta^{\sigma}}}{2}+1 \right) \ln \omega - \omega^{-1} \left[ \sum_{i | z_i^{\sigma} = 1} \left(\ln \delta^{\sigma}_{i} + \frac{1}{\delta^{\sigma}_{i}} \right)  + \frac{\underline{\tau}_{\delta^{\sigma}}}{2}\right].
\end{eqnarray*}

We sample from this non-standard posterior via random walk Metropolis-Hastings (RWMH) algorithm. Notice that the posterior is well-defined when $\omega > 0$. As a result, we use a truncated Normal distribution for the proposal distribution. Let us denote by $N_{+}\left(\mu, \mu^{-}, \sigma^{2}\right)$ the truncated normal distribution with left truncation point $\mu^{-}$. Then the proposal distribution $q(\vartheta | \omega^{j-1})$ follows $N_{+}\left(\omega^{j-1}, 0, c^{2}\right)$, whose density is given by
\[
f\left(\vartheta \mid \omega^{j-1}, 0, c^{2} \right) = \frac{ \phi \left( \frac{\vartheta - \omega^{j-1}}{c}\right) }{\Phi \left( \frac{\omega^{j-1}}{c}\right)}  \boldsymbol{1}_{\vartheta > 0},
\]
where $\phi$ and $\Phi$ are PDF and CDF of the standard Normal distribution. This proposal distribution ensures the posterior draw of $\omega$ is strictly bounded below by 0. This following adaptive RWMH Algorithm is based on \cite{atchade2005} and \cite{griffin2016}, who adaptively adjust the random walk step-size to keep the acceptance rate around certain desirable percentage.

\begin{algorithm}\label{appalgo:benchmark.hetero.v.delta.sigma} 
	{\normalfont (Adaptive RWMHAlgorithm for $\omega$)} \\
	Set $\omega^{0}$ to its prior mean 1 and initial random walk step-size $c^0 = 1$. For $j = 1,..., N_{MCMC}$, conditional on the posterior draws of other parameters,
	\begin{enumerate}[noitemsep]
		\item Draw $\vartheta$ from the proposal distribution $N_{+}\left(\omega^{j-1}, 0, c^{j}\right)$.
		\item Set $\omega^{j} = \vartheta$ with probability
		\[
		\alpha\left(\vartheta | \omega^{j-1} \right) = \min \left\{1, \exp \left[ \left( \ln p(\vartheta | z^{\sigma}, \delta^{\sigma}) - \ln p(\omega^{j-1} | z^{\sigma}, \delta^{\sigma}) \right) - \left( \ln \Phi \left( \vartheta / c^j \right) - \ln \Phi \left( \omega^{j-1}/c^j\right)\right) \right] \right\}
		\]
		and $\omega^{j} = \omega^{j-1}$ otherwise.
		\item Update the random walk step-size for the next iteration, 
		\[
			\log c^{j+1}=g \left \{\log c^{j} + j^{-p} \left[ \alpha\left(\vartheta | \omega^{j-1} \right)  - \alpha^*\right]\right \}
		\]
		where $0.5< p \leq 1$, $\alpha^*$ is the target acceptance rate, and
		$$
		g(x)=\min (|x|, \bar{x}) \cdot \operatorname{sgn}(x)
		$$
		where $\bar{x}=10$ is user-specific large number. Following \cite{griffin2016}, we set $p = 0.55$.
	\end{enumerate}
\end{algorithm}

Given the opacity of the posterior, it is difficult to derive a theoretically optimal acceptance rate in the RWMH algorithm. Instead, we set $ \alpha^*$ to  $0.44$, which is the optimal acceptance rates for the univariate sampling suggested by \cite{rosenthal2011}.

With a posterior draw of $\omega$, we can find the values for the hyperparameters of $\delta_i^{\sigma}$ by (\ref{eq:benchmark.hetero.hp.delta.sigma}), which will be used in the next block.

\noindent {\bf Conditional Posterior of $(z,\delta)$.} Posteriors for $(z_i^\alpha,\delta_i^\alpha)$ and $(z_i^\rho,\delta_i^\rho)$ are the same as in (\ref{appeq:benchmark.homo.post.z.alpha}, \ref{appeq:benchmark.homo.post.delta.alpha}) and (\ref{appeq:benchmark.homo.post.z.rho}, \ref{appeq:benchmark.homo.post.delta.rho}), except for a slight change in the formulas to capture the heteroskedasticity:
\begin{eqnarray*}
	\bar{v}_{\delta^\alpha_i} & = & \left(v_{\delta^\alpha}^{-1} + T\sigma_i^{-2} \right)^{-1} , \quad
	\bar{\delta}_i^\alpha  =  \bar{v}_{\delta^\alpha_i} \sigma_i^{-2} \sum_{t=1}^T \widecheck{y}_{it}^\alpha \\
	\widecheck{y}^\alpha_{it} & = & y_{it} - \alpha - (\rho + \delta_i^\rho) y_{it-1}, \quad
	\sigma_i  = \sigma \sqrt{\delta_i^\sigma}.
\end{eqnarray*}
and
\begin{eqnarray*}
	\bar{v}_{\delta_i^\rho} & = & \left( v_{\delta^\rho}^{-1} + \sigma_i^{-2}\sum_{t=1}^T y_{it-1}^2 \right)^{-1} , \quad
	\bar{\delta}_i^\rho  =  \bar{v}_{\delta_i^\rho} \sigma_i^{-2} \sum_{t=1}^T y_{it-1} \widecheck{y}_{it}^\rho\\
	\widecheck{y}^\rho_{it} & = & y_{it} - \alpha - \delta_i^\alpha - \rho y_{it-1}, \quad
	\sigma_i  =  \sigma \sqrt{\delta_i^\sigma}.
\end{eqnarray*}

We now turn to draws from $z_{i}^\sigma,\delta_{i}^\sigma | (\beta, q^\sigma, \sigma^2, \delta_{i}^\alpha, \delta_{i}^\rho, \nu_{\delta^\sigma}, \tau_{\delta^{\sigma}})$.
Define
$$\widecheck{y}^\sigma_{it} =y_{it} - (\alpha + \delta_i^\alpha) - (\rho + \delta_i^\rho) y_{it-1}.$$
If $z_i^\sigma=1$, then $\delta_i^{\sigma}$ has an inverse gamma distribution. The marginalized likelihood of $(Y_{i}|\beta,\sigma^2, \delta_{i}^\alpha, \delta_{i}^\rho, v_{\delta^\sigma}, z_i^\sigma)$ is:
\begin{eqnarray*}
	\lefteqn{p(Y_{i}|\beta, \sigma^2, \delta_{i}^\alpha, \delta_{i}^\rho, v_{\delta^\sigma}, z_i^\sigma=1)}
	\\
	& = &
	\int_{-\infty}^{\infty} p(Y_{i} | \beta, \sigma^2, \delta_{i}^\alpha, \delta_{i}^\rho, \delta_i^\sigma, z_i^\sigma=1) p(\delta_i^\sigma| v_{\delta^\sigma}, z_i^\sigma=1) d \delta_i^{\sigma} \\
	& = &
	\int_{-\infty}^{\infty} \left(2\pi \sigma^2 \delta_i^{\sigma} \right)^{-\frac{T}{2}} \exp \left( - \frac{\sum_{t=1}^T \widecheck{y}_{it}^{\sigma^2}}{2\sigma^2 \delta_i^{\sigma}} \right) \frac{ (v^{-1}_{\delta^{\sigma}} + 1)^{\left(v^{-1}_{\delta^{\sigma}} + 2\right)} }{ \Gamma(v^{-1}_{\delta^{\sigma}} + 2) }  \left(\frac{1}{\delta^{\sigma}_i} \right)^{v^{-1}_{\delta^{\sigma}} +  3} \exp \left(-\fra{ v^{-1}_{\delta^{\sigma}} +  1}{\delta^{\sigma}_i }\right) d \delta_i^{\sigma}
	\\
	& = & \left( 2\pi \sigma^2 \right)^{-\frac{T}{2}} \frac{\Gamma \left(\frac{\bar{\nu}_{\delta^{\sigma}}}{2} \right)}{\Gamma \left( v^{-1}_{\delta^\sigma} + 2\right)} \frac{\left( v^{-1}_{\delta^\sigma} + 1\right)^{v^{-1}_{\delta^\sigma} +2}}{\left(\frac{\bar{\tau}_{\delta_i^\sigma}}{2}\right)^{\frac{\bar{\nu}_{\delta^{\sigma}}}{2}}}.
\end{eqnarray*}
where $\bar{\nu}_{\delta^{\sigma}} = 2v^{-1}_{\delta^{\sigma}} +4 + T$ and $\bar{\tau}_{\delta_i^\sigma} = 2v^{-1}_{\delta^{\sigma}} + 2 +  \sigma^{-2} \sum_{t=1}^T \widecheck{y}_{it}^{\sigma^2} $.

For $z_i^\sigma = 0$, we have $\delta_{i}^{\sigma} = 1$ and
\be
	p(Y_{i}|\beta, \sigma^2, \delta_{i}^\alpha, \delta_{i}^\rho, z_i^\sigma=0) = p(Y_{i}|\beta, \sigma^2, z_i^\sigma=0)
	=  (2\pi \sigma^2)^{-\frac{T}{2}} \exp \left( - \frac{1}{2\sigma^2} \sum_{t=1}^T \widecheck{y}_{it}^{\sigma^2} \right).
\ee
Thus the posterior odds of $z_i^\sigma = 1$ vs. $z_i^\sigma = 0$ are
\be
K_i^\sigma
= \frac{q^\sigma}{1-q^\sigma} 
\frac{\Gamma \left(\frac{\bar{\nu}_{\delta^{\sigma}}}{2} \right)}{\Gamma \left( v^{-1}_{\delta^\sigma} + 2\right)} \frac{\left( v^{-1}_{\delta^\sigma} + 1\right)^{v^{-1}_{\delta^\sigma} +2}}{\left(\frac{\bar{\tau}_{\delta_i^\sigma}}{2}\right)^{\frac{\bar{\nu}_{\delta^{\sigma}}}{2}}}
\exp \left( \frac{1}{2\sigma^2} \sum_{t=1}^T \widecheck{y}_{it}^{\sigma^2} \right).
\ee

\noindent
Based on the posterior odds $K_i^\sigma$ we can draw $z_{i}^\sigma$ as follows:
\be
z_{i}^\sigma | (Y_{i},\beta,q^\sigma,\sigma^2,\delta_{i}^\alpha, \delta_{i}^\rho,  v_{\delta^\sigma}) = \left\{ \begin{array}{ll} 1 & \mbox{with prob. } \frac{K_i^\sigma}{K_i^\sigma+1} \\
	0 & \mbox{with prob. } \frac{1}{K_i^\sigma+1} \end{array} \right. .
\label{appeq:benchmark.hetero.post.z.sigma}
\ee
If $z_{i}^\sigma =1$, the conditional posterior of $\delta_{i}^{\sigma}$ is given by
\begin{eqnarray*}
	p(\delta_i^{\sigma} | Y_{i},z_i^\sigma,\beta,\sigma^2,\delta_{i}^\alpha,\delta_{i}^\rho, v_{\delta^\sigma})
	& \propto &
	\left( \frac{1}{\delta_i^{\sigma}} \right)^{\frac{T}{2}} \exp \left( - \frac{\sum_{t=1}^T \widecheck{y}_{it}^{\sigma^2}}{2\sigma^2 \delta_i^{\sigma}} \right) \left(\frac{1}{\delta_i^{\sigma}}\right)^{v^{-1}_{\delta_{\sigma}} + 3} \exp \left(-\fra{v^{-1}_{\delta_{\sigma}} + 1}{\delta_i^{\sigma} }\right)
	\\
	& \propto &
	\left(\frac{1}{\delta_i^{\sigma}}\right)^{\frac{T}{2}+ v^{-1}_{\delta_{\sigma}} + 3} \exp \left(-\frac{ \sigma^{-2} \sum_{t=1}^T \widecheck{y}_{it}^{\sigma^2}  + 2v^{-1}_{\delta_{\sigma}} + 2}{2\delta_i^{\sigma} }\right).
\end{eqnarray*}
If $z_{i}^\sigma = 0$ then $\delta_i^{\sigma} = 1$. Overall, we obtain
\be
\delta_i^{\sigma} | (Y_{i},z_i^\sigma,\beta,\sigma^2,\delta_i^\alpha, \delta_i^\rho,  v_{\delta^\sigma}) \sim
\left\{ \begin{array}{ll} \text{IG} \left( \frac{\bar{\nu}_{\delta^{\sigma}}}{2},  \frac{\bar{\tau}_{\delta_i^\sigma}}{2} \right)
	& \mbox{if } z_i^\sigma = 1 \\
	1 & \mbox{if } z_i^\sigma = 0 \end{array} \right.
\label{appeq:benchmark.hetero.post.delta.sigma}
\ee

\noindent {\bf Conditional posterior of $\sigma^2$:} Same as (\ref{appeq:benchmark.homo.post.sigma}), with the exception that $\bar{\tau}_\sigma$ is defined as
\[
\bar{\tau}_\sigma  =  \underline{\tau}_\sigma + \sum_{i=1}^N \left[Y_{i}-X_{i}(\beta + \delta_i) \right]' \left[Y_{i}-X_{i}(\beta + \delta_i) \right] /  \delta_i^{\sigma}.
\]

\subsubsection{A Homogeneous Version}

The model is obtained from (\ref{eq:model.benchmark}) by setting $\delta^\alpha_i = \delta^\rho_i = 0$ and $\delta^\sigma_i = 1$ while maintaining $\alpha$, $\rho$ and $\sigma^2$ as unknown parameters.
\begin{equation}
	y_{it} = \alpha + \rho y_{it-1} + \sigma u_{it} = x_{it}'\beta + \sigma u_{it}, \label{eq:model.benchmark.3.2}
\end{equation}
where $\beta = [\alpha \; \rho]'$, and $x_{it} = [1 \; y_{it-1}]'$. The posterior computations reduce to a Gibbs sampler with two blocks.

\noindent {\bf Conditional Posterior of $\beta = [\alpha \; \rho]'$}: Same as (\ref{appeq:benchmark.homo.post.beta}) with $\delta_i=0$.

\noindent {\bf Conditional posterior of $\sigma^2$:} Same as (\ref{appeq:benchmark.homo.post.sigma}) with $\delta_i=0$.

\subsubsection{A Fully Heterogeneous Version}

The model is obtained from (\ref{eq:model.benchmark}) by setting $q=1$ which in turn implies that $z_i=1$ for all $i$. The posterior computations reduce to a Gibbs sampler with five blocks.

\noindent {\bf Conditional Posterior of $\beta = [\alpha \; \rho]'$}: Same as (\ref{appeq:benchmark.homo.post.beta}).

\noindent {\bf Conditional Posterior of $v_\delta = [v_{\delta^\alpha} \; v_{\delta^\rho}]'$}: Same as (\ref{appeq:benchmark.homo.post.vdetla}).

\noindent {\bf Conditional Posterior of $v_{\delta^{\sigma}}$}: Same as Algorithm \ref{appalgo:benchmark.hetero.v.delta.sigma}.

\noindent {\bf Conditional Posterior of $\delta = [\delta_\alpha \; \delta_\rho \; \delta_\sigma]'$:} Same as (\ref{appeq:benchmark.homo.post.delta}) and  (\ref{appeq:benchmark.hetero.post.delta.sigma}) with $z_i^\sigma=1$.

\noindent {\bf Conditional Posterior of $\sigma^2$:} Same as (\ref{appeq:benchmark.homo.post.sigma}).

\clearpage
\section{Posterior Sampling for Variants of Model $M_2$}

A more elaborate specification for the income process takes the following form
\begin{eqnarray}
	y_{it} & = & H'_{it} (  \alpha + \delta^{\alpha}_{i} ) + s_{it} + \sigma_{u,t} \sqrt{\delta_{i,u}^\sigma} u_{it}  \label{eq:empirical.ss.model}\\
	s_{it} & = & (\rho+\delta_i^\rho) s_{it-1} + \sigma_{\epsilon,t} \sqrt{\delta_{i,\epsilon}^\sigma} \epsilon_{i,t}. \nonumber
\end{eqnarray} 
where $\alpha = [\alpha_0 \; \alpha_1]'$, $\delta^{\alpha}_{i} = [\delta^{\alpha}_{i0} \; \delta^{\alpha}_{i1}]'$, and $H_{it} = [1 \; h_{it}]'$.

\noindent {\bf Priors for Heterogeneous Parameters.} For $(\delta^{\alpha}, \delta^{\rho})$ we use a (multivariate) Normal distribution for the continuous part and let
\be
\delta_i^l|q,v_{\delta^l} \sim \left\{ \begin{array}{ll} {\cal N}(0,v_{\delta^l}) & \mbox{with prob. } q \\
	0                                          & \mbox{with prob. } 1-q \end{array} \right. \quad l \in \{ \alpha,\rho\}.
\ee
Note that $v_{\delta^\alpha}$ is a $3 \times 3$ matrix.
For $(\delta_{i,u}^{\sigma}, \delta_{i,\epsilon}^{\sigma})$ we use an Inverse Gamma (IG) distribution instead of a Normal distribution to capture the discrepancies:
\be
\delta_{i,m}^\sigma|q, \nu_{\delta_m^{\sigma}}, \tau_{\delta_m^{\sigma}}  \sim \left\{ \begin{array}{ll} IG\left( \frac{\nu_{\delta_m^{\sigma}}}{2},  \frac{\tau_{\delta_m^{\sigma}}}{2} \right)  & \mbox{with prob. } q \\
	1                                          & \mbox{with prob. } 1-q \end{array} \right. \quad m \in \{ u, \epsilon\}.
\ee
We impose that $\mathbb{E}[\delta_{i,m}^\sigma]=1$ and reparameterize the prior distribution in terms of its variance $v_{\delta_m^\sigma}$ which leads to
\be
\nu_{\delta_m^\sigma} = 2v^{-1}_{\delta_m^\sigma} + 4, \quad \tau_{\delta_m^\sigma} = 2v^{-1}_{\delta_m^\sigma} + 2.
\ee

\noindent {\bf Priors for Homogeneous Parameters.} We consider the following prior distributions for the homogeneous parameters
\begin{eqnarray*}
	\alpha &\sim& {\cal N}(0,\underline{v}_\alpha), \quad
	\rho \sim {\cal N}(0,\underline{v}_\rho), \quad
	\sigma_{u, t}^2 \sim IG \left( \frac{\underline{\nu}_{\sigma_{u}}}{2}, \frac{\underline{\tau}_{\sigma_{u}}}{2} \right), \quad 
	\sigma_{\epsilon, t}^2 \sim IG \left( \frac{\underline{\nu}_{\sigma_{\epsilon}}}{2}, \frac{\underline{\tau}_{\sigma_{\epsilon}}}{2} \right), \quad q \sim B(a,b) \\
	\quad v_{\delta^\alpha} & \sim & \mathcal{W}^{-1} \left( \underline{\nu}_{\delta^{\alpha}},  \underline{\Psi}_{\delta^{\alpha}} \right),
	\quad v_{\delta^\rho} \sim IG \left( \frac{\underline{\nu}_{\delta^{\rho}}}{2},  \frac{\underline{\tau}_{\delta^{\rho}}}{2} \right),
	\quad v_{\delta_{u}^{\sigma}} \sim  IG \left( \frac{\underline{\nu}_{\delta_{u}^{\sigma}}}{2},  \frac{\underline{\tau}_{\delta_{u}^{\sigma}}}{2} \right),
	\quad v_{\delta_{\epsilon}^{\sigma}} \sim  IG \left( \frac{\underline{\nu}_{\delta_{\epsilon}^{\sigma}}}{2},  \frac{\underline{\tau}_{\delta_{\epsilon}^{\sigma}}}{2} \right) \\
	\mu_{s_0} &\sim& {\cal N}(\underline{\mu}_{s_0}, \underline{v}_{s_0}), 
	\quad v_{s_0} \sim IG  \left( \frac{\underline{\nu}_{s_0}}{2},  \frac{\underline{\tau}_{s_0}}{2} \right),
	\nonumber
\end{eqnarray*}
where $B(a,b)$ is the Beta distribution and $\mathcal{W}^{-1} \left( \underline{\nu}_{\delta^{\alpha}},  \Phi_{\delta^{\alpha}} \right)$ is the inverse Wishart distribution. We collect the homogeneous parameters into the vector
\[
\theta = [\alpha', \rho, \sigma_{u}^{2'}, \sigma_{\epsilon}^{2'}, q, v_{\delta^\alpha},v_{\delta^\rho}, v_{\delta_u^\sigma}, v_{\delta_\epsilon^\sigma}, \mu_{s_0}, v_{s_0}]'
\]
and the hyperparameters that index the various prior distributions into the vector
\[
\lambda = [\underline{v}_\alpha, \underline{v}_\rho, \underline{\nu}_{\sigma_u}, \underline{\tau}_{\sigma_u}, \underline{\nu}_{\sigma_\epsilon}, \underline{\tau}_{\sigma_\epsilon},  a, b,  \underline{\nu}_{\delta^{\alpha}},  \underline{\Psi}_{\delta^{\alpha}}, 
\underline{\nu}_{\delta^{\rho}}, \underline{\tau}_{\delta^{\rho}}, 
\underline{\nu}_{\delta_u^{\sigma}}, \underline{\tau}_{\delta_u^{\sigma}},
\underline{\nu}_{\delta_\epsilon^{\sigma}}, \underline{\tau}_{\delta_\epsilon^{\sigma}}, 
\underline{\mu}_{s_0}, \underline{v}_{s_0}, \underline{\nu}_{s_0}, \underline{\tau}_{s_0} ].'
\]

We can sample from the joint posterior of $ \left(\alpha, \rho, q, v_{\delta^\alpha},v_{\delta^\rho}, v_{\delta_u^\sigma}, v_{\delta_\epsilon^\sigma}, \delta, z, \sigma_{u}^{2}, \sigma_{\epsilon}^{2}, s \right)$ using a Gibbs sampling algorithm with eight blocks.

\noindent {\bf Conditional posterior of $\alpha$.} Let $\widecheck{y}^\alpha_{it} =  y_{it} - H'_{it} \delta^{\alpha}_{i} - s_{it}$ and $\sigma^{u}_{it} =  \sigma_{u,  t} \sqrt{\delta^{\sigma}_{i, u}}$. Then
\begin{eqnarray*}
	p \left(\alpha | \rho, s,  \sigma_{u}^2, \delta^{\sigma}_{u} \right)
	& \propto & \exp \left[- \sum_{i=1}^N \sum_{t=1}^T  \fra{\left(\widecheck{y}^\alpha_{it}  - H'_{it} \alpha \right)^2 }{2\sigma^{u^2}_{it}} \right] \exp \left[ - \frac{1}{2} \left(\alpha -  \underline{\mu}_{\alpha}\right)' \underline{v}^{-1}_\alpha \left(\alpha -  \underline{\mu}_{\alpha}\right) \right]\\
	& \propto & \exp \left[ - \frac{1}{2} ( \alpha -  \bar{\alpha})' \bar{v}^{-1}_{\alpha} ( \alpha -  \bar{\alpha}) \right],
\end{eqnarray*}
where
\[
\bar{v}_{\alpha} = \left( \underline{v}_{\alpha}^{-1} + \sum_{i=1}^{N} \sum_{t=1}^{T} \sigma^{u^{-2}}_{it} H_{it} H'_{it} \right)^{-1}, \quad \bar{\alpha} = \bar{v}_{\alpha} \left(  \underline{v}_{\alpha}^{-1}  \underline{\mu}_{\alpha} + \sum_{i=1}^{N} \sum_{t=1}^{T} \sigma^{u^{-2}}_{it}  H_{it} \widecheck{y}^\alpha_{it} \right) .
\]
This implies
\be
\alpha | \rho, s,  \sigma_{u}^2, \delta^{\sigma}_{u} \sim N \left( \bar{\alpha}, \bar{v}_{\alpha} \right).
\ee

\noindent {\bf Conditional posterior of $\rho$.} Let $\sigma^{\epsilon}_{it} =  \sigma_{\epsilon,  t} \sqrt{\delta^{\sigma}_{i, \epsilon}}$. Then
\begin{eqnarray*}
	p \left(\rho | s,  \sigma_{\epsilon}^2, \delta^{\sigma}_{\epsilon} \right)
	& \propto & \exp \left[- \sum_{i=1}^N \sum_{t=1}^T  \fra{\left(\tilde{s}_{it} - \rho s_{it-1} \right)^2 }{2\sigma^{\epsilon^2}_{it}} \right] \exp \left[ - \frac{\left(\rho - \mu_{\rho}\right)^2}{2\underline{v}_\rho} \right] \\
	& \propto & \exp \left[ - \frac{( \rho -  \bar{\rho})^2}{2\bar{v}_{\rho}} \right],
\end{eqnarray*}
where
\[
\bar{v}_{\rho} = \left( \underline{v}_{\rho}^{-1} + \sum_{i=1}^{N} \sum_{t=1}^{T} \sigma^{\epsilon^{-2}}_{it}  s^2_{i t-1} \right)^{-1}, \quad \bar{\rho} = \bar{v}_{\rho} \left( \underline{v}_{\rho}^{-1} \underline{\mu}_{\rho} + \sum_{i=1}^{N} \sum_{t=1}^{T} \sigma^{\epsilon^{-2}}_{it}  s_{it-1} \tilde{s}_{it} \right), \quad \tilde{s}_{it} = s_{it} - \delta_{i}^{\rho} s_{it-1}.
\]
This implies
\be
\rho | \left(s,  \sigma_{\epsilon}^2, \delta^{\sigma}_{\epsilon} \right) \sim N \left( \bar{\rho}, \bar{v}_{\rho} \right).
\ee

\noindent {\bf Conditional Posterior of $(q^\alpha, q^\rho, q^{\sigma_u}, q^{\sigma_\epsilon})$}: Same as (\ref{appeq:benchmark.homo.post.q}), for $l \in \{\alpha,\rho,\sigma_u, \sigma_\epsilon\}$.

\noindent {\bf Conditional Posterior of $(v_{\delta^\alpha}, v_{\delta^\rho})$}: Posterior of $v_{\delta^\rho}$ is the same as (\ref{appeq:benchmark.homo.post.vdetla}) while the posterior of $v_{\delta^\alpha}$ is now a multivariate distribution.
\begin{eqnarray*}
	p(v_{\delta^\alpha} | z^\alpha, \delta^\alpha)
	& \propto & \prod_{i=1}^N \left[  |v_{\delta^{\alpha}}|^{-\frac{1}{2}} \exp \left( - \frac{1}{2} \delta_{i}^{\alpha'} v^{-1}_{\delta^\alpha} \delta_{i}^{\alpha} \right)\right]^{z_i^{\alpha}} 
	|v_{\delta^{\alpha}}|^{-\frac{\underline{\nu}_{\delta^{\alpha}} + 3 + 1}{2}} 
	\exp \left[ -\fra{1}{2} \text{tr} \left( \Psi_{\delta^{\alpha}} v^{-1}_{\delta^\alpha} \right)\right] \\
	& \propto &	|v_{\delta^{\alpha}}|^{-\frac{\underline{\nu}_{\delta^{\alpha}} + \psi(z^{\alpha})+ 3 + 1}{2}}  \exp \left\{ -\fra{1}{2} \text{tr} \left[ \left( \underline{\Psi}_{\delta^{\alpha}} + \sum_{i | z_i^l = 1} \delta_{i}^{\alpha}  \delta_{i}^{\alpha'} \right)  v^{-1}_{\delta^\alpha} \right]\right\}.
\end{eqnarray*}
This implies
\be
v_{\delta^\alpha} | (z^\alpha, \delta^\alpha) \sim \mathcal{W}^{-1} \left(\bar{\nu}_{\delta^\alpha}, \bar{\Psi}_{\delta^{\alpha}} \right),
\ee
where
\begin{eqnarray*}
	\bar{\nu}_{\delta^\alpha} & = & \underline{\nu}_{\delta^\alpha} + \psi(z^\alpha) \\
	\bar{\Psi}_{\delta^{\alpha}} & = & \underline{\Psi}_{\delta^{\alpha}} + \sum_{i | z_i^l = 1} \delta_{i}^{\alpha}  \delta_{i}^{\alpha'}.
\end{eqnarray*}

\noindent {\bf Conditional Posterior of $(v_{\delta_u^{\sigma}}, v_{\delta_\epsilon^{\sigma}})$}: Same as Algorithm \ref{appalgo:benchmark.hetero.v.delta.sigma}.

\noindent {\bf Conditional Posterior of $(z_i^\rho,\delta_i^\rho)$}: Posteriors $(z_i^\rho,\delta_i^\rho)$ are similar to (\ref{appeq:benchmark.homo.post.z.rho} and \ref{appeq:benchmark.homo.post.delta.rho}), except for changes in the formulas to capture the heteroskedasticity:
\be
K_i^\rho
= \frac{q^\rho}{1-q^\rho} \left( \frac{v_{\delta^\rho}}{\bar{v}_{\delta_i^\rho}} \right)^{-\frac{1}{2}} \exp \left(  \frac{\bar{\delta}_i^{\rho^2}}{2\bar{v}_{\delta_i^\rho}} \right).
\ee
with
\begin{eqnarray*}
	\bar{v}_{\delta_i^\rho} & = & \left( v_{\delta^\rho}^{-1} + \sum_{t=1}^T  \sigma^{\epsilon^{-2}}_{it} s_{it-1}^2 \right)^{-1} , \quad
	\bar{\delta}_i^\rho  =  \bar{v}_{\delta_i^\rho} \sum_{t=1}^T  \sigma^{\epsilon^{-2}}_{it} s_{it-1} \widecheck{s}_{it}^\rho\\
	\widecheck{s}^\rho_{it} & = & s_{it}  - \rho s_{it-1}, \quad
	\sigma^{\epsilon}_{it}  = \sigma_{\epsilon,t} \sqrt{\delta_{i,\epsilon}^\sigma}.
\end{eqnarray*}


\noindent {\bf Conditional Posterior of $(z_{i,u}^\sigma, \delta_{i,u}^\sigma)$ and $(z_{i,\epsilon}^\sigma, \delta_{i,\epsilon}^\sigma)$} are similar to (\ref{appeq:benchmark.hetero.post.z.sigma} and \ref{appeq:benchmark.hetero.post.delta.sigma}) with moderate changes in the formulas.

For $m \in \{u, \epsilon \}$, the posterior odds of $z_{i,m}^\sigma = 1$ vs. $z_{i,m}^\sigma = 0$ are
\be
K_{i,m}^{\sigma}
= \frac{q^{\sigma_m}}{1-q^{\sigma_m}} 
\frac{\Gamma \left(\frac{\bar{\nu}_{\delta_m^{\sigma}}}{2} \right)}{\Gamma \left( v^{-1}_{\delta_m^\sigma} + 2\right)} \frac{\left( v^{-1}_{\delta_m^\sigma} + 1\right)^{v^{-1}_{\delta_m^\sigma} +2}}{\left(\frac{\bar{\tau}_{\delta_{i,m}^\sigma}}{2}\right)^{\frac{\bar{\nu}_{\delta_m^{\sigma}}}{2}}}
\exp \left( \frac{1}{2} \sum_{t=1}^T \sigma^{m^{-2}}_{it} \widecheck{y}_{it}^{\sigma_m^2} \right).
\ee
where
\begin{eqnarray*}
	\bar{\nu}_{\delta_m^{\sigma}} & = & 2v^{-1}_{\delta_m^{\sigma}} +4 + T, \\
	\bar{\tau}_{\delta_{i,m}^\sigma} & = & 2v^{-1}_{\delta_m^{\sigma}} + 2 + \sum_{t=1}^T \sigma^{m^{-2}}_{it} \widecheck{y}_{it}^{\sigma_m^2}, \\
	\widecheck{y}_{it}^{\sigma_u}  & = & y_{it} - H'_{it} (\alpha + \delta^{\alpha}_{i}) - s_{it}, \\
	\widecheck{y}_{it}^{\sigma_\epsilon}  & = & s_{it} - (\rho+\delta_i^\rho) s_{it-1}.
\end{eqnarray*}

Based on the posterior odds $K_{i,m}^{\sigma}$ we can draw $z_{i,m}^\sigma$ as follows:
\be
z_{i,m}^\sigma | (Y_{i}, s_{i},q^{\sigma_m},\sigma_m^2, \delta_{i}^\alpha, \rho, \delta_{i}^\rho,  v_{\delta_m^\sigma}) = \left\{ \begin{array}{ll} 1 & \mbox{with prob. } \frac{K_{i,m}^{\sigma}}{K_{i,m}^{\sigma}+1} \\
	0 & \mbox{with prob. } \frac{1}{K_{i,m}^{\sigma}+1} \end{array} \right. .
\ee

Overall, we obtain
\be
	\delta_{i,m}^{\sigma} | (Y_{i}, s_i, z_{i,m}^\sigma, \sigma_m^2,\delta_i^\alpha, \delta_i^\rho,  v_{\delta_m^\sigma}) \sim
\left\{ \begin{array}{ll} \text{IG} \left( \frac{\bar{\nu}_{\delta_m^{\sigma}}}{2},  \frac{\bar{\tau}_{\delta_{i,m}^\sigma}}{2} \right)
	& \mbox{if } z_{i,m}^\sigma = 1 \\
	1 & \mbox{if } z_{i,m}^\sigma = 0 \end{array} \right.
\ee

\noindent {\bf Conditional Posterior of $\mu_{s_0}$ and $v_{s_0}$}:  
\be
	\mu_{s_0} | (s_{i0}, v_{s_0}) \sim {\cal N}(\bar{\mu}_{s_0}, \bar{v}_{s_0}),
\ee
where 
\begin{eqnarray*} 
	\bar{v}_{s_0} &=& \left( \underline{v}^{-1}_{s_0} + N v^{-1}_{s_0} \right)^{-1} \\
	\bar{\mu}_{s_0} & = & \bar{v}_{s_0} \left( \underline{v}^{-1}_{s_0} \underline{\mu}_{s_0} + v^{-1}_{s_0} \sum_{i=1}^N s_{i0} \right).
\end{eqnarray*}

\be
	v_{s_0} | (s_{i0}, \mu_{s_0}) \sim \text{IG} \left(\frac{\bar{\nu}_{s_0}}{2}, \frac{\bar{\tau}_{s_0, i}}{2} \right),
\ee
where
\begin{eqnarray*}
	\bar{\nu}_{s_0} & = & \underline{\nu}_{s_0} + N \\
	\bar{\tau}_{s_0,i} & = & \underline{\tau}_{s_0} + \sum_{i=1}^N \left(  s_{i0} - \mu_{s_0} \right)^2.
\end{eqnarray*}

\noindent {\bf Conditional Posterior of $s_{it}$ and $(z^{\alpha}_{i}, \delta^{\alpha}_{i})$}: 
We choose to draw $s_{it}$ and $\delta^{\alpha}_{i}$ jointly with the following representation:
\[
	\widecheck{y}^\beta_{it} = y_{it} - H'_{it} \alpha = X'_{it} \beta_i + \sigma^{u}_{it} u_{it},
\]
and its matrix form
\[
	\widecheck{Y}^\beta_{i} = X_{i} \beta_i + \Sigma^{\frac{1}{2}}_{i,u} u_{i},
\]
where the design matrix $X_{i}$ contains $H_{it}$ and dummies for $s_{it}$:
\be
	X_{i} =
	\begin{bmatrix}
		1 & 1  & 1 & 0 & \cdots & 0\\
		1 & 2  & 0 & 1 & \cdots & 0\\
		\vdots & \vdots  & \vdots & \vdots & \ddots & \vdots\\
		1 & T  & 0 & 0 & \cdots & 1\\
	\end{bmatrix}
\ee
$\beta_i = [\delta^{\alpha}_{i0}, \; \delta^\alpha_{i1}, \; s_{i1}, \; ... \;, s_{iT}]'$ and $\Sigma_{i,u} = \delta^{\sigma}_{i, u} \diag \left(\sigma^{2}_{u,1}, \sigma^{2}_{u,2},...,\sigma^{2}_{u,T} \right)$. 

We construct the prior for $\beta_i$ to account for the sparse assumption in $\delta^{\alpha}$, which given by 

\begin{eqnarray} 
	\beta_i |  (\rho, \delta^{\rho}_i, z_i^\alpha, q^{\alpha}, v_{\delta^{\alpha}}, D_{i,\epsilon}) &\sim&  N \left( 
	\begin{bmatrix}
		0_{2 \times 1} \\
		0_{T \times 1} \\
	\end{bmatrix}, 
	\begin{bmatrix}
		z_i^\alpha v_{\delta^{\alpha}} & 0_{2 \times T} \\
		0_{T \times 2} & \underline{V}_{s_i} \\
	\end{bmatrix} 
	\right) \\
	z_i^\alpha|q^{\alpha} &=& \left\{ \begin{array}{ll}  0 & \mbox{with prob. } 1-q^\alpha\\ 1 & \mbox{with prob. } q^\alpha \end{array} \right. \; , \nonumber
\end{eqnarray}

The $(t,\tau)$ elements of the $T \times T$ prior covariance matrix $\underline{V}_{s_i}$ can be calculated as follows:
\begin{eqnarray*} 
	\underline{V}_{s_i}(t,t) &=& (\rho + \delta^{\rho}_{i})^2 \underline{V}_{s_i}(t-1,t-1) + \sigma^2_{\epsilon,t} \delta^{\sigma}_{i,\epsilon}, \quad t=1,\ldots,T \\
	\underline{V}_{s_i}(t,\tau) &=& (\rho + \delta^{\rho}_{i})^{|t - \tau|} \underline{V}_{s_i}\big(\min(t,\tau),\min(t,\tau)\big), \quad
	t=1,\ldots,T \; \mbox{and} \; \tau \not=t,
\end{eqnarray*}
with the understanding that $\underline{V}_{s_i}(0,0)=v_{s_0}$.


Under $z_i^\beta = 1$, $\beta_i$ is set to jointly follow a normal distribution. The marginalized likelihood of $Y_{i}|( \rho, \delta^{\rho}, v_{\delta^{\alpha}}, \sigma^2_u, \sigma^2_e, z_i^\beta = 1)$ can be obtained by rearranging Bayes Theorem:
\be
p(Y_{i}|  \rho, \delta^{\rho}, v_{\delta^{\alpha}}, \sigma^2_u,  \sigma^2_e, z_i^\beta=1) =
\frac{p(Y_{i}|   \rho, \delta^{\rho}, v_{\delta^{\alpha}}, \sigma^2_u,  \sigma^2_e, \beta_i,  z_i^\alpha=1)p(\beta_i| \rho, \delta^{\rho}, v_{\delta^{\alpha}}, \sigma^2_u,  \sigma^2_e, z_i^\alpha=1)}{p(\beta_i | Y_{i},  \rho, \delta^{\rho}, v_{\delta^{\alpha}}, \sigma^2_u,  \sigma^2_e, z_i^\alpha=1)}.
\ee
It is given by
\begin{eqnarray*}
	p(Y_{i}| \rho, \delta^{\rho}, v_{\delta^{\alpha}}, \sigma^2_u,  \sigma^2_e, z_i^\beta=1)
	& = & \frac{ (2\pi )^{-\frac{T}{2}} |\Sigma_{i,u}|^{-\frac{1}{2}} \exp \left( - \frac{1}{2} \widecheck{Y}^{\beta'}_{i} \Sigma^{-1}_{i,u} \widecheck{Y}^{\beta}_{i} \right) \left(2 \pi\right)^{-\frac{T+2}{2}}  |\underline{v}_{\beta_i}|^{-\frac{1}{2}} }
	{\left(2 \pi\right)^{-\frac{T+2}{2}}  |\bar{v}_{\beta_i}|^{-\frac{1}{2}}  \exp \left( - \frac{1}{2}\bar{\beta}_i' \bar{v}^{-1}_{\beta_i} \bar{\beta}_i \right) } \\
	& = & (2\pi )^{-\frac{T}{2}} |\Sigma_{i,u}|^{-\frac{1}{2}}  \left( \frac{|\underline{v}_{\beta_i} | }{|\bar{v}_{\beta_i}|} \right)^{-\frac{1}{2}} \exp \left(- \frac{1}{2} \widecheck{Y}^{\beta'}_{i} \Sigma^{-1}_{i,u} \widecheck{Y}^{\beta}_{i} + \frac{1}{2}\bar{\beta}_i' \bar{v}^{-1}_{\beta_i} \bar{\beta}_i \right),
\end{eqnarray*}
where $\bar{\beta}_i$ and $\bar{v}_{\beta_i}$ are posterior mean and variance of $\beta_i$:
\[
\bar{v}_{\beta_i} =  \left(\underline{v}_{\beta_i}^{-1} + X_{i}' \Sigma_{i,u}^{-1} X_{i} \right)^{-1}, \quad
\bar{\beta}_i  =  \bar{v}_{\beta_i} \left( X_{i}' \Sigma_{i,u}^{-1} \widecheck{Y}^{\beta}_{i}  \right).
\]
For $z_i^\beta = 0$ we have $\beta_i' = [0, \; 0, \; 0, \; s_{i1}, \; ... \;, s_{iT}]$. Then,
\begin{eqnarray*}
	p(Y_{i}| \rho, \delta^{\rho}, \sigma^2_u, \sigma^2_e, z_i^\beta=0)
	& = & \frac{ (2\pi )^{-\frac{T}{2}} |\Sigma_u|^{-\frac{1}{2}} \exp \left( - \frac{1}{2} \widecheck{Y}^{\beta'}_{i} \Sigma^{-1}_{i,u} \widecheck{Y}^{\beta}_{i} \right) \left(2 \pi \underline{V}_{s_i} \right)^{-\frac{1}{2}} }
	{\left(2 \pi \overline{V}_{s_i} \right)^{-\frac{1}{2}} \exp \left( - \frac{1}{2}\bar{s}_i' \bar{v}^{-1}_{s_i} \bar{s}_i \right) } \\
	& = & (2\pi )^{-\frac{T}{2}} |\Sigma_u|^{-\frac{1}{2}}  \left( \frac{|\underline{V}_{s_i} | }{|\overline{V}_{s_i}|} \right)^{-\frac{1}{2}} \exp \left(- \frac{1}{2} \widecheck{Y}^{\beta'}_{i} \Sigma^{-1}_{i,u} \widecheck{Y}^{\beta}_{i} + \frac{1}{2}\bar{s}_i' \bar{v}^{-1}_{s_i} \bar{s}_i \right),
\end{eqnarray*}
where $\bar{s}_i$ and $\overline{V}_{s_i}$ are posterior mean and variance of $s_i$ when $\delta^{\alpha}_{i} = 0$:
\[
\overline{V}_{s_i} =  \left(\underline{V}_{s_i}^{-1} +  \Sigma^{-1}_{i,u} \right)^{-1}, \quad
\bar{s}_i  =  \overline{V}_{s_i} \left( \Sigma^{-1}_{i,u} \widecheck{Y}^{\beta}_{i}  \right).
\]

We conclude that the posterior odds of $z_i^\beta = 1$ vs. $z_i^\beta = 0$ are
\be
K_i^\beta
=  \frac{q^\beta}{1-q^\beta}  \left( \frac{|\underline{v}_{\beta_i}| |\overline{V}_{s_i}|}{ |\bar{v}_{\beta_i}| |\underline{V}_{s_i} |} \right)^{-\frac{1}{2}}  \exp \left(  \frac{1}{2}\bar{\beta}_i' \bar{v}^{-1}_{\beta_i} \bar{\beta}_i - \frac{1}{2}\bar{s}_i' \overline{V}^{-1}_{s_i} \bar{s}_i \right).
\ee
We can draw $z_{i}^\beta$ as follows:
\be
z_{i}^\beta|(Y_{i}, \rho, \delta^{\rho}, v_{\delta^{\alpha}}, \sigma^2_u, \sigma^2_e) = \left\{ \begin{array}{ll} 1 & \mbox{with prob. } \frac{K_i^\beta}{K_i^\beta+1} \\
	0 & \mbox{with prob. } \frac{1}{K_i^\beta+1} \end{array} \right. .
\ee
Moreover,
\be
\beta_i|(Y_{i},z_i^\beta, \rho, \delta^{\rho}, v_{\delta^{\alpha}}, \sigma^2_u, \sigma^2_e) \sim
\left\{ \begin{array}{ll} N \left( \bar{\beta}_i,\bar{v}_{\beta_i} \right)
	& \mbox{if } z^\beta_i = 1 \\
	 \delta^{\alpha}_{i} = 0 \text{ and } s_i \sim N \left( \bar{s}_i,\overline{V}_{s_i} \right)  & \mbox{if } z^\beta_i = 0 \end{array} \right. .
\ee

\noindent {\bf Conditional Posterior of $s_{i0}$}:  We sample $s_{i0}$ using simulation smoother conditional on the posterior draws of $s_{i1}$ and other parameters:
\be
	s_{i0}| (s_{i1}, \mu_{s_0}, v_{s_0}, \rho, \delta^{\rho}, \sigma^2_u) \sim {\cal N}(\bar{s}_{i, 0|1}, P_{i, 0|1}),
\ee
where 
\begin{eqnarray*}
	\bar{s}_{i, 0 | 1} & = & \bar{s}_{i, 0 | 0}  + P_{i, 0|0} \Phi'_{i1} P_{i, 1 | 0} ^{-1}  \left( s_{i1} - \Phi_{i1} \bar{s}_{i, 0 | 0} \right), \\
	P_{i, 0 | 1} & = &  P_{i, 0 | 0} - P_{i, 0|0} \Phi'_{i1} P_{i, 1 | 0} ^{-1} \Phi_{i1} P_{i, 0|0}, \\
	P_{i, 1|0}  & = & \Phi_{i1}'  P_{i, 0|0} \Phi_{i1} + \sigma^{\varepsilon^2}_{i0}, \\
	\Phi_{i1} & = & \rho + \delta^{\rho}_i, \\
	\bar{s}_{i, 0 | 0} & = & \underline{\mu}_{s_0}, \\
	P_{i, 0|0}  & = & \underline{v}_{s_0}.
\end{eqnarray*}

\noindent {\bf Conditional Posterior of $( \sigma_{u}^2, \sigma_{\epsilon}^2 )$}: Posteriors for $( \sigma_{u}^2, \sigma_{\epsilon}^2 )$ are similar to (\ref{appeq:benchmark.homo.post.sigma}), except for a slight change in the formulas to reflect the time-variation in variance:
\be
\sigma_{u, t}^2| (y, \delta) \sim \text{IG} \left(\frac{\bar{\nu}_{\sigma_u}}{2}, \frac{\bar{\tau}_{\sigma_{u,t}}}{2} \right),
\ee
where
\begin{eqnarray*}
	\bar{\nu}_{\sigma_u} & = & \underline{\nu}_{\sigma} + N, \\
	\bar{\tau}_{\sigma_{u,t}} & = & \underline{\tau}_{\sigma_u} + \sum_{i=1}^N \left[ y_{it} -  H'_{it} \left(\alpha + \delta^{\alpha}_i \right)- s_{it}\right]^2 / \delta^\sigma_{i,u},
\end{eqnarray*}
and 
\be
\sigma_{\epsilon, t}^2|( s, \rho, \delta) \sim \text{IG} \left(\frac{\bar{\nu}_{\sigma_\epsilon}}{2}, \frac{\bar{\tau}_{\sigma_{\epsilon,t}}}{2} \right),
\ee
where
\begin{eqnarray*}
	\bar{\nu}_{\sigma_\epsilon} & = & \underline{\nu}_{\sigma} + N, \\
	\bar{\tau}_{\sigma_{\epsilon,t}} & = & \underline{\tau}_{\sigma_\epsilon} + \sum_{i=1}^N \left[ s_{it} - \left(\rho + \delta_{i}^{\rho}\right) s_{it-1} \right]^2 / \delta^\sigma_{i,\epsilon}.
\end{eqnarray*}

\clearpage 
\section{Additional Empirical Results}
\label{appsec:empirics}

Table \ref{apptab:empirics.core.group} shows the numbers of units that share the same posterior median, i.e., how long the flat segments in Figure~\ref{fig:empirics.post.hip.hetsk.T20} are. $N$ is the number of individuals in the subsample and $N_l$ is the size of core $l$ groups, $l \in \{\alpha, \rho, \sigma\}$. 

\begin{table}[!h]
	\caption{Size of Core Groups}
	\label{apptab:empirics.core.group}
	\begin{center}
		\begin{tabular}{ c c | c c c c c c c c}
		 \toprule \toprule
			$\tau$ & $N$ & $N_\alpha$ & $N_\alpha / N$ & $N_\rho$ & $N_\rho / N$ & $N_{\sigma_u}$ & $N_{\sigma_u} / N$ & $N_{\sigma_e}$ & $N_{\sigma_e} / N$ \\ \midrule
			1988 & 82 & 74 & 90.24\% & 82 & 100\% & 0 & 0\% & 0 & 0\% \\
			1991 & 88 & 32 & 36.36\% & 87 & 98.86\% & 0 & 0\% & 0 & 0\% \\ \bottomrule
		\end{tabular}
	\end{center}
\end{table}

\begin{figure}[h!]
	\caption{Prior and Posterior Densities for $q$, SS-HIP-Hetsk, $T=20$}
	\label{appfig:empirics.post.q.hip.hetsk.T20}
	\begin{center}
		\scalebox{.9}{%
			\begin{tabular}{cc}
				$\tau = 1988$ & $\tau = 1991$ \\
				\includegraphics[width=0.45\textwidth]{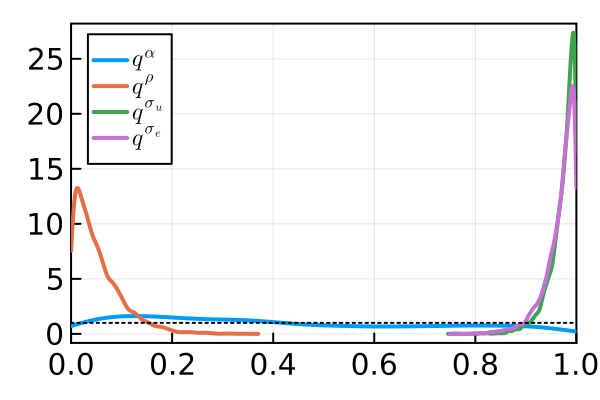} & 
				\includegraphics[width=0.45\textwidth]{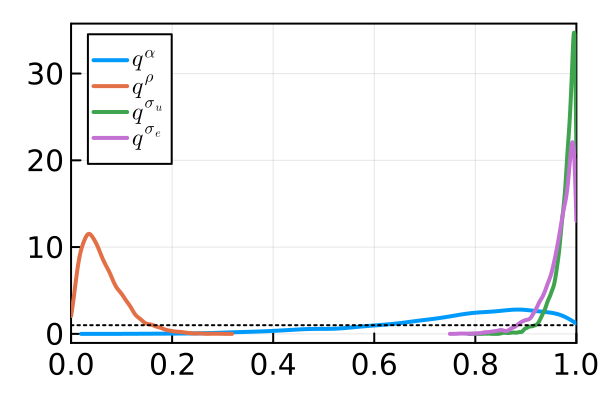}
			\end{tabular}
		}
	\end{center}
	{\footnotesize {\em Notes}: We overlay a Beta(1,1) prior density for $q^l_i$ (black, dotted) and posterior densities for $q^\alpha$ (blue, solid), and $q^\rho$ (red, solid).}\setlength{\baselineskip}{4mm}
\end{figure}

In Figure~\ref{appfig:empirics.post.q.hip.hetsk.T20} we plot prior densities, which we take to be uniform, and posterior densities for the $q$ parameters that determines the probability of deviating from the core group. The posterior mode for $q^\rho$ is close to zero for both samples, which is consistent with the $\rho_i$ estimates being identical for most units, as could be seen in Figure~\ref{fig:empirics.post.hip.hetsk.T20}. The posterior density for $q^\alpha$ stayed close to the prior density and is fairly flat. It has a small mode near 0.1 for the 1988 sample, and a small mode near 0.9 for the 1991 sample.

\clearpage

\begin{figure}[h!]
	\caption{Post Prob. of Unit $i$ Belonging to Core Group, SS-HIP-Hetsk, $T=20$}
	\label{appfig:empirics.post.z.hip.hetsk.T20}
	\begin{center}
		\begin{tabular}{ccc}
			& $\tau = 1988$ & $\tau = 1991$ \\
			\rotatebox{90}{\hspace*{0.35in} $\alpha$ Group} &
			\includegraphics[width = 0.35\textwidth]{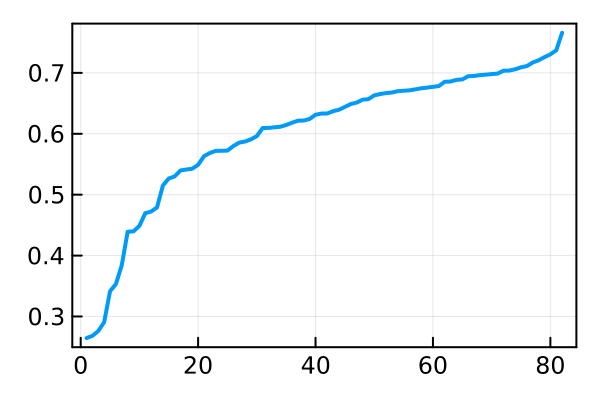} &
			\includegraphics[width=0.35\textwidth]{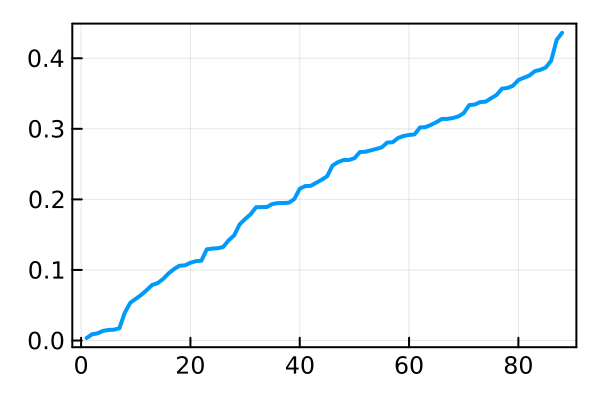} \\
			
			\rotatebox{90}{\hspace*{0.35in} $\rho$ Group} &
			\includegraphics[width=0.35\textwidth]{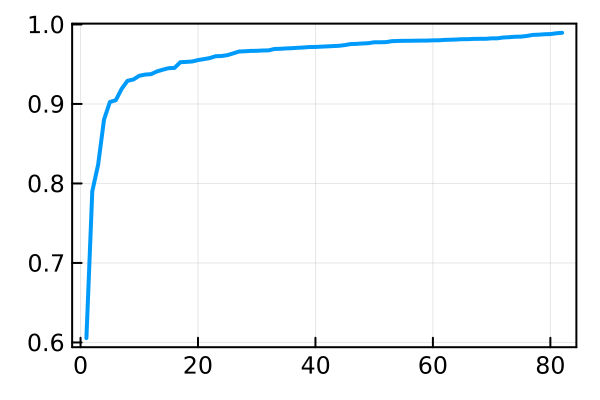} & 
			\includegraphics[width=0.35\textwidth]{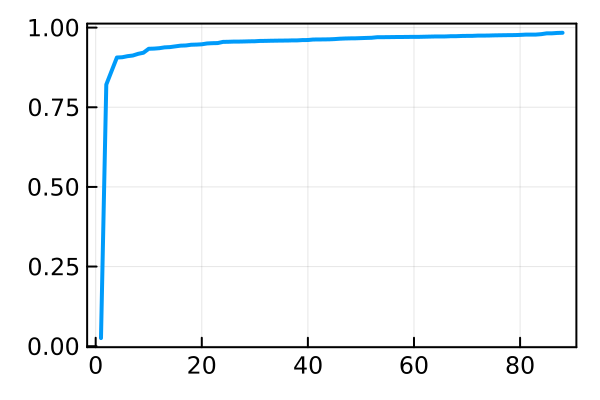} \\
			
			\rotatebox{90}{\hspace*{0.35in} ${\sigma}_{u}$ Group} &
			\includegraphics[width=0.35\textwidth]{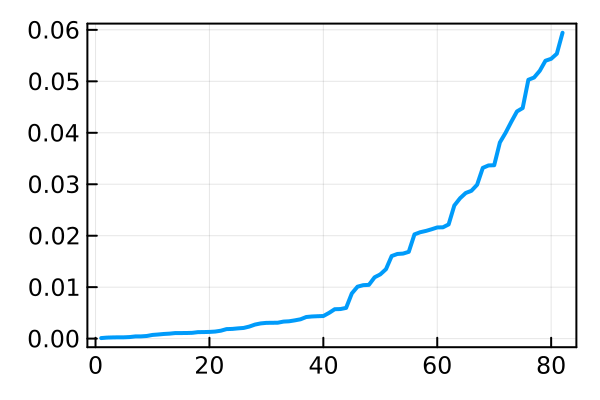} & 
			\includegraphics[width=0.35\textwidth]{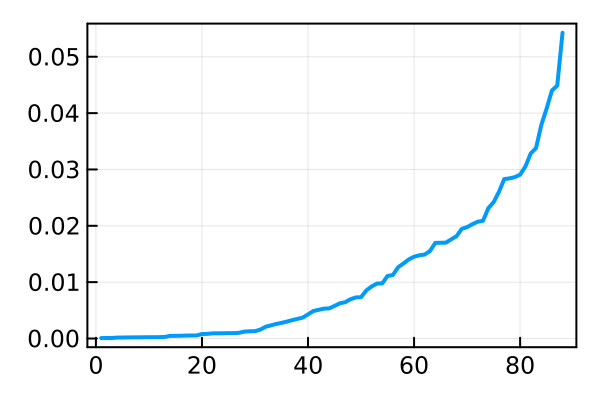}\\
			
			\rotatebox{90}{\hspace*{0.35in} $\sigma_{\epsilon}$ Group} &
			\includegraphics[width=0.35\textwidth]{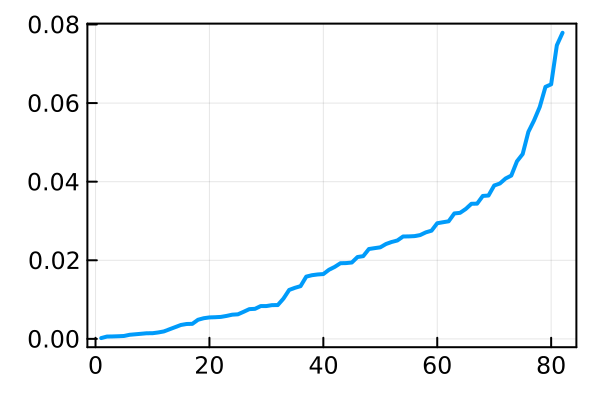} & 
			\includegraphics[width=0.35\textwidth]{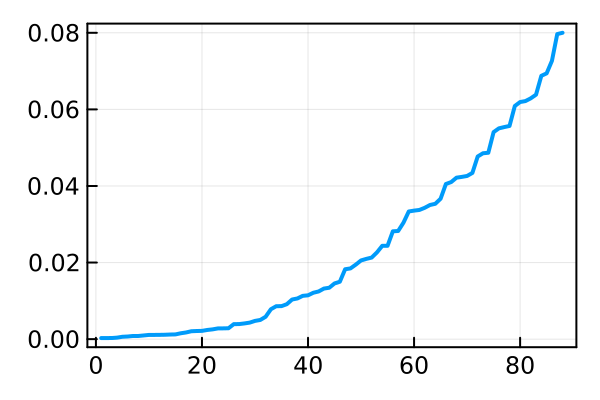}
		\end{tabular}
	\end{center}
\end{figure}

In Figure~\ref{appfig:empirics.post.z.hip.hetsk.T20} we plot for each unit the posterior probability of belonging to the core group, defined as $\mathbb{P}\{z_i^l=0|Y_{1:N,1:T}\}$.

\clearpage

\begin{figure}[h!]
	\caption{Posterior Distribution of $\hat{s}_{i0}$ and $\hat{s}_{iT}$, SS-HIP-Hetsk, $T=20$}
	\label{appfig:empirics.post.s}
	\begin{center}
		\begin{tabular}{ccc}
			& $\tau = 1988$ & $\tau = 1991$ \\
			\rotatebox{90}{\hspace*{0.525in}  $\hat{s}_{i0}$} &
			\includegraphics[width = 0.4\textwidth]{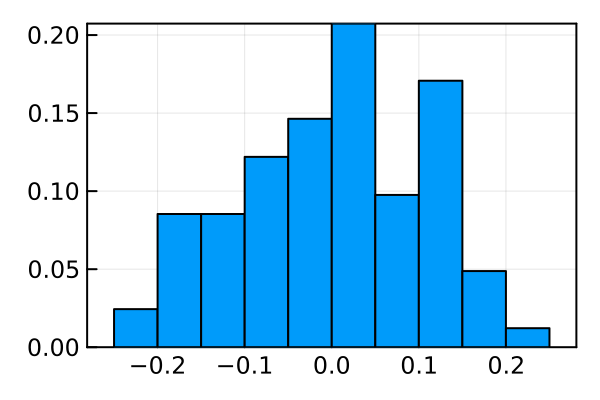} &
			\includegraphics[width = 0.4\textwidth]{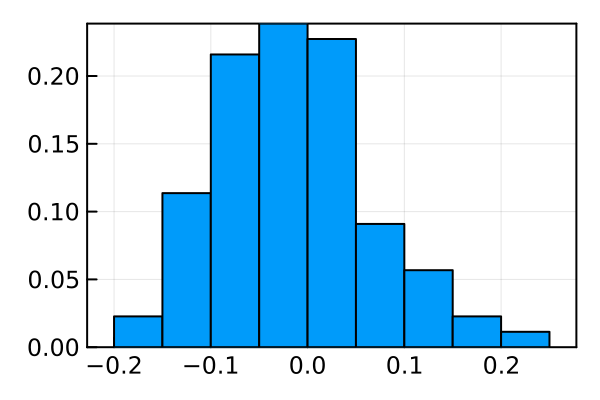} \\		
			\rotatebox{90}{\hspace*{0.525in}  $\hat{s}_{iT}$} &
			\includegraphics[width = 0.4\textwidth]{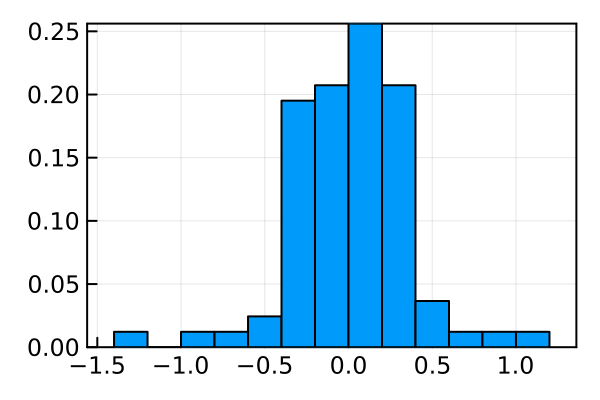} &
			\includegraphics[width = 0.4\textwidth]{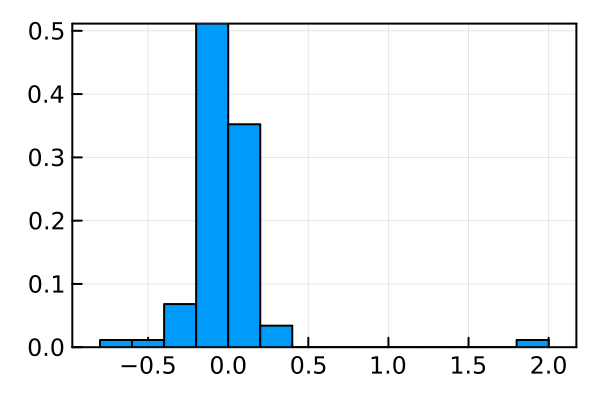}
		\end{tabular}
	\end{center}
	{\footnotesize {\em Notes}: $\hat{s}_{i0}$ and $\hat{s}_{iT}$ are posterior mean estimates.}\setlength{\baselineskip}{4mm}
\end{figure}

In Figure~\ref{appfig:empirics.post.s} we plot the cross-sectional distribution of the posterior mean estimates $\hat{s}_{i0}$ and $\hat{s}_{iT}$.

\begin{table}[h!]
	\caption{Probability Distributions of Experience in the Starting Period}
	\label{apptab:empirics.experience.dist}
	\begin{center}
		\begin{tabular}{cccccccc}
			\toprule
			Subsample & Mean & Std & Min & 25\% & Median & 75\% & Max\\ 
			\midrule 
			1988 & 10.27 & 4.22 & 2 & 7 & 10 & 13 & 20 \\ 
			1991 & 9.39 & 3.92 & 2 & 7 & 9 & 11 & 20 \\ 
			\bottomrule
		\end{tabular}
	\end{center}
\end{table}	

Table \ref{apptab:empirics.experience.dist} provides descriptive statistics of the distribution of years of experience in the starting period of the two samples considered in the main text.

\end{appendix}

\end{document}